%% file: Main.tex
\documentclass[journal=jpclcd,layout=traditional, manuscript=letter]{achemso}

\usepackage[version=3]{mhchem} 
\usepackage{caption}
\usepackage{amssymb,amsmath,amsthm,amsfonts,amsbsy,mathrsfs, nccmath}
\usepackage{subcaption}
\usepackage{array}
\newcolumntype{M}[1]{>{\centering\arraybackslash}m{#1}}
\usepackage{tabularx}
\usepackage[dvipsnames]{xcolor}
\usepackage{multirow}
\usepackage{textcomp}
\usepackage[normalem]{ulem}
\usepackage{bm,bbm}
\usepackage{xr-hyper}
\usepackage{hyperref}
\hypersetup{
    colorlinks=true,
    linkcolor=blue,
    citecolor=blue}
\setkeys{acs}{doi = true}

\usepackage{tikz}
\usepackage{pgfplots}
\usepackage{pgfplotstable}
\usetikzlibrary{shapes.geometric, arrows}
\usetikzlibrary{patterns}
\usetikzlibrary{positioning}
\usetikzlibrary{external}
\usetikzlibrary{matrix,calc,fit}

\pgfplotsset{compat=1.7}

\definecolor{myturq}{RGB}{157,209,199}
\definecolor{mypurple}{RGB}{145, 128, 172}
\definecolor{mylilac}{RGB}{217, 189, 216}
\definecolor{myorange}{RGB}{229, 133, 121}
\definecolor{mybabyblue}{RGB}{138, 177, 210}
\tikzstyle{startstop} = [rectangle, rounded corners, line width = 0.6pt, minimum width=3cm, minimum height=1cm,text centered, draw=black, fill=red!30]
\tikzstyle{long_startstop} = [rectangle, rounded corners, line width = 0.6pt, minimum width=4cm, minimum height=1cm, text width=4cm, text centered, draw=black, fill=red!30]
\tikzstyle{io} = [trapezium, trapezium left angle=70, trapezium right angle=110, line width = 0.6pt, minimum width=3cm, minimum height=1cm, text centered, draw=black, fill=blue!30]
\tikzstyle{process} = [rectangle, line width = 0.6pt, minimum width=3cm,  text width=3cm, minimum height=1cm, text centered, draw=black, fill=orange!30]
\tikzstyle{short_process} = [rectangle, line width = 0.6pt, minimum width=2.5cm,  text width=2.5cm, minimum height=1cm, text centered, draw=black, fill=orange!30]
\tikzstyle{long_process} = [rectangle, minimum width=4cm, line width = 0.6pt, minimum height=1cm, text centered, text width=4cm, draw=black, fill=orange!30]
\tikzstyle{very_long_process} = [rectangle, minimum width=5cm, line width = 0.6pt, minimum height=1cm, text centered, text width=5cm, draw=black, fill=orange!30]
\tikzstyle{decision} = [diamond, line width = 0.6pt, minimum width=3cm, minimum height=1cm, text centered, aspect=3, draw=black, fill=green!30]
\tikzstyle{arrow} = [thick,->,>=stealth]

\author{Shivam Sharma}
\affiliation[uiuc]
{Department of Mechanical Science and Engineering, University of Illinois at Urbana-Champaign, IL  61801, USA}
\author{Chenhaoyue Wang}
\affiliation[ucla-mse]
{Department of Materials Science and Engineering, University of California, Los Angeles, CA 90095, USA}
\author{Hsuan Ming Yu}
\affiliation[ucla-mse]
{Department of Materials Science and Engineering, University of California, Los Angeles, CA 90095, USA}
\author{Amartya S. Banerjee}
\email{asbanerjee@ucla.edu}
\affiliation[ucla-mse]
{Department of Materials Science and Engineering, University of California, Los Angeles, CA 90095, USA}

\title[P2C3]{Dirac Fermions and Flat Bands in Phosphorus Carbide Nanotubes: Structural and Quantum Phase Transitions in a Quasi-One-Dimensional Material}


\input{preamble}

\begin{document}

\begin{tocentry}





\includegraphics[scale=0.27,trim={7.5cm 0cm 7cm 0cm},clip]{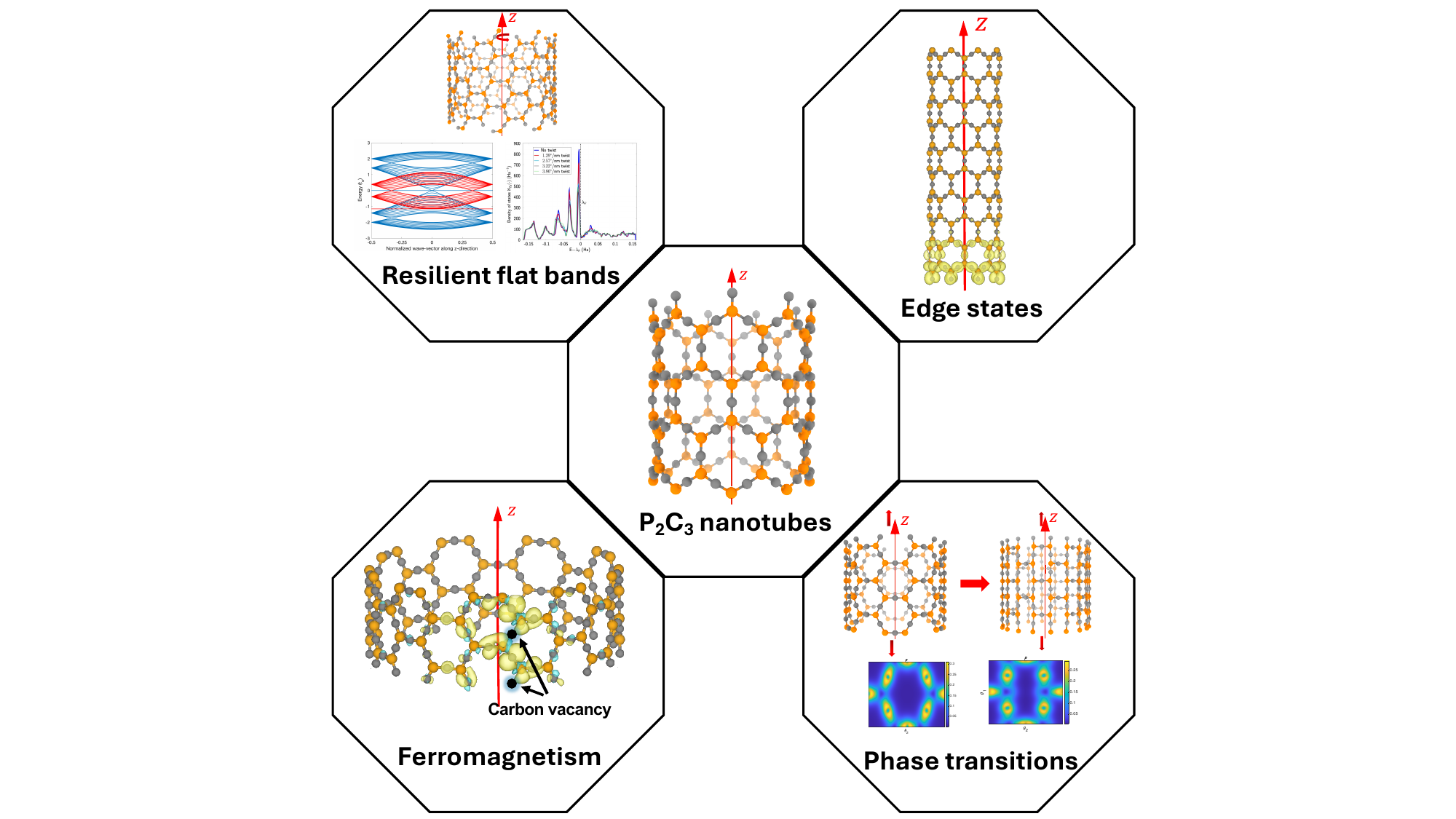}
\end{tocentry}
\begin{abstract}
\noindent Chemically realistic quasi-one-dimensional (1D) materials in which Dirac fermions and highly degenerate flat bands coexist intrinsically at the Fermi level are exceedingly rare, while representing a highly desirable platform for correlated and topological quantum phenomena. Here, using specialized symmetry-adapted first-principles calculations we predict a new class of nanomaterials --- phosphorus carbide nanotubes (\ce{P2C3}NTs) --- obtained by rolling monolayer \ce{P2C3}, a two-dimensional material shown in a previous letter to host ``double Kagome bands''. Both armchair and zigzag \ce{P2C3}NTs are stable at room temperature and feature the rare coexistence of Dirac crossings and multiple flat bands at the Fermi level inherited from the underlying honeycomb–Kagome lattice, with the flat bands resilient to elastic deformations. Under large strain, the structure transforms from honeycomb–Kagome to ``brick-wall,'' accompanied by multiple coupled structural and quantum phase transitions. We also uncover localized edge states, spin splitting from vacancies and dopants, and strain-tunable magnetism. Together, these results establish \ce{P2C3}NTs as a chemically specific and mechanically tunable 1D material platform with potential applications in quantum hardware and spintronics.
\end{abstract}
A significant amount of contemporary materials research is directed towards the discovery, synthesis, and characterization of nanomaterials and nanostructures featuring exotic electronic states. Such materials can manifest remarkable and unusual physical properties, leading to promising applications in quantum technologies, spintronic devices, and next-generation microelectronics \citep{tokura2017emergent, de2021materials, giustino20212021, basov2017towards, aiello2022chirality, keimer2017physics, bauer2020quantum, ball2017quantum, mas20112d}. Two well-known examples of such electronic states, contrasting conventional parabolic dispersion in common semiconductors, are ones exhibiting linear dispersion\citep{novoselov2005two} (e.g. Dirac cones in graphene) and ones without dispersion (i.e. electronic \emph{flat bands}, e.g. in Kagome lattices \citep{li2018realization}). The former is associated with massless fermions with high carrier mobility, leading to unconventional electronic \citep{neto2009electronic}, transport \citep{peres2010colloquium},
optical \citep{falkovsky2008optical} and topological properties \citep{gomes2012designer,liu2011quantum}. The latter is often associated with infinitely massive Fermions with quenched kinetic energies and spatially localized electronic states that interact in the strongly correlated regime. Such interaction leads to fascinating electronic phases with collective properties \citep{derzhko2015strongly},  e.g. superconductivity \citep{balents2020superconductivity,iglovikov2014superconducting,peotta2015superfluidity},   ferromagnetism \citep{pollmann2008kinetic,pons2020flat}, Wigner crystallization \citep{wu2007flat}, and the fractional quantum Hall effect \citep{tang2011high,sun2011nearly}.

Bulk and nanomaterials featuring the Kagome lattice host both these types of electronic states, thus driving a wide variety of interesting properties associated with these materials, and leading to a proliferation of studies on them in recent years \citep{ortiz2021superconductivity,kim2023monolayer,wang2023quantum,ortiz2020cs,jiang2021unconventional,PhysRevLett.127.046401, yu2024carbon}. Generally, the Dirac crossings and flat bands in Kagome materials do not appear simultaneously at the Fermi level \citep{du2017quadratic, uebelacker2011instabilities, barreteau2017bird, ortiz2021superconductivity,kim2023monolayer,wang2023quantum}  --- instead, their electronic structure usually exhibits  \emph{quadratic band touching} points. However, from the perspective of applications, the simultaneous presence of Dirac cones and flat bands at the Fermi level can be particularly intriguing: Dirac cones contribute to the emergence of electronic edge states, while flat bands promote strongly correlated behavior, rendering the material multifunctional. Despite this promise, chemically realistic materials in which Dirac points and highly degenerate flat bands coexist intrinsically at the Fermi level remain exceedingly rare. In most known Dirac materials, such as graphene and carbon nanotubes, the electronic structure near the Fermi level is dominated by linearly dispersing bands with a vanishing density of states. Conversely, flat-band systems based on Kagome or related lattices typically exhibit quadratic band-touching points or flat bands that are energetically displaced from the Fermi level. As a result, achieving the coexistence of Dirac fermions and dispersionless electronic states at the Fermi level generally requires fine-tuning through external fields, chemical modification, or artificial lattice engineering. 

Motivated by this challenge, a number of studies have explored the Honeycomb-Kagome (HK) and Lieb lattices, which do exhibit coexisting Dirac dispersions and flat bands, at least in idealized tight-binding models \citep{mizoguchi2020square, jiang2019topological}. In particular, the HK lattice (also referred to as the Honeycomb Splitgraph \citep{ma2020spin}) can be envisioned as a combination of conventional honeycomb and Kagome lattices, and its electronic states can be deduced accordingly   \citep{mizoguchi2020square,mizoguchi2021square,ma2020spin}. While realizations of the Lieb lattice in realistic 2D materials remain scarce, a number of recent contributions have computationally explored planar materials of the type \ce{A2B3}, that structurally feature the HK lattice. Most thermodynamically stable materials proposed in this category  constitute metal oxides \citep{mellaerts2021two,hashmi2020ising,zhang2017intrinsic,wang2022first} or carbides \citep{wang2017new,ji2017planar,pan2018half,liu2018hexagonal}, with some of these materials having  been associated with fascinating properties such as non-trivial topological phases and half metallicity. Although these materials exhibit Dirac cones near the Fermi level, they either lack a flat band at the Fermi level or exhibit one that is misplaced. Such deviations arise because the specific chemical characteristics of the atoms involved --- including, orbital hybridization states and different on-site energies --- can cause differences in the electronic structure from idealized tight-binding models of HK lattices.

A notable exception to these limitations was reported in a previous letter to this journal by Huang et al.~\citep{huang2018double}, who introduced a new form of phosphorus carbide (\ce{P2C3}) featuring so-called ``double Kagome bands.'' In this two-dimensional material with HK geometry, the out-of-plane $p_z$ orbitals of both phosphorus and carbon atoms combine to produce the rare coexistence of Dirac fermions and strongly correlated dispersionless electronic states directly at the Fermi level. In this letter, we extend this work by studying \ce{P2C3} nanotubes (\ce{P2C3}NTs), thereby exploring a far less studied class of nanomaterials (i.e., quasi-one-dimensional or 1D nanostructures) with fascinating properties.  From the perspective of quantum materials design, the realization of such Dirac--flat-band coexistence in a quasi-one-dimensional geometry is particularly appealing, as reduced dimensionality enhances electronic interactions and enables strong coupling between mechanical deformation, electronic structure, and emergent quantum phases. Our findings unveil a range of noteworthy electronic, structural, and magnetic characteristics unique to these nanotubes. While \ce{P2C3} nanotubes are yet to be synthesized, we anticipate that our computational and theoretical investigations will provide impetus for future experimental efforts \citep{hu2021metallic,li2022efficient}. Indeed, both phosphorus and carbon are already well-known for their ability to form a large number of elemental allotropes, and a number of varieties of stable monolayer phosphorus carbide have also been investigated in recent years \citep{guan2016two, kistanov2020point, ma2021phase, chen2022two}. Relatedly, some studies have suggested the possibility of creating quasi-one-dimensional forms of such compounds \citep{shcherbinin2020two, kistanov2020starfish, shcherbinin2020dynamical}. These previous findings not only highlight the growing interest in low-dimensional materials with unique properties, but also lend support to the experimental synthesizability of \ce{P2C3}NTs in the near future, thereby bolstering the relevance of the current work (also see Supplementary Information).

\begin{figure}[ht!]
\centering
  \subfloat[]{\includegraphics[trim={13cm 6cm 13cm 6cm}, clip, width=.27\linewidth]{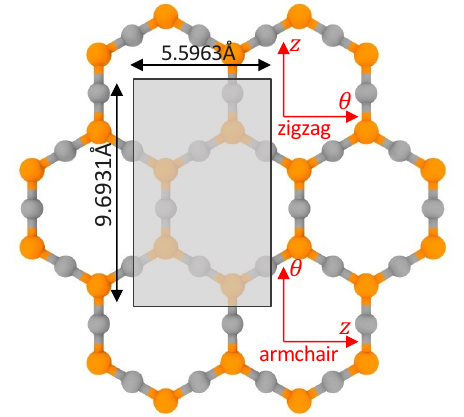}\label{fig:p2c3_2d}} 
    \subfloat[]{\includegraphics[trim={7cm 0.5cm 6cm 0.5cm}, clip, width=.23\linewidth]{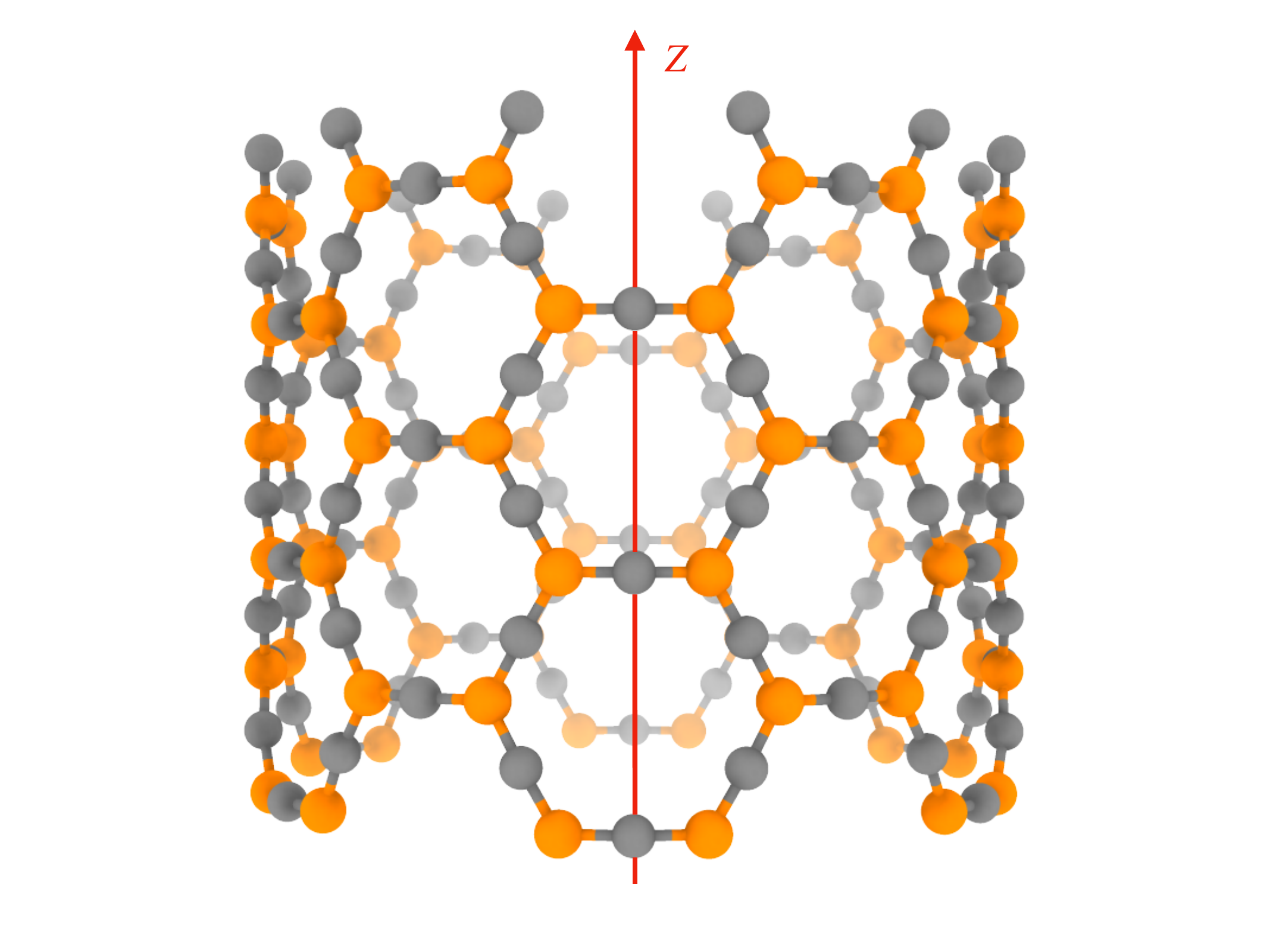}\label{fig:p2c3_armchair}} 
    \subfloat[]{\includegraphics[trim={9cm 0.5cm 8cm 0.5cm}, clip,scale = 0.17]{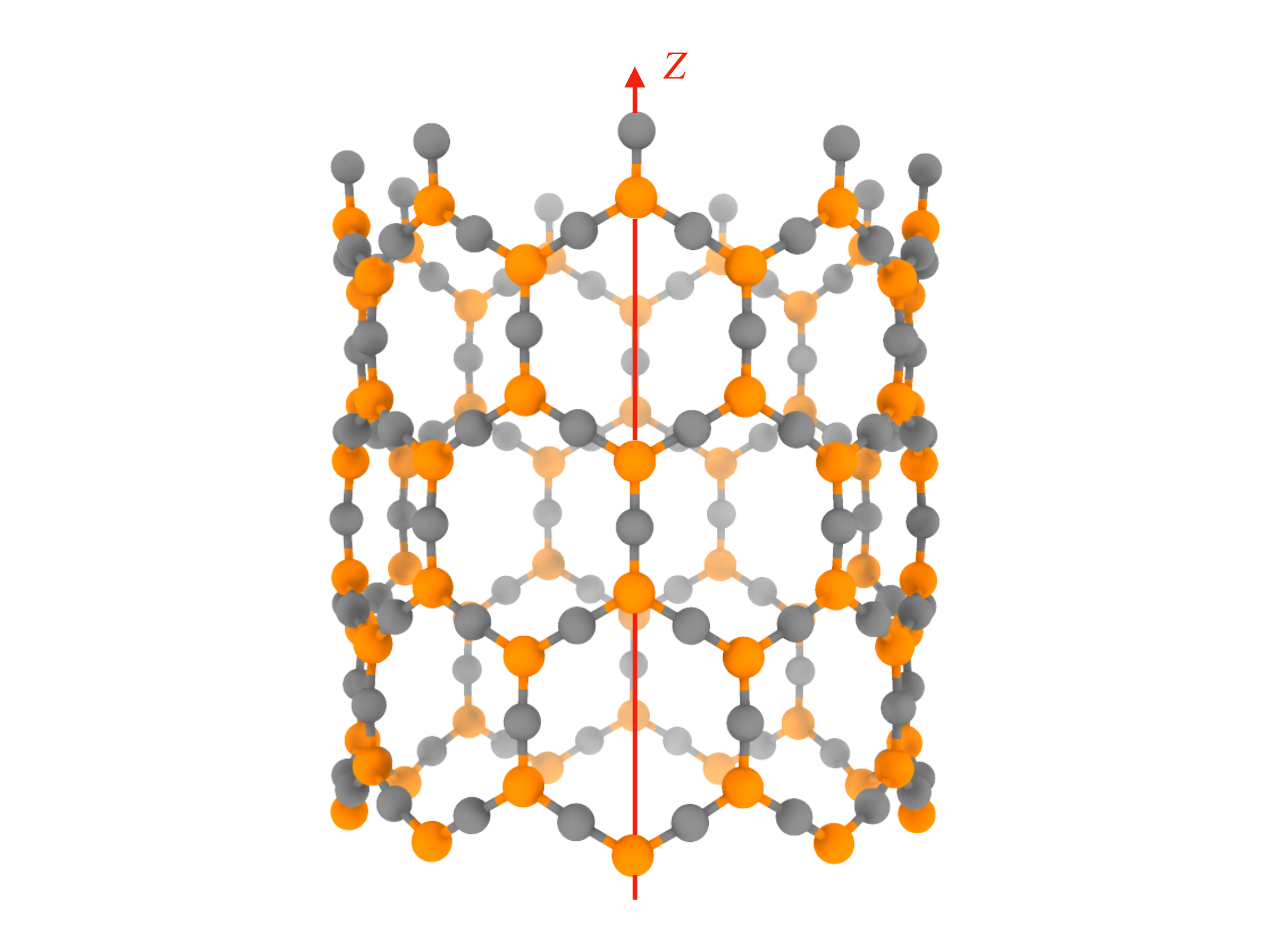}\label{fig:p2c3_zigzag}}
        \subfloat[]{\includegraphics[trim={1cm 6cm 1cm 7cm}, clip,scale = 0.28]{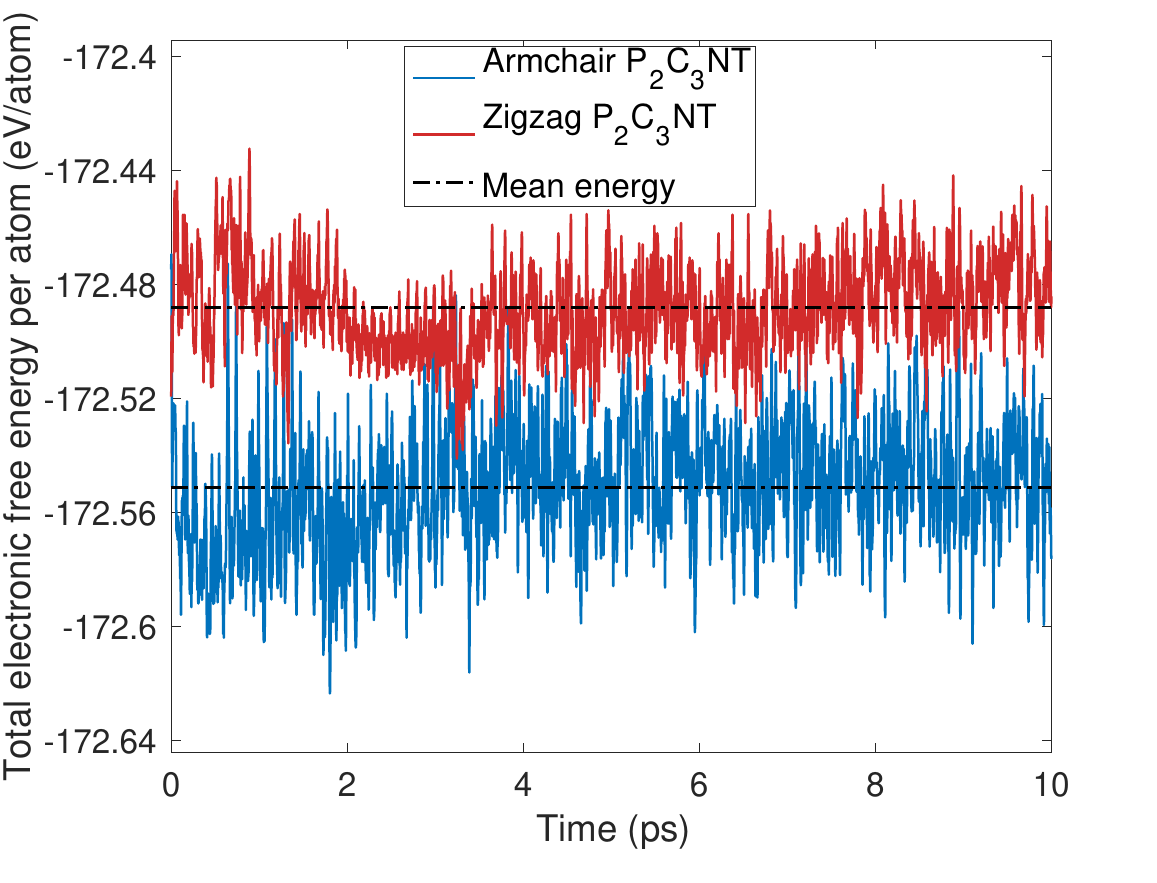}\label{fig:md_sim}}  \\
\subfloat[]{\scalebox{0.30}
{\begin{tikzpicture}
\begin{axis}[
width=\textwidth,
xlabel={Radius (nm)},
ylabel={Cohesive Energy (eV)},
legend style={at={(.04,.04)},anchor=south west,font=\sffamily\large,cells={anchor=east},row sep=1pt,only marks, draw},
label style={font=\sffamily\huge},
tick label style={font=\sffamily\huge},
xmin=5,
xmax=35,
xtick={5, 10, 15, 20, 25, 30, 35},
xticklabels={0.5, 1.0, 1.5, 2.0, 2.5, 3.0, 3.5},
ymin=-5.357,
ymax=-5.350,
ytick ={-5.357, -5.356, -5.355, -5.354, -5.353, -5.352, -5.351,-5.350},
y tick label style={
        /pgf/number format/.cd,
        fixed,
        fixed zerofill,
        precision=3,
        /tikz/.cd
    }
]
\addplot[color=blue, dashed, ultra thick, mark size=5.0pt, mark=asterisk, mark options={solid, blue}] table[x index=0,y index=1]{./figures/P2C3_zigzag_cohesive_energy.txt};
\addplot[color=red, dashed, ultra thick, mark size=5.0pt, mark=o, mark options={solid, red}] table[x index=0,y index=1]{./figures/P2C3_armchair_cohesive_energy.txt};
\legend{{$\;$Zigzag \ce{P2C3}NT}, {$\;$Armchair \ce{P2C3}NT}};
\coordinate (insetPosition) at (rel axis cs:0.95,0.35);
\end{axis}
    \begin{axis}[title=\Large{Cohesive energy of CNTs},xlabel={Radius (nm)},
        ylabel={Cohesive Energy (eV)},
        at={(insetPosition)},anchor={outer south east},
        y tick label style={
        /pgf/number format/.cd,
        fixed,
        fixed zerofill,
        precision=3,
        /tikz/.cd
    }]
    \addplot[color=teal, dashed, ultra thick, mark size=3.0pt, mark=asterisk, mark options={solid, teal}] table[x index=0,y index=1]{./figures/cohesive_energy_cnt_zigzag.txt};
    \addplot[color=orange, dashed, ultra thick, mark size=3.0pt, mark=o, mark options={solid, orange}] table[x index=0,y index=1]{./figures/cohesive_energy_cnt_armchair.txt};
    \legend{{$\;$Zigzag CNT}, {$\;$Armchair CNT}};
    \end{axis}
\end{tikzpicture}
}\label{fig:cohesive}}
\subfloat[]{\scalebox{0.3}
{\begin{tikzpicture} 
\begin{axis}[
width=\textwidth,
xlabel={Axial strain (\%)},
ylabel={$U_{\text{stretch}}(\epsilon)$ (eV/nm)},
legend style={at={(0.15,0.8)},anchor=west,font=\sffamily\large,cells={anchor=west},row sep=1pt, draw},
label style={font=\sffamily\huge},
tick label style={font=\sffamily\huge},
xmin=-4,
xmax =4,
xtick ={-4,-3,-2,-1,0,1,2,3,4},
ymin=-0.5,
ymax=4,
ytick ={-0.5,0,1,2,3,4},
]
\addplot[color=blue, draw=none, ultra thick, mark size=5.0pt, mark=asterisk, mark options={solid, blue}, only marks] table[x index=0,y index=1]{./figures/P2C3_zigzag_extentional_stiffness_go12.txt};
\addplot[color=blue, dashed, ultra thick] table[x index=0,y index=1]{./figures/P2C3_zigzag_extentional_stiffness_fit_go12.txt};
\addplot[color=red, draw=none, ultra thick, mark size=5.0pt, mark=o, mark options={solid, red}, only marks] table[x index=0,y index=1]{./figures/P2C3_armchair_extentional_stiffness_go12.txt};
\addplot[color=red, dashed, ultra thick] table[x index=0,y index=1]{./figures/P2C3_armchair_extentional_stiffness_fit_go12.txt};
\legend{{$\;$Zigzag \ce{P2C3}NT, Radius = 1.07 nm}, {$\;$ Parabolic fit}, {$\;$Armchair \ce{P2C3}NT, Radius = 1.85 nm}, {$\;$ Parabolic fit}};
\end{axis}
\end{tikzpicture}
}\label{fig:stretch}}
 \subfloat[]{\scalebox{0.3}
 {\begin{tikzpicture} 
 \begin{loglogaxis}[
 width = \textwidth,
 xlabel={$\beta$ (radians/nm)},
 ylabel={$U_{\text{twist}}(\beta)$ (eV/nm)},
 legend style={at={(0.42,0.14)},
 anchor=west, font=\sffamily\normalsize,fill=none,cells={anchor=west, align=left},row sep=0pt,},
 xtick={ 0.002, 0.003 , 0.006,0.012},
 xticklabels={ 0.002, 0.003, 0.006,0.012},
 scaled y ticks = false,
 scaled x ticks = false,
 label style={font=\sffamily\huge},
 tick label style={font=\sffamily\huge},
 x tick label style={yshift=-5pt},
 xmin=0.0010,xmax=0.013,
 ymax=2,
 ]
\addplot[color=blue, draw=none, ultra thick, mark size=5.0pt, mark=asterisk, mark options={solid, blue}, only marks,] table[x index=0,y index=1]{./figures/P2C3_zigzag_torsional_stiffness_go12.txt};
\addplot[color=blue, dashed, ultra thick] table[x index=0,y index=1]{./figures/P2C3_zigzag_torsional_stiffness_fit_go12.txt};
\addplot[color=red, draw=none, ultra thick, mark size=5.0pt, mark=o, mark options={solid, red}, only marks,] table[x index=0,y index=1]{./figures/P2C3_armchair_torsional_stiffness_go12.txt};
\addplot[color=red, dashed, ultra thick] table[x index=0,y index=1]{./figures/P2C3_armchair_torsional_stiffness_fit_go12.txt};
 \legend{{$\;$Zigzag \ce{P2C3}NT, Radius = 0.8019 nm}, {$\;$St.~line fit: $q = 1.975,\,c = 2965.76$ eV-nm}, 
 {$\;$Armchair \ce{P2C3}NT, Radius = 1.3885 nm}, {$\;$St.~line fit: $q = 1.9763,\,c = 41597$ eV-nm}};
 \end{loglogaxis}
 \end{tikzpicture}
 }\label{fig:twist}}
 \caption{ (a) Pristine 2D \ce{P2C3} lattice showing the roll-up direction, $\theta$ for the zigzag and armchair nanotubes and $z$ is in the direction of nanotube's axis. Two types of \ce{P2C3}NTs is investigated in this work: (b) Armchair $(n,n)$ and (c) Zigzag $(n,0)$ nanotubes, where, $n$ is the cyclic  group order about the tube axis. (d) System energy variation over ab initio molecular dynamics (AIMD) trajectories at temperatures $315$ K for an $(12,12)$ armchair (blue) and a  $(15,0)$ zigzag (red) \ce{P2C3}NTs. Dashed line denotes the mean energy. (e) Cohesive energy of zigzag and armchair \ce{P2C3}NTs. (f) Extensional energy per unit length as a function of axial strain for two representative \ce{P2C3}NTs. Dotted curves indicate parabolic fits of the data to an ansatz of the form $U_{\text{stretch}}(\epsilon) = c\times\epsilon^2$. (g) Twist energy per unit length as a function of angle of twist per unit length for two representative nanotubes (both axes logarithmic). Dotted lines indicate straight line fits of the data to an ansatz of the form $U_{\text{twist}}(\beta) = c\times\beta^q$. The exponent $q$ is nearly $2.0$ in both cases, suggesting linear elastic behavior.}
 \label{fig:structure_mechanical}
 \end{figure}
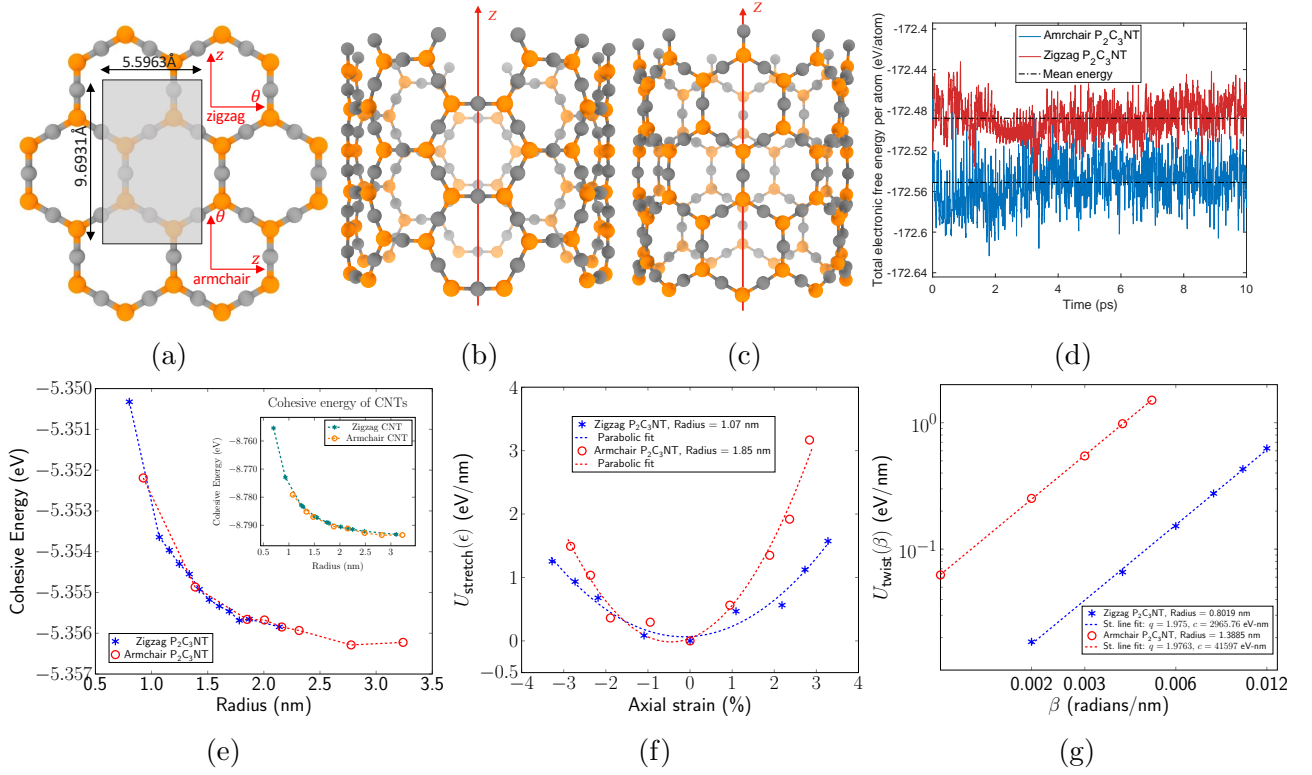

We arrive at \ce{P2C3}NTs through a ``roll-up''  operation \citep{Dumitrica_James_OMD}, in which   structurally relaxed planar \ce{P2C3} \citep{huang2018double} (with hexagonal lattice parameters $a = b =  0.569$ nm) is folded into seamless cylinders. The hexagonal unit cell in the planar structure comprises two phosphorous and three carbon atoms that are strategically positioned at the corners of the hexagonal lattice and the center of the edges of the hexagons, respectively (see Fig.~\ref{fig:structure_mechanical}). This ensures that the $p_z$ orbitals of \ce{P} and \ce{C} atoms are oriented radially outward in the resulting nanotubes and can overlap to feature HK-like electronic bands in a quasi-one-dimensional setting.  Further details of the orbital origin of these bands in \ce{P2C3}NTs are described below and in the supplementary information (SI). The resulting nanotubes are classified by nonnegative integers $(n,m)$, i.e., the chirality indices of the tube, which specify the direction of rolling. Here, we have exclusively studied armchair $(n,n)$ and zigzag $(n,0)$ tubes (see Fig.~\ref{fig:p2c3_armchair} \& \ref{fig:p2c3_zigzag}), in their pristine and distorted states. Our computational studies are enabled by a recently developed suite of real-space first principles simulation techniques, that take advantage of the cyclic and helical symmetries inherent to 1D nanostructures \citep{banerjee2021ab, banerjee2016cyclic, yu2022density, ghosh2019symmetry, pathrudkar2022machine, PhysRevB.103.035101}. Exploitation of global symmetries allows this framework to efficiently simulate pristine or deformed nanotubes (of any chirality) using Kohn-Sham Density Functional Theory (KS-DFT) \citep{hohenberg1964inhomogeneous, kohn1965self}, while considering only a few atoms in the computational unit cell. Our investigations of mechanical and electronic properties using these techniques generally used $5$ or $10$ representative atoms (i.e., just one or two formula units of \ce{P2C3}) in the symmetry-adapted unit cells (Fig.~\ref{fig:p2c3_2d}). Typically, such simulations also employed a discretization of the reciprocal space associated with the helical symmetry, which we refer to as $\eta$-point sampling (in analogy to \emph{k-point sampling} in solid state systems).  Notably, many of such calculations would require an impractically large number of atoms in the computational unit cell if conventional first-principles methods (e.g. plane-wave-based approaches), were used \citep{banerjee2021ab, yu2022density}. Simulation cells containing more atoms were employed for ab initio molecular dynamics and magnetism calculations.

Additional symmetry-related parameters for the simulations are as follows. The screw-transformation (or helical symmetry operation) used to describe the nanotubes has an associated pitch of $\tau = 0.55963$ nm \;and $0.96931$ nm, for  undeformed armchair and zigzag nanotubes, respectively. Changes in $\tau$ allow examination of the effects of uniaxial extensions and compressions on the material. Concurrently, a scalar parameter $0 \leq \alpha < 1$ represents applied twist to the structure, with $\beta = 2\pi\alpha/\tau$ denoting the twist per unit length. The nanotubes are also associated with cyclic symmetry about the tube axis, with the rotation angle $\Theta = 2\pi/\mathfrak{N}$ being directly related to the tube chirality indices (e.g., $n=\mathfrak{N}$ for zigzag $(n,0)$ and armchair $(n,n)$ tubes). 

For the majority of our computations, we employed Helical DFT, a finite difference based implementation of symmetry adapted Kohn-Sham density functional theory \citep{banerjee2021ab, yu2022density, pathrudkar2022machine}. We used $12^\text{th}$ order finite differences with a vacuum padding of $10$ Bohr in the nanotube radial direction. The Perdew-Wang parametrization \citep{perdew1992accurate} of the local density approximation \citep{kohn1965self} was chosen as the exchange-correlation functional. Furthermore, norm conserving pseudopotentials \citep{troullier1991efficient, hamann2013optimized}, and $1$ mHa of smearing using the Fermi–Dirac distribution were chosen.  Self-consistent field iteration convergence was achieved via the Periodic Pulay scheme \citep{banerjee2016periodic}.  To reduce computational burden, Helical DFT simulations were conducted in three successive phases, with increasing levels of discretization fineness \citep{yu2024carbon} (see SI for further details). Additionally to augment Helical DFT results, the Quantum Espresso \citep{giannozzi2009quantum,giannozzi2020quantum} and SPARC \citep{xu2021sparc, ghosh2017sparc1, ghosh2017sparc2, PhysRevB.103.035101} codes were used for calculations involving projected density of states (pDOS), magnetism effects due to  vacancies and dopants, 2D \ce{P2C3} sheets, and some ab-initio molecular dynamics runs.

We used our simulations to analyze various structural properties of \ce{P2C3}NTs and to assess their stability. The cohesive energy depicted in Fig.~\ref{fig:cohesive} shows monotonically decreasing behavior from $-5.350$eV to $-5.457$eV as the radius of zigzag and armchair nanotubes is increased from $0.5$ nm to $3.5$ nm. This is consistent with the expectation that tubes of larger radius are energetically more favorable due to the reduced elastic bending energy of the 2D sheet. In contrast to these numbers, the corresponding cohesive energy value for a conventional phosphorene nanotube \citep{cai2016thermal, guan2016prediction}  of approximately $0.7$ nm radius is about $-4.22$ eV, and it is about $-8.77$ eV for a conventional carbon nanotube \citep{dresselhaus2000carbon} (CNT) of similar radius (Fig.~\ref{fig:cohesive} inset). These intermediate cohesive energy values of \ce{P2C3}NTs are strongly suggestive of their synthesizability. The bending modulus of the planar \ce{P2C3} sheet comes out to be $0.140$4 eV and $0.1520$ eV when the sheet is bent in the armchair and zigzag directions, respectively. These values indicate that the sheet bending modulus of \ce{P2C3} is approximately one-tenth of that observed for conventional graphene, estimated to be around $1.5$ eV \citep{ghosh2019symmetry}. The phonon stability of \ce{P2C3} sheets has been investigated earlier \citep{huang2018double} and no unstable modes were found. Based on band-folding considerations \citep{zhang2007phonon, saito1998raman, jishi1993phonon}, such calculations are also likely to be indicative of the stability of \ce{P2C3}NTs at $0$ K, especially given the low elastic energies associated with folding \ce{P2C3} sheets into tubes. To assess the stability of \ce{P2C3}NTs at finite temperature, we performed AIMD calculation at  room temperature for several nanotubes (both zigzag and armchair varieties). Supercells containing $60$ or more atoms --- i.e., several copies of the unit cell in the cyclic and axial directions --- were employed, in order to capture potential long-wavelength instabilities.  The tubes were observed to be stable throughout the simulation duration (up to $10$ ps). The variation of the system's energy for two representative \ce{P2C3}NTs is shown in Fig.~\ref{fig:md_sim}.

The kinetic stability of the \ce{P2C3}NTs investigated in this letter is  a promising sign of their synthesizability. Earlier Huang et al. \citep{huang2018double} have demonstrated a methodology to grow phosphorus carbide monolayer on silver (\ce{Ag}) (111) substrate. They found that the lattice mismatch between \ce{P2C3} and Ag (111) substrate is less than $1.6\%$ and the adhesion energy between them is $-4.73$ eV/atom. After synthesizing \ce{P2C3}, the target etching of the silver layer can cause the 2D material to curl up and result in \ce{P2C3}NTs \citep{schmidt2001thin} as illustrated in the supplementary information (Fig.~S7). The lower bending stiffness of \ce{P2C3} sheets in comparison to graphene and phosphorene will likely make it easier for the material to fold up into nanotubes.

To investigate mechanical properties of \ce{P2C3}NTs, particularly their response to torsional and uniaxial strain, we performed Helical DFT simulations with variations in the symmetry group parameters used to define the nanotube \citep{banerjee2021ab}. From these simulations, the energy per unit length of the deformed system, $U_{\text{deformed}}(x)$ may be calculated as a function of the strain parameter $x$, and the corresponding stiffness $k$ may be obtained as: $\displaystyle k = \frac{\partial^{2} U_{\text{deformed}}(x)}{\partial x^{2}}\bigg\rvert_{x = 0}$. Further details of calculating $U_{\text{deformed}}(x)$ and the associated stiffness parameter, for torsional and uniaxial strains are provided in the SI. For torsional simulations, we imposed twists of up to about $\beta = 4.5^{\circ}$ per nanometer, the acknowledged limit of linear response for conventional CNTs \citep{Dumitrica_James_OMD}. Our investigation shows that the twisting deformation energy for \ce{P2C3}NTs is very nearly quadratically dependent on the twist per unit length,  i.e., these tubes also exhibit linear elastic behavior within this range (Fig. \ref{fig:twist}). In particular, we estimate from the data that the torsional stiffness constant ($k_{\text{twist}}$) for a zigzag \ce{P2C3}NT of radius about $0.80$ nm is $207.53$ eV/nm, while it is $956.10$ eV/nm for an armchair nanotube of radius about $1.4$ nm. To compare these numbers with  those of conventional CNTs, we note that  CNTs are known to show behavior consistent with continuum theory  \citep{yu2022density, Dumitrica_James_OMD}, wherein $k_{\text{twist}}$ depends on tube radius cubically \citep{timoshenko1968elements}. We use this fact and first principles data \citep{yu2022density}  to estimate that $k_{\text{twist}}$ values are $3021.2$ eV/nm and $15809$ eV/nm for armchair and zigzag CNTs of similar radii, respectively. Along similar lines, we also carried out axial stiffness calculations (further details in SI) while constraining the strain to $\pm 3.3 \%$. In this range, the deformation energy displays a quadratic trend (Fig.~\ref{fig:stretch}), consistent with linear response. In particular, the extensional stiffness ($k_{\text{stretch}}$) for an armchair \ce{P2C3}NT of radius $1.85$ nm is about $2711.3$ eV/nm and it is about $1257.4$ eV/nm for a zigzag  tube of radius $1.07$ nm. To compare these numbers against $k_{\text{stretch}}$ values of CNTs, we once again utilized scaling laws obtained from continuum theory and first principles data \citep{timoshenko1968elements, yu2022density}. We estimated that armchair and zigzag CNTs of similar radii as the \ce{P2C3}NTs above, are expected to have $k_{\text{stretch}}$ values equal to $13318.3$ eV/nm and $7678.1$ eV/nm for  armchair and zigzag varieties, respectively.  Overall, these stiffness calculations imply that \ce{P2C3}NTs are significantly more compliant to torsional and axial strains, when compared to their conventional carbon counterparts. In turn, these findings imply lower values of (thickness normalized) Young's and shear moduli of \ce{P2C3} sheets, when compared to graphene.

\begin{figure}[ht!]
\centering
\subfloat[]{
{
\includegraphics[scale =0.3,trim={1.cm 6.5cm 1.5cm 6cm},clip]{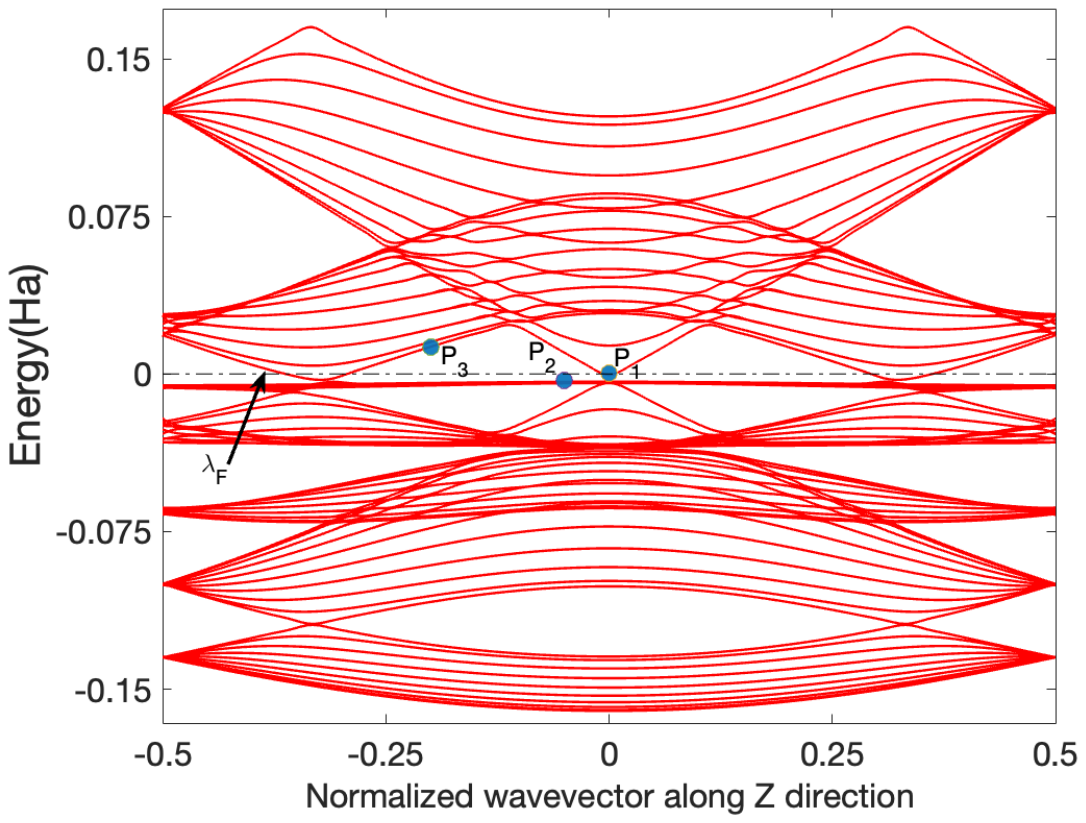}
}\label{fig:banddiag_armchairfull_DFT}}
\subfloat[]{
{
\includegraphics[scale =0.3,trim={1.cm 6.5cm 1.5cm 6cm},clip]{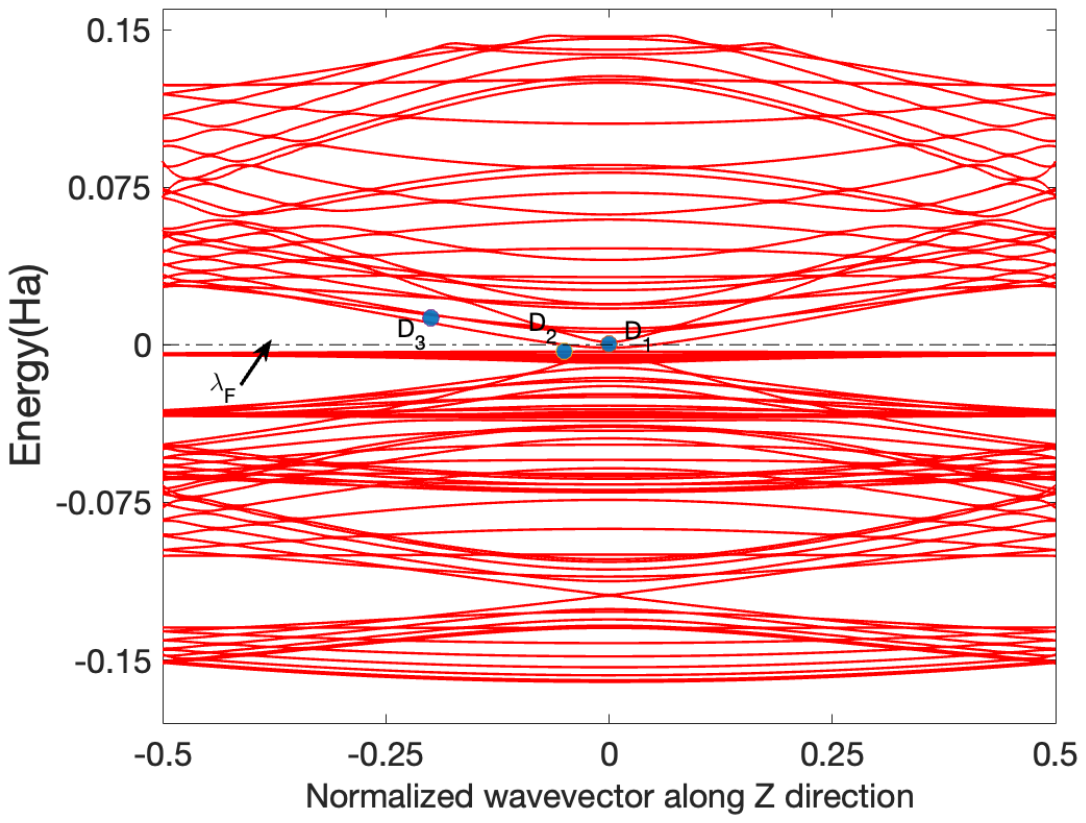}
}\label{fig:banddiag_zigzagfull_DFT}}\\
\subfloat[]{
{
\includegraphics[scale =0.5,trim={10cm 3.5cm 10cm 3.5cm},clip]{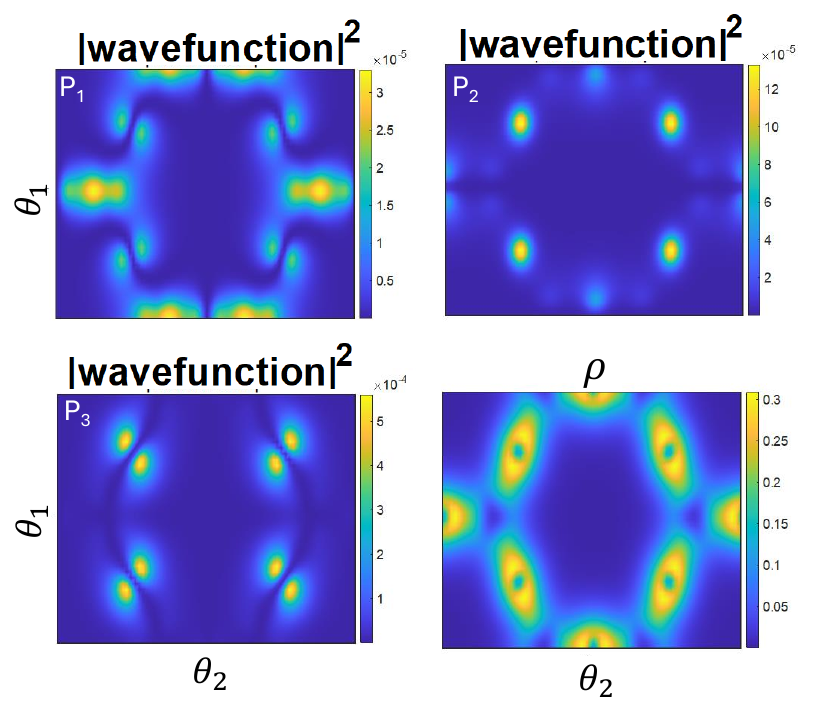}
}\label{fig:wave_func_armchair}}
\subfloat[]{
{
\includegraphics[scale =0.5,trim={10.cm 3.5cm 10cm 3.5cm},clip]{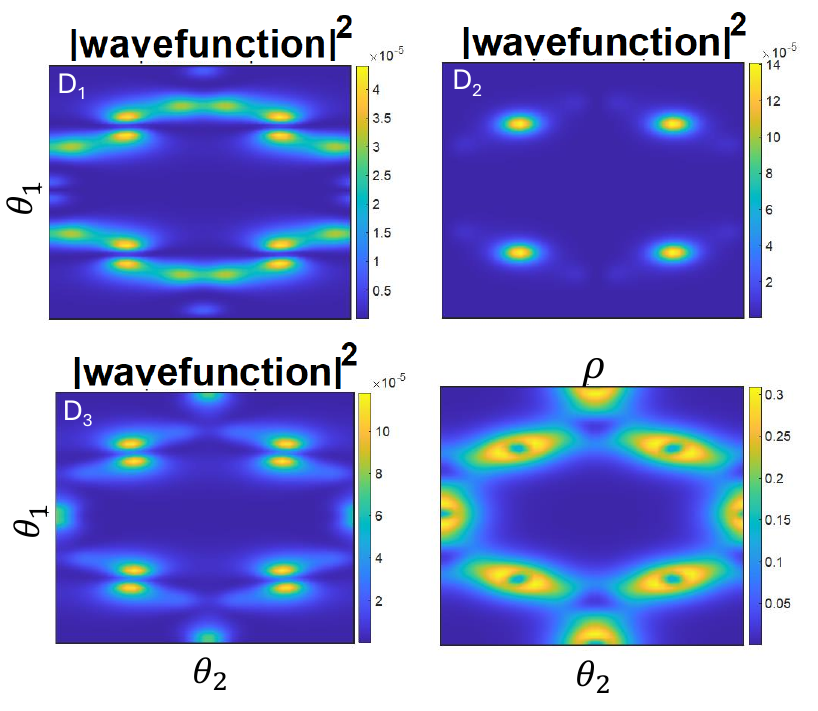}
}\label{fig:wave_func_zigzag}}
\caption{(a) and (b) show the band diagram of  undeformed $(9,9)$ armchair and $(12,0)$ zigzag  \ce{P2C3}NTs, respectively. The Fermi level $\lambda_F$ corresponds to the x-axis. (c) shows the electronic states (square of absolute value of wavefunction) associated with $P_1$, $P_2$ and $P_3$  points shown in (a). The bottom right panel is electron density. (d) shows the square of the wavefunction for the electronic states associated with the $D_1$, $D_2$ and $D_3$ points shown in (b). The bottom right panel is the electron density. A slice of electronic fields at an average radial distance of the atoms in the computational domain is shown in each case. $\theta_1,\theta_2$ denote helical coordinates that parametrize the tube surface at a fixed radial distance.}
\label{fig:electronic_structure}
\end{figure}

\begin{figure}[ht!]
  \subfloat[]{\includegraphics[scale=0.28,trim={9cm 2cm 9cm 2cm},clip]{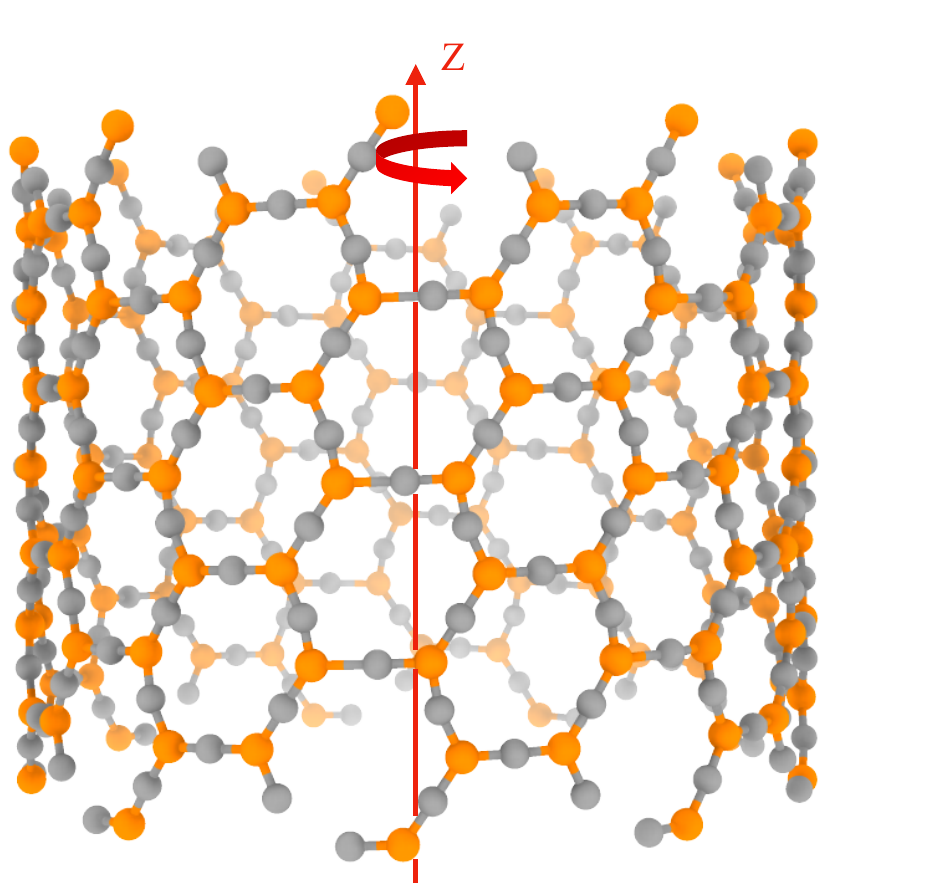}\label{fig:twisted_armchair_NT}}\;
\subfloat[]{
{
\includegraphics[scale =0.3,trim={1.cm 6.5cm 1.5cm 6cm},clip]{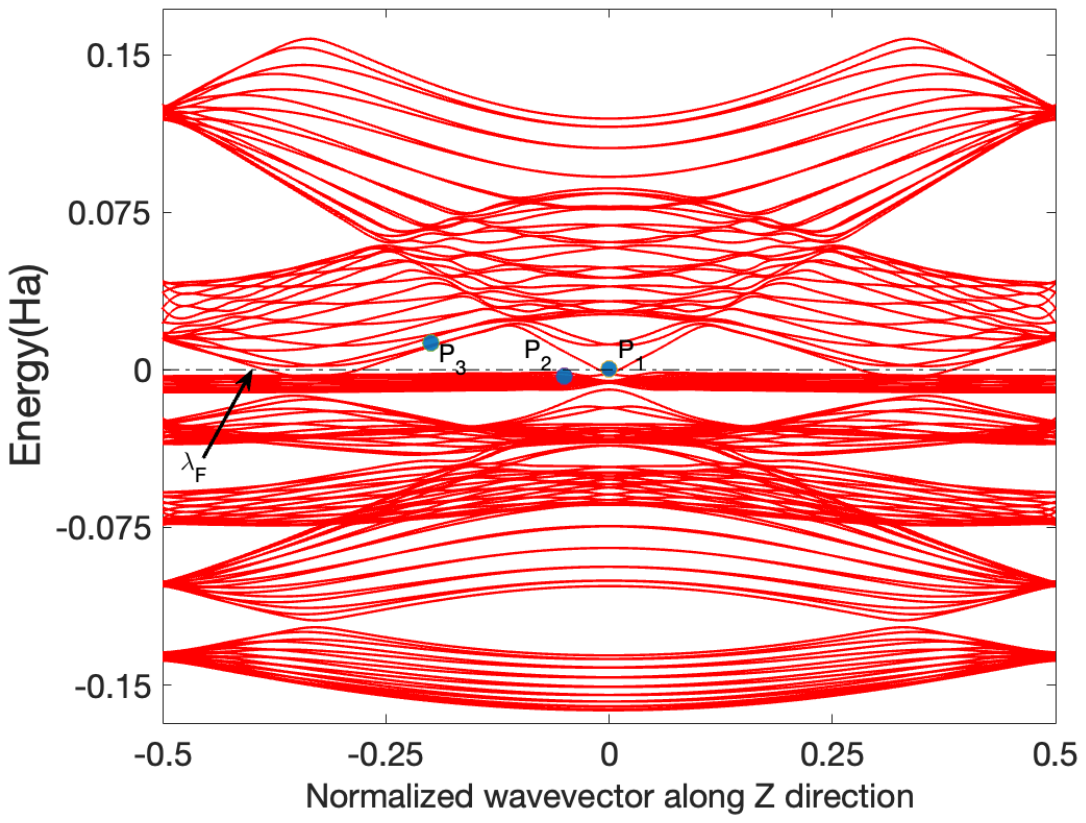}
}\label{fig:banddiag_armchairfull_DFT_twist}} \\
\subfloat[]{\scalebox{0.3}
{\begin{tikzpicture} 
\begin{axis}[
width=\textwidth,
xlabel=\huge{E$-\lambda_\text{F}$ (Ha)},
ylabel=\huge{Density of states $\aleph_{T_{\text{e}}}(\cdot)$ ($\text{Ha}^{-1}$)},
legend pos = {north west}, 
legend style={at={(0.090,0.82)},anchor=west,font=\sffamily, fill=none,cells={anchor=west},row sep=1pt},
y label style={xshift=-10pt},
x label style={xshift=-10pt},
label style={font=\sffamily\huge},
tick label style={font=\sffamily\huge},
xmin=-0.167, xmax=0.162,
ymin = -0.05, ymax=900,
ytick={0,100,200,300,400,500,600,700,800,900},
yticklabels={0,100,200,300,400,500,600,700,800,900},
xtick={-0.15,-0.1,-0.05,0,0.05,0.1,0.15},
xticklabels={-0.15,-0.1,-0.05,0,0.05,0.1,0.15},
x tick label style={yshift=-5pt},
y tick label style={xshift=-5pt},
]
\addplot[line width=0.90mm, blue,] table[x index=0,y index=1]{./figures/armchair_go9_21kpts_DOS.txt};
\addplot[line width=0.70mm, red,] table[x index=0,y index=1]{./figures/armchair_go9_twist_op002_dos.txt};
\addplot[line width=0.50mm, cyan,] table[x index=0,y index=1]{./figures/armchair_go9_twist_op004_dos.txt};
\addplot[line width=0.30mm, black,] table[x index=0,y index=1]{./figures/armchair_go9_twist_op005_dos.txt};
\addplot[line width=0.20mm, green,] table[x index=0,y index=1]{./figures/armchair_go9_twist_op006_dos.txt};
\addplot[mark=none, black, dashed, ultra thick] coordinates{(0, -0.05) (0, 900)}; 
\node[] at (axis cs: 0.010,350) {\huge{$\lambda_{\text{F}}$}};
\legend{\LARGE{$\;$No twist}, \LARGE{$\;$ $1.29^\circ/$nm twist},\LARGE{$\;$ $2.57^\circ/$nm twist},\LARGE{$\;$ $3.22^\circ/$nm twist},\LARGE{$\;$ $3.86^\circ/$nm twist}};
\end{axis}
\end{tikzpicture}}
\label{fig:DOS_twist_armchir}}
\subfloat[]{
{
\includegraphics[scale =0.5,trim={10.cm 3.5cm 10cm 4cm},clip]{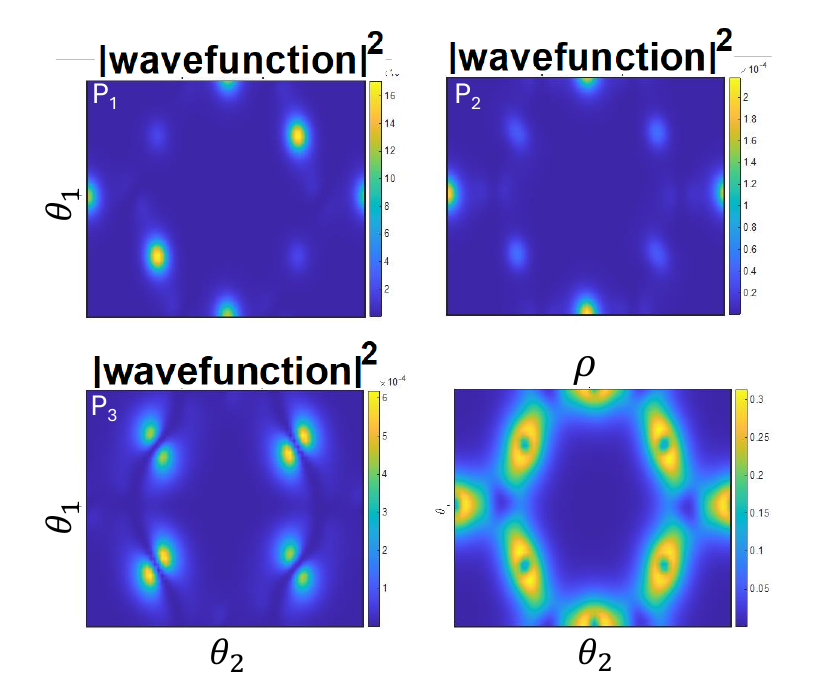}
}\label{fig:wave_func_armchair_twisted}}
\caption{(a) Twisted armchair \ce{P2C3}NT. (b) Band diagram of $(9,9)$ twisted armchair \ce{P2C3}NT at $3.86^\circ$/nm. The Fermi level $\lambda_F$ corresponds to the x-axis.  (c) Electronic density of states (eDOS) plot for different rate of twists. (d) shows the electronic states (square of absolute value of wavefunction) associated with the points $P_1$, $P_2$ and $P_3$ points shown in (b). Bottom right panel shows the electronic density $\rho$.  A slice of electronic fields at an average radial distance of the atoms in the computational domain is shown. $\theta_1,\theta_2$ denote helical coordinates that parametrize the tube surface at a fixed radial distance.}
\label{fig:armchair_density_VBM_twisted}
\end{figure}

The symmetry adapted first principles calculations described above reveal that all  pristine \ce{P2C3}NTs are metallic. Remarkably, we observe that a \ce{P2C3}NT with cyclic group order $\mathfrak{N}$ (with $\mathfrak{N} = n$ for zigzag $(n,0)$ and armchair $(n,n)$ tubes), possesses $2\mathfrak{N}$ nearly degenerate flat bands very close to the Fermi level (Fig.~\ref{fig:electronic_structure}). There is also an associated sharp peak in the electronic density of states (eDOS), suggesting the easy availability of electron-rich states in these tubes (see Fig.\ref{fig:DOS_twist_armchir} and Fig.~S8 in SI). { These flat bands originate due to destructive interference of electron hopping within the Kagome-like structure formed by C and P atoms, specifically involving the $p_z$ and $p_{xy}$ orbitals. Since this is a geometric frustration effect, standard KS-DFT under the local  density approximation (LDA) can adequately describe the electronic structure without needing on-site Coulomb correction, $U$. For bands near the Fermi level, the flat band bandwidth ($W$) along the 1D Brillouin zone is approximately $35 - 78$ meV for the armchair $(9,9)$ nanotube, which is somewhat higher than the $13 - 22$ meV observed for the zigzag $(12,0)$ nanotube (electronic smearing for these calculations is $1$ mHa, i.e., $27.2$ meV). Plausible $U$ values for p-orbitals in P and C atoms is typically $U \sim 1-3$ eV, making $U/W \gg 1$.} Overall, these features are  suggestive that \ce{P2C3}NTs are likely to be a notable instance of quasi-one-dimensional materials that are inclined to display strongly correlated electronic states which are often associated with fascinating properties such as superconductivity and flat-band ferromagnetism \citep{chen2018ferromagnetism}. Moreover, to our knowledge, the type of Dirac and flat-band coexistence at the Fermi level seen in  \ce{P2C3}NTs, has not been previously reported in a chemically realistic quasi-one-dimensional material.

As described above, in addition to the flat bands, the band structures of \ce{P2C3}NTs also feature Dirac cones. In particular, for pristine armchair \ce{P2C3}NTs, the flat band near the Fermi level ($\lambda_F$) touches a lower Dirac point near the gamma point ($\eta = 0$), and is separated from the upper Dirac point with a minute gap ($\sim 3.2$ mHa for a $(9,9)$ nanotube; Fig.~\ref{fig:banddiag_armchairfull_DFT}). Two other sets of Dirac bands near $\eta = \pm \frac{1}{3}$, which touch the flat band near $\lambda_F$, thereby making the tubes metallic, are also present. {These crossings between the flat bands and Dirac points at $\eta = \pm \frac{1}{3}$ are accidental rather than being symmetry-protected or the effect of curvature.} Additionally, a family of quasi-flat bands reminiscent of the band structure of Kagome lattices with next-nearest neighbor interactions \citep{barreteau2017bird} are also present (in the energy range $35.2 - 70.4$ mHa in Fig.~\ref{fig:banddiag_armchairfull_DFT}). Similarly, zigzag \ce{P2C3}NT also exhibit flat bands with Dirac points crossing near $\lambda_F$. However, kagome-like Dirac points which appear in the armchair nanotube are folded to the $\eta = 0$ point (labeled $D_1$ in Fig.~\ref{fig:banddiag_zigzagfull_DFT}) while rolling the sheet to form zigzag nanotubes. Overall, these arrangements of Dirac points at $\eta = \pm \frac{1}{3}$ and $\eta = 0$  are reminiscent of the electronic structure of conventional armchair and zigzag CNTs, respectively. 

To elucidate the orbital source {and hybridization} of the electronic band structure of the \ce{P2C3}NTs studied here, we computed the projected density of states (pDOS). This allows us to estimate the contributions of the different orbitals of \ce{P} and \ce{C} atoms participating in the formation of energy bands near the Fermi level (Fig.~S1). From these calculations, it is evident that many of the electronic features of \ce{P2C3}NTs  largely originate from the participation of $\pi$ electrons derived from radially oriented $p_z$  ($l = 1, m_l = 0$)  orbitals of both \ce{C} and \ce{P} atoms. Specifically, the flat bands arise from the individual $\pi$-electrons of \ce{C} atoms, while the Dirac points situated at $\eta = 0$ near the Fermi level (in both armchair and zigzag tubes) are derived from the mixture of $p_z$ orbitals of both \ce{P} and \ce{C} atoms (see Fig. S1 (a) \& (b)). Thus, the $p_z$ orbitals in the nanotube create a bipartite honeycomb split graph lattice \citep{ma2020spin}. Indeed, the split graph operation applied to the bipartite honeycomb lattice introduces additional sites at the center of each edge, resulting in the lattice depicted in Fig.~\ref{fig:p2c3_2d}. {This hybridization between $p_z$ orbitals of P and C atoms lifts the degeneracy and separates the flat band from the Dirac cone at the $\Gamma$-point as shown in Fig.~\ref{fig:banddiag_armchairfull_DFT}}. On the other hand, in-plane $p_{xy}$ orbitals of \ce{C} atoms form a Kagome lattice resulting in Kagome-like bands. These groups of bands include the aforementioned quasi-flat bands in the $35.2 - 70.4$ mHa range, and additional Dirac bands positioned near the Fermi level (at $\eta = \pm 1/3$ in armchair nanotubes and at $\eta = 0$ in zigzag nanotubes). {These $p_{xy}$ bands do not hybridize with the $p_z$ flat band and form an accidental crossing at $\eta = \pm 1/3$ and $0$ in armchair and zigzag nanotubes, respectively.} Overall, the \ce{P2C3}NTs bands are the direct sum of Honeycomb-Kagome bands and Kagome bands (also see Fig. S1(c)). 

The above analysis is also consistent with direct visualization of the spatial distribution of the wavefunctions (Fig.~\ref{fig:wave_func_armchair} \& \ref{fig:wave_func_zigzag}). Considering the band diagrams of two representative armchair and zigzag  \ce{P2C3}NTs (Fig.~\ref{fig:banddiag_armchairfull_DFT} \& \ref{fig:banddiag_zigzagfull_DFT}), we see that points $P_2$ and $D_2$ on the flat bands show the electrons localized to  $p_z$-type atomic orbitals of C atoms(top right of Figs.~\ref{fig:wave_func_armchair},  \ref{fig:wave_func_zigzag}). The lobes of these orbitals are pointing perpendicular to the $\theta_1-\theta_2$ plane, i.e., along the radial direction in the tube. In contrast, points $P_3$ and $D_3$ on the Dirac bands correspond to electronic states with in-plane $p_{xy}$ orbital characteristics (bottom left of Figs.~\ref{fig:wave_func_armchair},  \ref{fig:wave_func_zigzag}). Finally, points $P_1$ and $D_1$, which lie  at the gamma point, are at the intersection of the Dirac cones and the flat bands, and are therefore associated with electronic states with both these characteristics. This is evident in the shapes of the corresponding orbitals (top left of Figs.~\ref{fig:wave_func_armchair},  \ref{fig:wave_func_zigzag}).  

Drawing insights from the pDOS calculations (see SI) and the discussion above, we constructed a next-nearest neighbor (NNN) symmetry-adapted tight binding (TB) model to capture the salient features of the electronic structure of \ce{P2C3}NTs (see Fig. S1). We utilized the Dresselhaus approach \citep{dresselhaus2000carbon} and expressed the $8$ bands TB Hamiltonian in terms of two sets of orthogonal orbitals ---  three radially oriented $p_z$ orbitals of \ce{C} atoms, along with two more from \ce{P} atoms and three in-plane $p_{xy}$ orbitals \ce{C} atoms.  The results of our TB calculations for pristine \ce{P2C3}NTs are illustrated in Fig. S2(a) \& (b).  It is evident from these figures that there is a remarkable qualitative agreement between these results and the first principles data presented earlier, thus lending support to our interpretation of the origin of the electronic features of \ce{P2C3}NTs.

\begin{figure}[ht!]
    \centering
       \subfloat[]{
{
\includegraphics[scale =0.46,trim={1cm 6.5cm 0cm 6.5cm},clip]{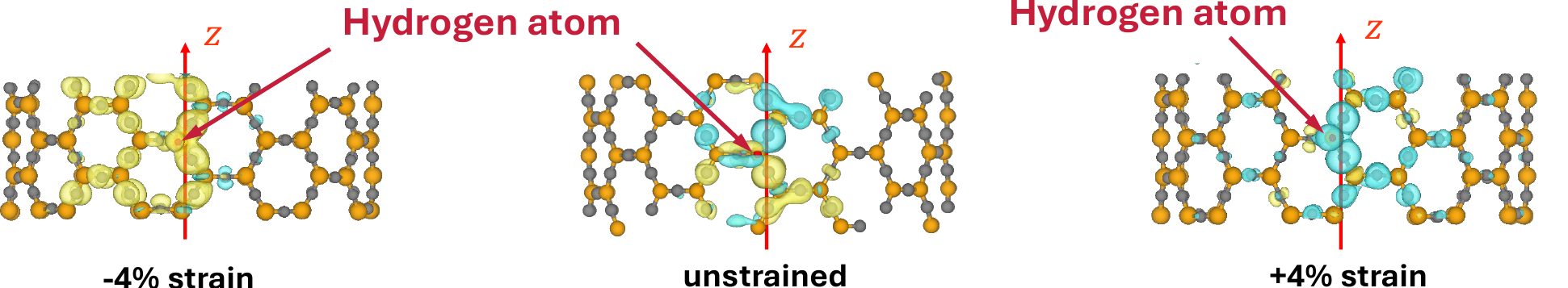}
}\label{fig:magnetization_vs_strain_ntb}}\\
   \subfloat[]{
{
\includegraphics[scale =0.3,trim={1.1cm 6cm 2.5cm 6.5cm},clip]{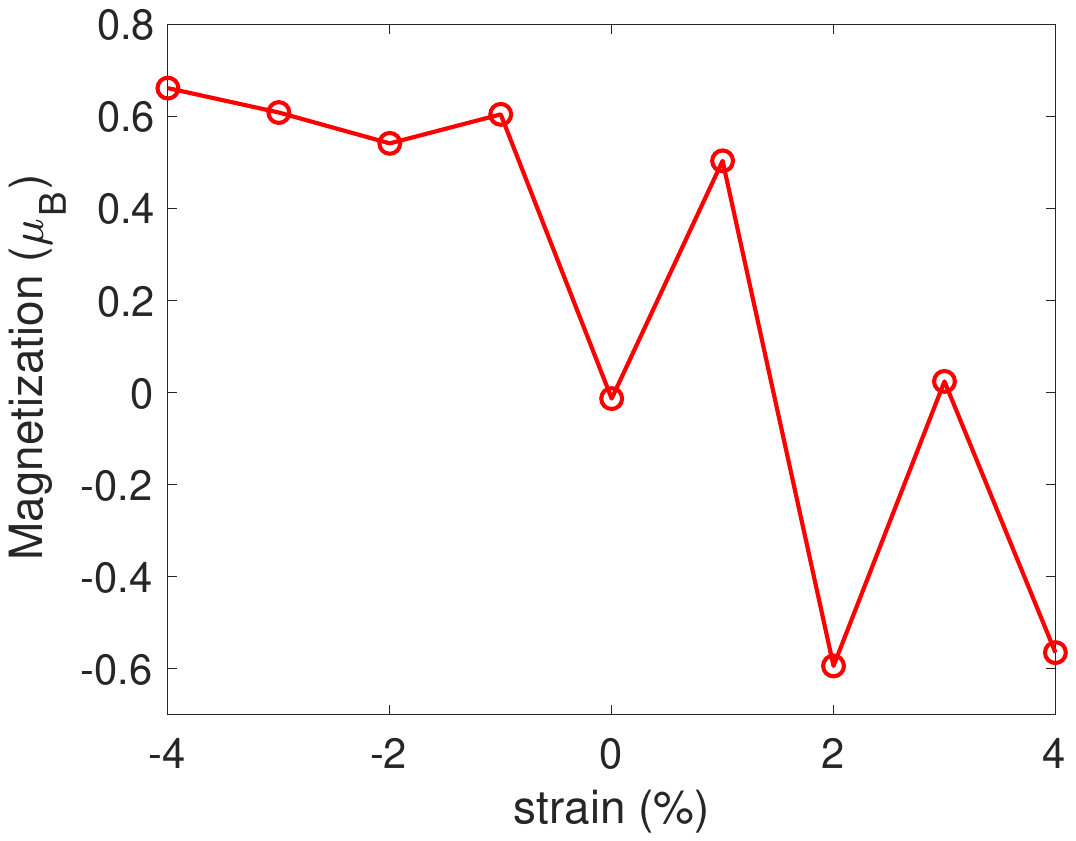}
}\label{fig:magnetization_vs_strain}}
 \subfloat[$-4\%$ strain]{
{
\includegraphics[scale =0.3,trim={4cm 6.5cm 5.cm 6.5cm},clip]{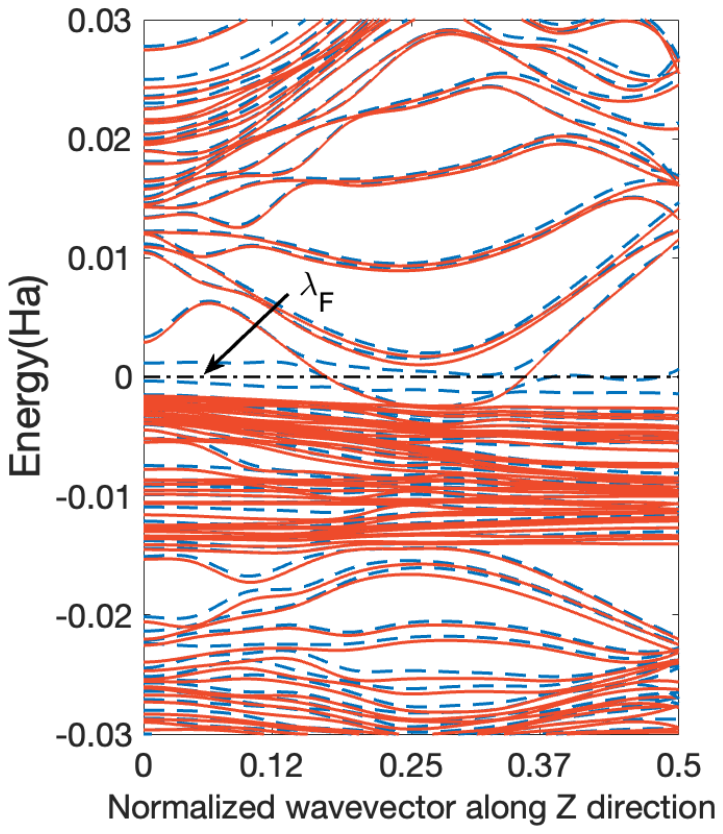}
}\label{fig:band_diagram_mag_compression}}
 \subfloat[unstrained]{{
\includegraphics[scale =0.3,trim={5.1cm 6.5cm 5.2cm 6.5cm},clip]{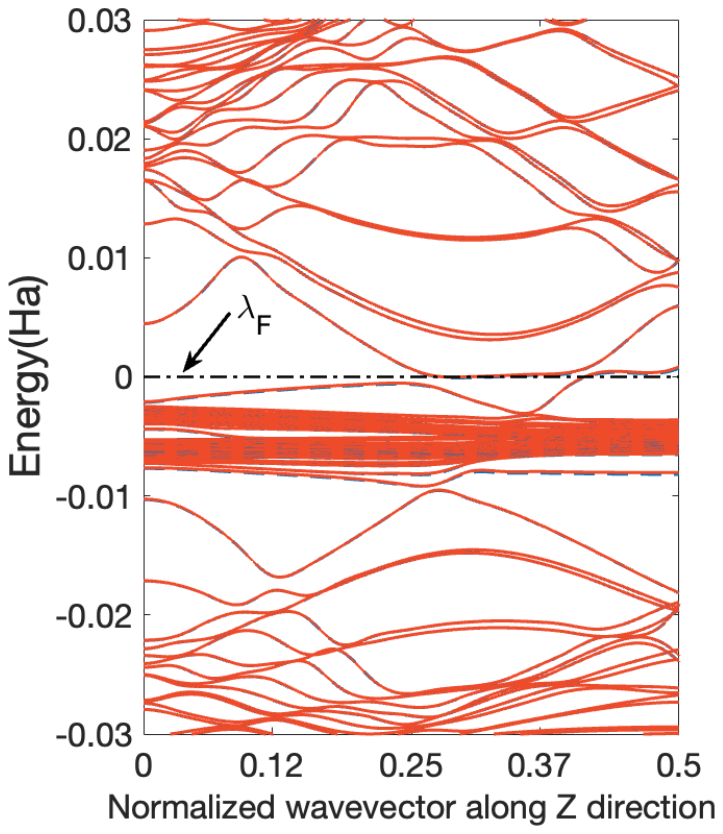}
}\label{fig:band_diagram_mag_pristine}}
 \subfloat[$+4\%$ strain]{{
\includegraphics[scale =0.3,trim={5.1cm 6.5cm 5.2cm 6.5cm},clip]{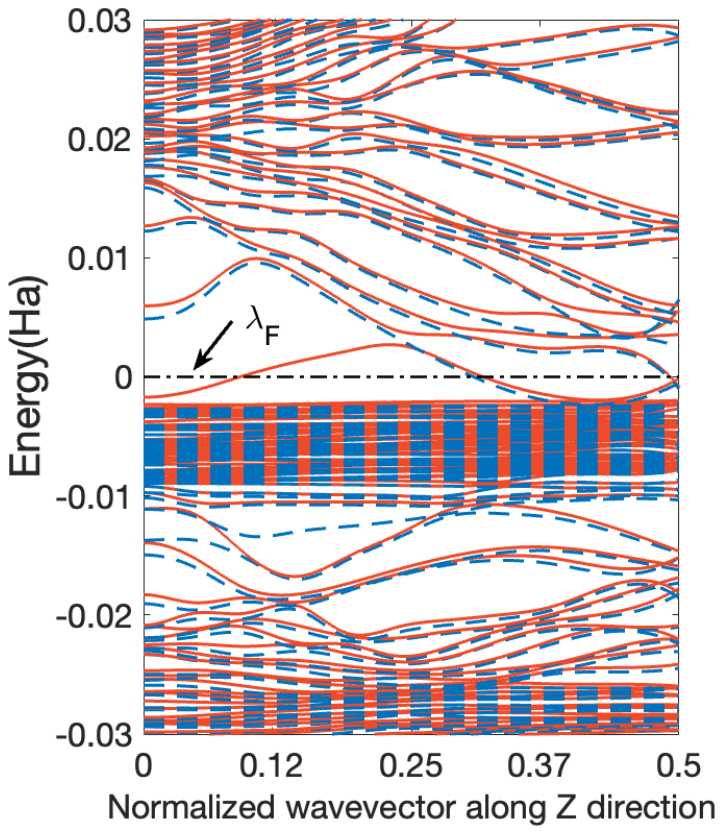}
}\label{fig:band_diagram_mag__tensile}}\\
\subfloat[]{\scalebox{0.3}
{\begin{tikzpicture} 
\begin{axis}[
width=\textwidth,
xlabel={E$-\lambda_\text{F}$ (Ha)},
ylabel={Density of states $\aleph_{T_{\text{e}}}(\cdot)$ ($\text{Ha}^{-1}$)},
legend pos = {north west}, 
legend style={at={(0.60,0.82)},anchor=west,font=\sffamily, fill=none,cells={anchor=west},row sep=1pt},
y label style={xshift=-10pt},
x label style={xshift=-10pt},
label style={font=\sffamily\huge},
tick label style={font=\sffamily\huge},
xmin=-0.03, xmax=0.03,
ymin = -0.05, ymax=6500,
ytick={0,1000,2000,3000,4000,5000,6000,7000,8000,9000},
yticklabels={0,1000,2000,3000,4000,5000,6000,7000,8000,9000},
xtick={-0.03,-0.02,-0.01,0,0.01,0.02,0.03},
x tick label style={yshift=-5pt},
y tick label style={xshift=-5pt},
]
\addplot[line width=0.90mm, red,] table[x index=0,y index=1]{./figures/DOS_vacancy_spin_neg_4pcnt.txt};
\addplot[line width=0.90mm, dashed,blue,] table[x index=0,y index=2]{./figures/DOS_vacancy_spin_neg_4pcnt.txt};
\addplot[mark=none, black, dashed, ultra thick] coordinates{(0, -0.05) (0, 9000)}; 
\node[] at (axis cs: 0.003,3000) {\huge{$\lambda_{\text{F}}$}};
\legend{\huge{$\;$Spin-up},\huge{$\;$Spin-down}};
\end{axis}
\end{tikzpicture}}
\label{fig:DOS_neg_strain}}
\subfloat[]{\scalebox{0.3}
{\begin{tikzpicture} 
\begin{axis}[
width=\textwidth,
xlabel={E$-\lambda_\text{F}$ (Ha)},
ylabel={Density of states $\aleph_{T_{\text{e}}}(\cdot)$ ($\text{Ha}^{-1}$)},
legend pos = {north west}, 
legend style={at={(0.60,0.82)},anchor=west,font=\sffamily, fill=none,cells={anchor=west},row sep=1pt},
y label style={xshift=-10pt},
x label style={xshift=-10pt},
label style={font=\sffamily\huge},
tick label style={font=\sffamily\huge},
xmin=-0.03, xmax=0.03,
ymin = -0.05, ymax=9800,
ytick={0,1000,2000,3000,4000,5000,6000,7000,8000,9000},
yticklabels={0,1000,2000,3000,4000,5000,6000,7000,8000,9000},
xtick={-0.03,-0.02,-0.01,0,0.01,0.02,0.03},
x tick label style={yshift=-5pt},
y tick label style={xshift=-5pt},
]
\addplot[line width=0.90mm, red,] table[x index=0,y index=1]{./figures/DOS_vacancy_spin_pos_4pcnt.txt};
\addplot[line width=0.90mm, dashed,blue,] table[x index=0,y index=2]{./figures/DOS_vacancy_spin_pos_4pcnt.txt};
\addplot[mark=none, black, dashed, ultra thick] coordinates{(0, -0.05) (0, 10000)}; 
\node[] at (axis cs: 0.003,4000) {\huge{$\lambda_{\text{F}}$}};
\legend{\huge{$\;$Spin-up},\huge{$\;$Spin-down}};
\end{axis}
\end{tikzpicture}}
\label{fig:DOS_pos_strain}}
    \caption{(a) Magnetization density isosurfaces for hydrogenated $(9,9)$ armchair \ce{P2C3}NT, where the hydrogen atom (red color) is attached to the phosphorous atom. The blue and yellow color clouds denotes the spin-down and spin-up electrons, respectively. (b) Evolution of the magnetization energy per unit cell with respect to the applied strain. (c),(d) and (e) show the band diagram for the $-4\% $ strain, unstrained and $+4\%$ strain cases, respectively. Spin-up and spin-down channels are represented by solid red and dashed blue lines, respectively. (f) and (g) show the spin DOS for $4\%$ compression and tensile strain, respectively.}
    \label{fig:magnetism}
\end{figure}
Flat bands with Coulomb interactions are often associated with magnetism \citep{chen2018ferromagnetism}. However, in most flat-band materials, the electrons remain unpolarized. In the past, vacancies and hydrogenation of graphene and CNTs have been shown to induce magnetic order in these materials \citep{bhatt2022various,banhart2011structural, ma2004magnetic,yazyev2008magnetism,park2003magnetism,yang2005ferromagnetism}.  Taking cue from these studies, in Fig.~\ref{fig:magnetism},  we show that a $(9,9)$ armchair \ce{P2C3}NT can exhibit magnetic characteristics  when a hydrogen atom is adsorbed by the phosphorus atom (one hydrogen atom per two periodic layers in axial direction considered). The presence of  the hydrogen atom distorts the nanotube in the radial direction and induces anisotropy in the bond lengths and angles in the hexagonal plaquette. This breaks the local lattice symmetry and lifts the degeneracy of the flat bands in undeformed nanotube (Fig.~\ref{fig:band_diagram_mag_pristine}),  leading to  nonzero magnetism. In the absence of external strain, the nanotube has contributions from both spin-up (yellow color isosurface) and spin-down (blue color isosurface) electrons, making the magnetic order ferrimagnetic-like with total magnetic moment of $-0.0133$ $ \mu_B$ (middle column of Fig.~\ref{fig:magnetization_vs_strain_ntb}). The spin-up and spin-down orbitals are largely localized on the carbon atoms closest to the hydrogen atom with the local magnetic moments being $0.109 \; \mu_B$ and $-0.116 \; \mu_B$.

Since structural distortion often plays an important role in tuning magnetism \citep{boukhvalov2008hydrogen, yang2009itinerant}, we next applied axial strain to the hydrogenated nanotube. Interestingly, under compressive strain the nanotube transitions from ferrimagnetic{-like} to ferromagnetic{-like} state, where most contribution to the magnetic order comes from the spin-up electrons (left panel of Fig.~\ref{fig:magnetization_vs_strain_ntb}, corresponding to $4\%$ compressive strain). The magnetization increases under compression and saturates to $0.66 \mu_B$ at $-4\%$ strain. Under tensile strain, the nanotube exhibits a dynamic interplay between antiferromagnetic{-like} and ferromagnetic{-like} behaviors across different strain levels (see Fig.~\ref{fig:magnetization_vs_strain}). In particular, the nanotube under $+1\%$ strain has majority spin-up states with $0.50\,\mu_B$ magnetization, after which the polarity switches to spin-down with $-0.59\,\mu_B$ magnetization, at $+2\%$ strain. Upon further increasing the strain to $+3\%$ the nanotube becomes antiferromagnetic{-like}, and finally, under $+4\%$ strain it turns back to ferromagnetic{-like} order with magnetization $-0.57\,\mu_B$. Correspondingly, a  high concentration of spin-down clouds is visible in the right column of Fig.~\ref{fig:magnetization_vs_strain_ntb}. The strain induced ferromagnetic{-like} behavior can also be seen from the band diagrams and spin-DOS of two extreme strain cases, i.e., $4\%$ compression (Fig.~\ref{fig:band_diagram_mag_compression} and \ref{fig:DOS_neg_strain}) and extension (Fig.~\ref{fig:band_diagram_mag__tensile} and \ref{fig:DOS_pos_strain}), where the energy bands splits into spin-up (red solid lines) and spin-down (dashed blue lines) channels. Overall, this shows a remarkable example of controllable magnetic behavior in \ce{P2C3}NTs where spin polarity can be changed with the application of strain.

{To verify the robustness of the predicted magnetic states, additional spin-polarized calculations were performed using different initial magnetic moments on the H atom while keeping all other computational parameters unchanged. For both tensile and compressive strain, all calculations initialized with a finite magnetic moment converge to the same magnetic solution with a net magnetization magnitude of approximately $0.99\,\mu B$ per periodic unit cell. Depending on the initial spin configuration, the final magnetization was found to appear with either positive or negative sign; however, these solutions correspond simply to globally spin-inverted states that are energetically equivalent within numerical precision, which is  expected in collinear spin KS-DFT calculations. Calculations initialized with zero magnetic moment converge instead to a nonmagnetic state that is slightly higher in energy than the spin-polarized solution. This indicates that the magnetic configuration represents the energetically favored ground state. Since the primitive periodic cell contains a single adsorption site, the present calculations probe the existence of a localized magnetic moment associated with hydrogen adsorption rather than a rigorously defined long-range magnetic ordering. The resulting spin polarization can therefore be interpreted as exhibiting ferromagnetic-like behavior within the unit cell. The relatively small energy difference between magnetic and nonmagnetic solutions suggests a soft, itinerant magnetic character, similar to defect-induced magnetism reported previously for hydrogenated carbon nanostructures \citep{yang2009itinerant}. The strain dependence of the magnetic moment arises from changes in the local electronic structure induced by deformation of the nanotube lattice. In particular, tensile or compressive strain modifies the P–H bond geometry and the hybridization between the hydrogen 1s orbital and neighboring phosphorus-carbon states, thereby altering the exchange splitting of the defect-derived electronic state responsible for the magnetic moment.}

The above discussed mechanism of generating unpaired spins without an external magnetic field may find applications in the emergent fields such as quantum hardware devices and spintronics. {More realistic and detailed calculations of such magnetic effects (e.g. explicit inclusion of Hubbard corrections), especially within the context of device applications, form the scope of future work.} A discussion of magnetism effects in \ce{P2C3}NTs induced from a carbon vacancy is discussed in the SI.  { Notably, the periodically hydrogenated and vacancy configurations discussed here represent an idealized limit to maintain 1D translational periodicity. In realistic quasi-1D systems, random hydrogen adsorption or ionic vacancies may break translational symmetry and can induce disorder scattering \cite{ostrovsky2006electron,PhysRevB.84.195418}.} Further studies of the effect of concentration of {random disorder}, dopant, torsional deformation and nanotube chirality on magnetic properties of \ce{P2C3}NTs are all interesting subjects for future investigation. 

Next, we investigated the general impact of applied strains to the electronic structure of \ce{P2C3}NTs,  and observed that both torsional and axial deformations disrupt the flat band degeneracy near the Fermi level. Increasing the applied strain leads to an increase in the energy width of the flat bands, accompanied by a decrease in flatness: {for the maximum elastic strains considered here, the bandwidth of the flat bands near the Fermi level is $13.6 - 168.7$ meV for the zigzag $(12,0)$  tube under compression and $87.1 - 182.3$ meV for the armchair $(9,9)$ tube under twist}. This is also demonstrated by the electronic density of states plots of twisted armchair and uniaxially compressed zigzag nanotubes shown in Fig.~\ref{fig:DOS_twist_armchir} and Fig. S9, respectively. The height of the sharp peak of the eDOS decreases with applied strain, and its width broadens near $\lambda_F$. Specifically, when torsional deformation is applied to the prototypical example of an armchair nanotube with group order $\mathfrak{N}=9$, the flat bands become slightly dispersive close to the $\eta =0$ point, exhibiting partial flatness (Fig.~\ref{fig:banddiag_armchairfull_DFT_twist}). Although the gap diminishes between the top Dirac point and the flat bands, a minute gap emerges between the flat bands and the lower Dirac points situated at $\eta =0$ and $\eta =\pm 1/3$. Interestingly,  the electronic state corresponding to the point $P_1$ in armchair nanotube band diagrams (Fig.~\ref{fig:wave_func_armchair}) has contribution from the $p_z$ orbitals of both P and C atoms in an unstrained state, but it is concentrated only at the $p_z$ atomic orbitals of carbon atoms under torsional strain ( $P_1$ in Fig.~\ref{fig:wave_func_armchair_twisted}). Simultaneously, the electronic state corresponding to point $P_2$ in the flat band is redistributed to other sets of carbon atoms (Fig.~\ref{fig:wave_func_armchair_twisted}). However, the state corresponding to the point $P_3$ does not show a significant change in spatial distribution. Similarly, uniaxial strain also induces some degree of dispersion in the flat bands. As illustrated in the band structure of $(12,0)$ zigzag nanotube under longitudinal compression of $3.28\%$ (Fig. S9(a)), the flat bands show comparatively no drastic changes near the $\eta =0$ point. No significant effect of compression is observed on the spatial distribution of the electronic states either (see Fig. S9(c)-(e)). The TB band structure for the twisted armchair nanotube and the uniaxially compressed zigzag nanotube is presented in supplementary information (Fig. S2(c) \& (d)) agrees well with the ab inito results. Overall, the flat bands in both nanotubes exhibit some dispersion under small strains but remain largely robust, likely maintaining any strongly correlated electronic states in the material. This likely stems from the robustness of the electronic states in \ce{P2C3} sheets themselves to elastic strains (Figs.~S5, S6). Interestingly, this resilient behavior is in sharp contrast to other proposed 1D materials with flat bands (e.g., Carbon Kagome nanotubes \citep{yu2024carbon}), where small strains can break the local symmetries  of the unit cell, thus introducing more noticeable dispersion into the flat bands. Usually, under such circumstances, the quadratic band touching point evolves into a pair of tilted Dirac cones \citep{yu2024carbon}.  This makes \ce{P2C3}NTs a realistic quasi-one dimensional material platform where stable and robust strongly correlated physics can be studied. 

\begin{figure}[ht!]
\centering
\subfloat[]{
{
\includegraphics[scale =0.25,trim={7cm 2cm 7cm 1.5cm},clip]{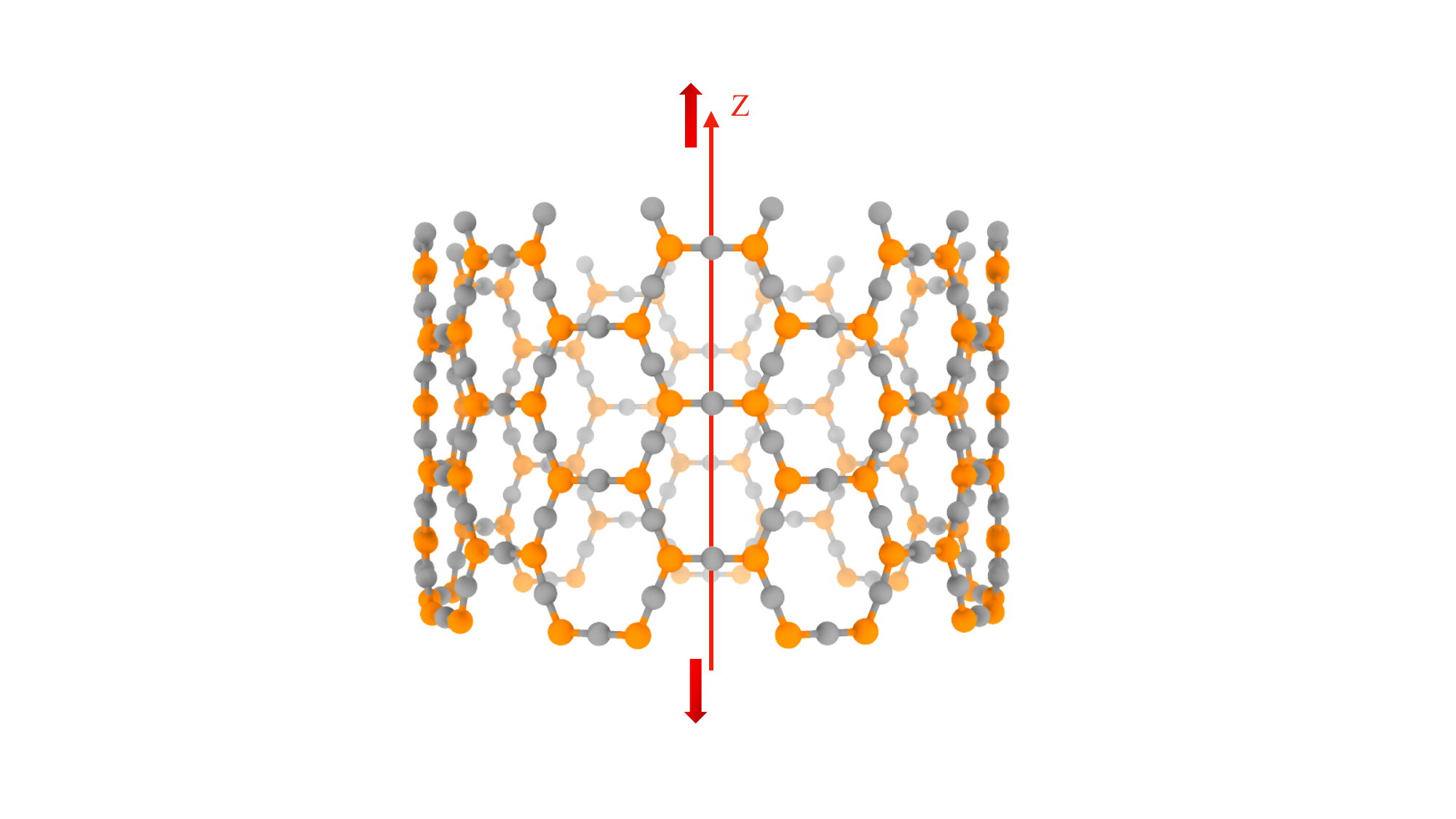}
}\label{fig:113def_ntube}}
\subfloat[]{{
\includegraphics[scale =0.34,trim={3cm 8cm 3.5cm 8cm},clip]{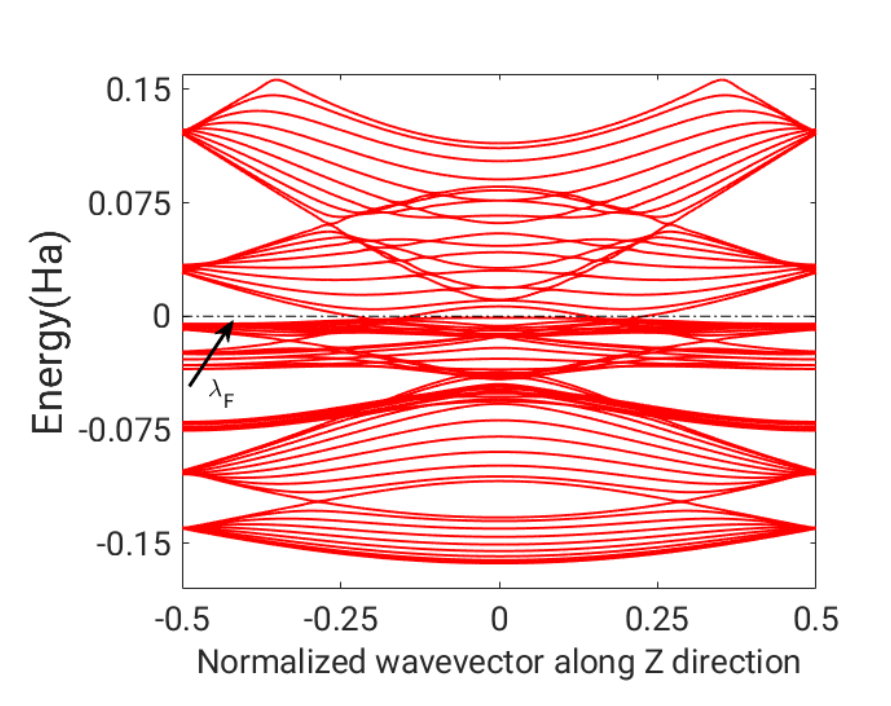}
}\label{fig:113def_ntube_banddiag}}\\
\hspace{0.5cm}\subfloat[]{
{
\includegraphics[scale =0.25,trim={7.5cm 2cm 7.5cm 1.5cm},clip]{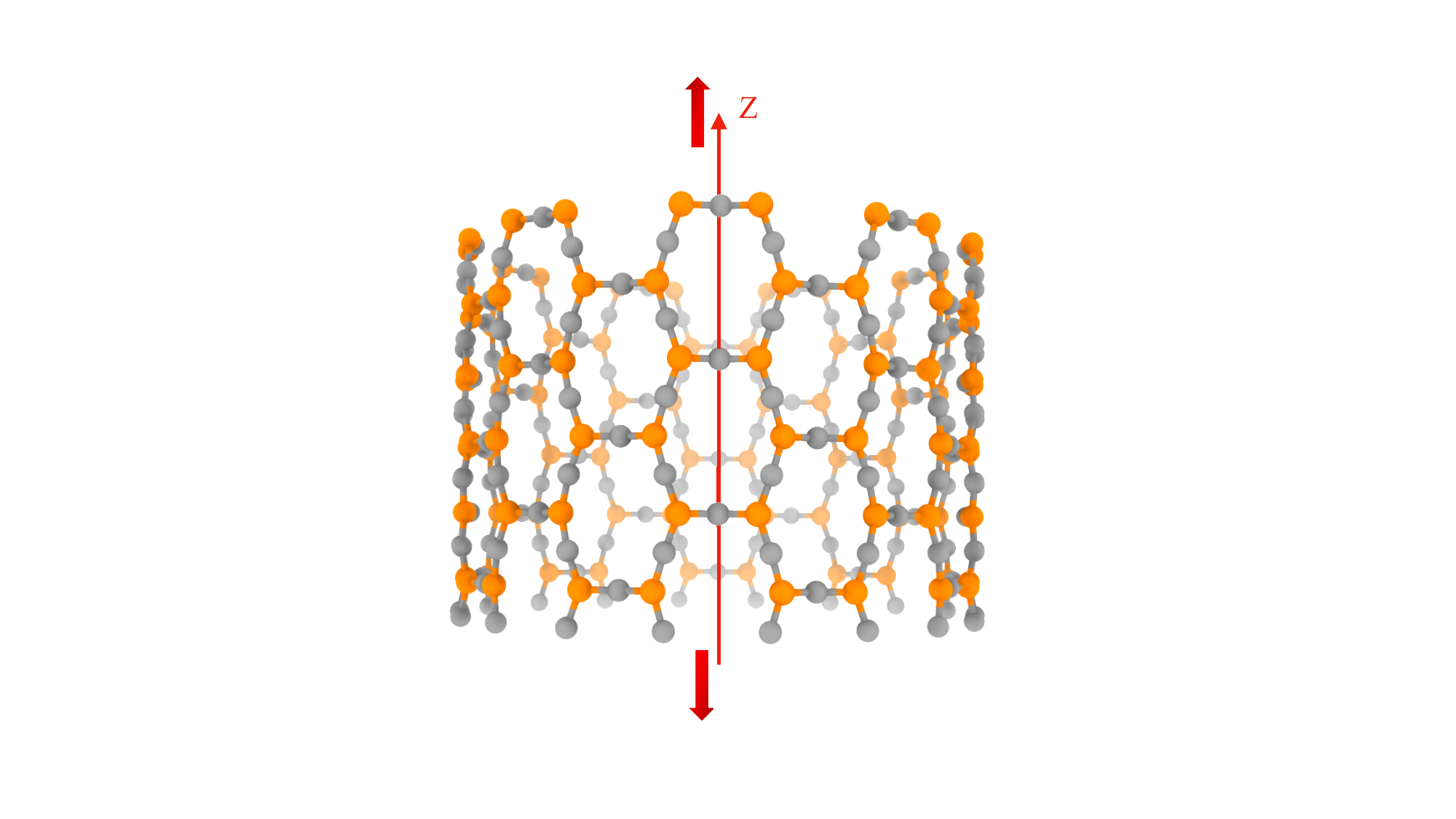}
}\label{fig:109def_ntube}} 
\subfloat[]{{
\includegraphics[scale =0.34,trim={2.5cm 8cm 2cm 8cm},clip]{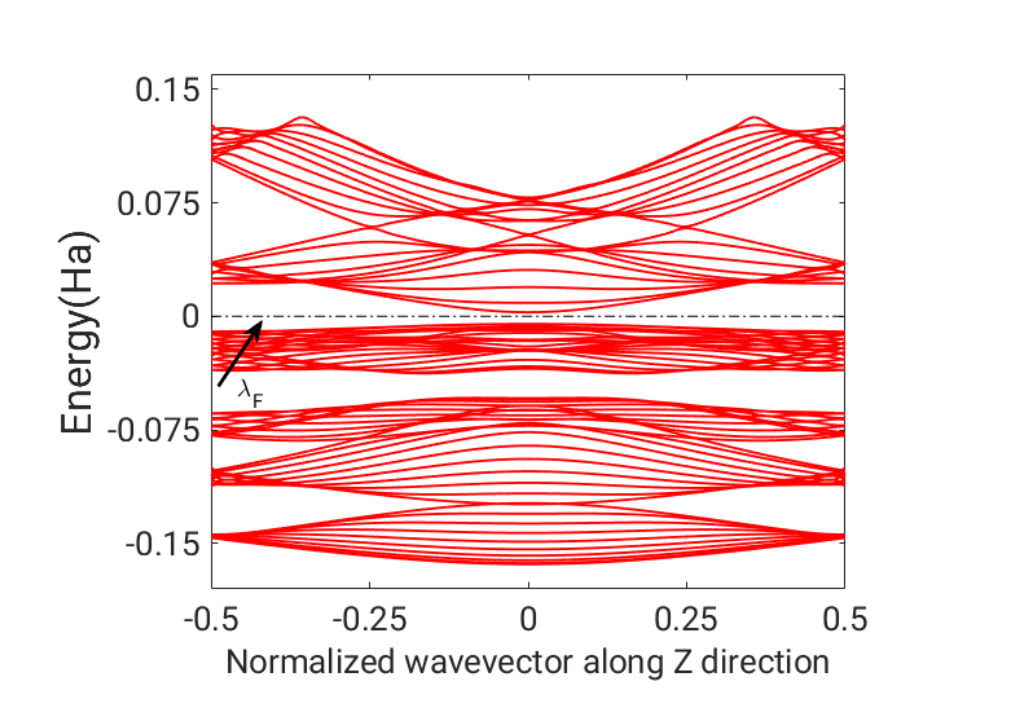}
}\label{fig:109def_ntube_banddiag}}\\
\hspace{0.cm}\subfloat[]{{
\includegraphics[scale =0.25,trim={7.5cm 2cm 6cm 2cm},clip]{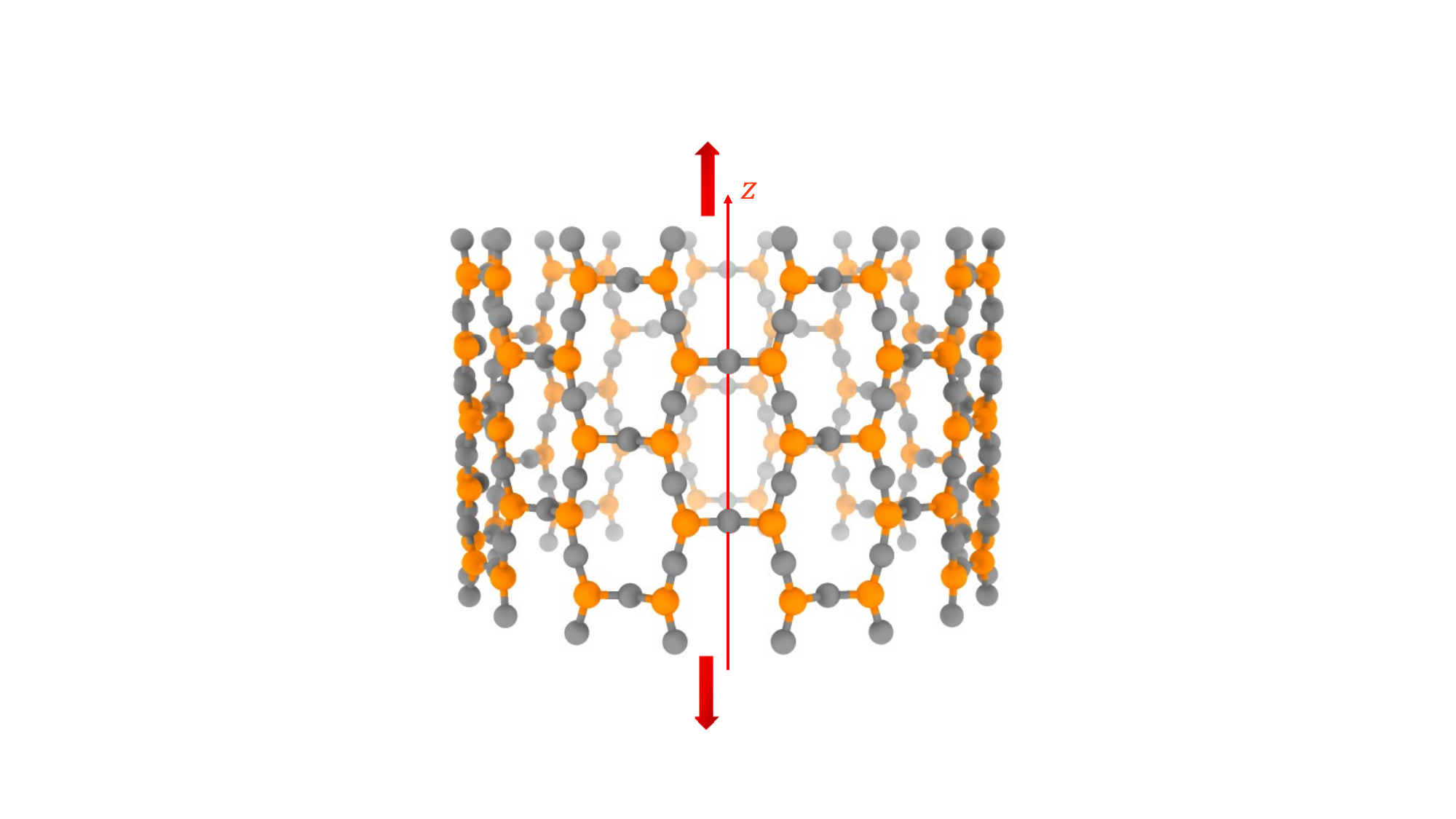}
}\label{fig:100def_ntube}} \hspace{0.cm}
\subfloat[]{{
\includegraphics[scale =0.24,trim={1.cm 6cm 2cm 6cm},clip]{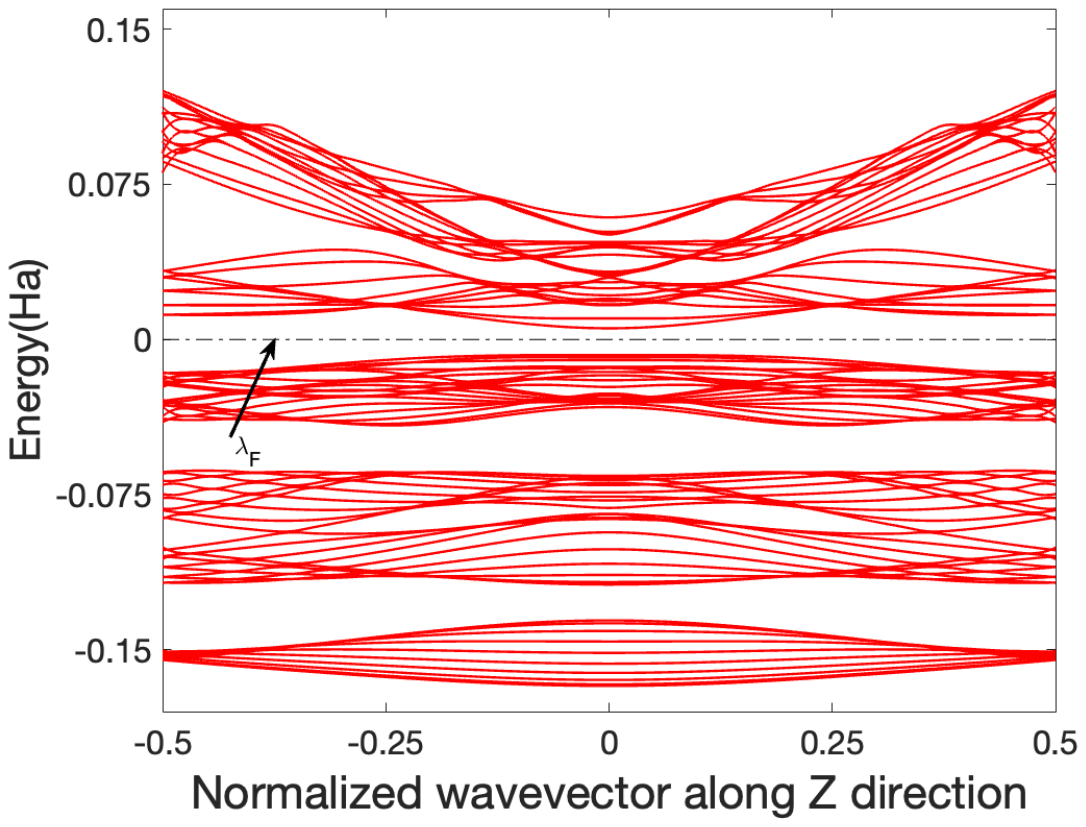}
}\label{fig:100def_ntube_banddiag}}\\
\hspace{0.85cm}\subfloat[]{{
\includegraphics[scale =0.25,trim={7.5cm 1cm 7.5cm 1.5cm},clip]{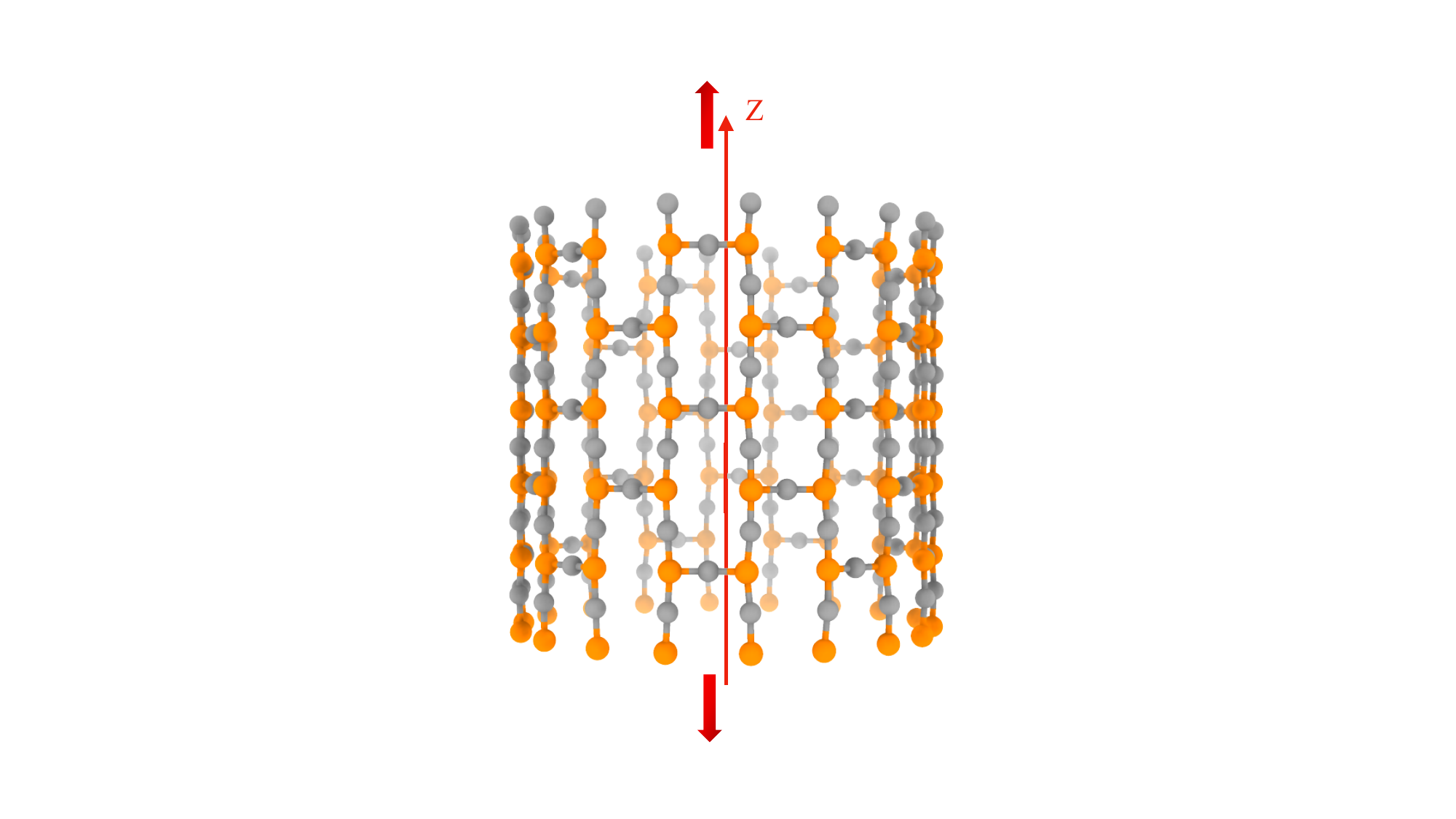}
}\label{fig:90def_ntube}}
\subfloat[]{{
\includegraphics[scale =0.34,trim={2.5cm 8cm 2.cm 8cm},clip]{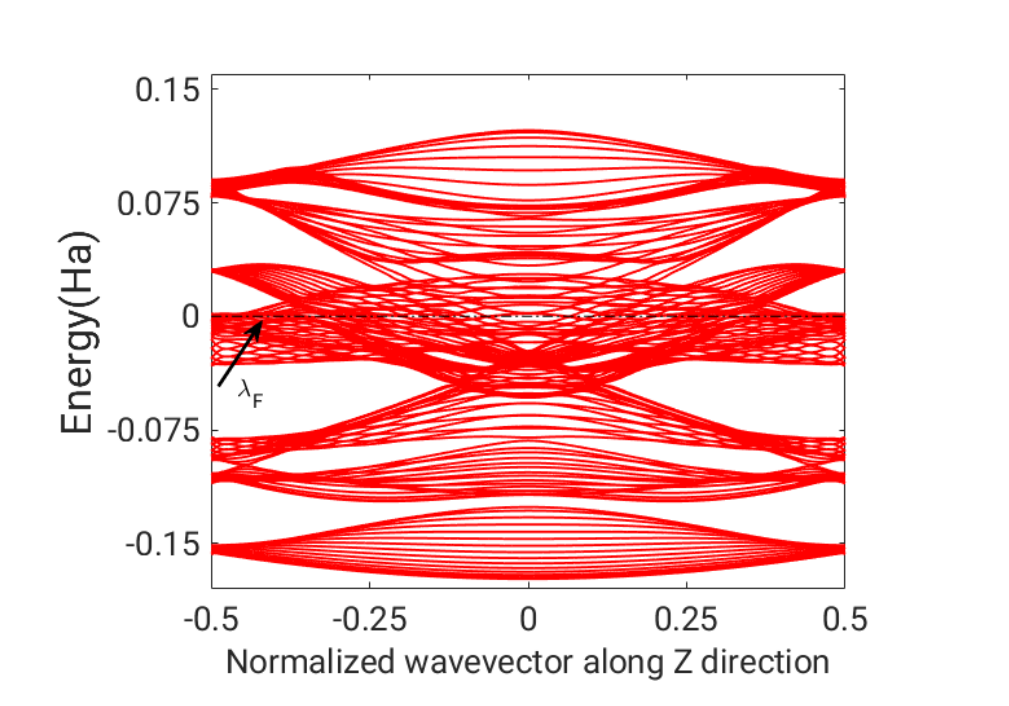}
}\label{fig:90def_ntube_banddiag}}\;
\caption{Structural and electronic phase transition of nanotube from pristine \ce{P2C3}NT with hexagonal unit cell to the brick wall \ce{P2C3}NT. Associated the band diagrams are also shown. The arrows represent the direction of applied strain. (a) \& (b) intermediate $6.35\%$, (c) \& (d) intermediate $12.34\%$, (e) \& (f) intermediate $18\%$ and (g) \& (h) brick wall $24.67\%$. }
\label{fig:indermediate_tubes}
\end{figure}

Continuing with our strain simulations, we next subjected the nanotubes to more extensive (inelastic) deformations, going up to the limit of failure. This leads to structural phase changes and triggers multiple quantum phase transitions. In particular, the armchair nanotube transforms into a ``brick-wall'' \citep{hou2015hidden} structure due to large tensile strain exerted along the tube axis (Fig.~\ref{fig:indermediate_tubes}). This results in shrinking of the nanotube along the radial direction, which is reminiscent of the Poisson effect. To study the transformation pathways in the nanotubes, we first 
strained and optimized the underlying  2D lattices using a ``freeze and relax'' strategy \citep{suwannakham2017dynamics} (details in the SI). Following this, we rolled up these structures to form nanotubes and carried out further structural relaxation to arrive at plausible low-energy transition states. Three transition points along the transformation pathway require special attention. At $6.35\%$ strain, the triple degeneracy point at $\eta =0$ is lifted and Dirac cones disappear (Fig.~\ref{fig:113def_ntube_banddiag}), thus signifying that this Dirac point is stable only for small deformation. Remarkably however, some of the dispersion-less states are still intact near the Fermi level. On further increasing the strain to $12.34\%$ the Dirac points of opposite vortices (Berry phase  $\pm \pi$) at $\eta = \pm 1/3$ annihilate at the time-reversal invariant point ($\eta = 0$) and open a gap near the Fermi energy (see band diagram in Fig.~\ref{fig:109def_ntube_banddiag}), suggesting transition from the metallic state to the insulating phase. This phenomenon of annihilation of Dirac nodes due to high structural anisotropy has been previously investigated in NNN TB models of graphene \cite{bernevig2013topological} and Kagome  lattices \citep{lim2020dirac,montambaux2018winding,jiang2019topological,montambaux2009merging}. In the brick-wall structure at $24.67\%$ strain, the bands become highly dispersive with multiple bands crossing the Fermi energy leading to another electronic phase transition from the insulating to the metallic state. This demonstrates that under large deformation \ce{P2C3}NTs show fascinating electronic state transitions. Such structural and electronic transitions have been theoretically studied \citep{hou2015hidden,hou2013hidden} in the literature, and have also been and experimentally explored in optical lattices \citep{tarruell2012creating}. \ce{P2C3}NTs provide a realistic material platform to explore such phenomena further. 

\begin{figure}[ht!]
\centering
\subfloat[]{
{
\includegraphics[scale =0.42,trim={1cm 5cm 1cm 4cm},clip]{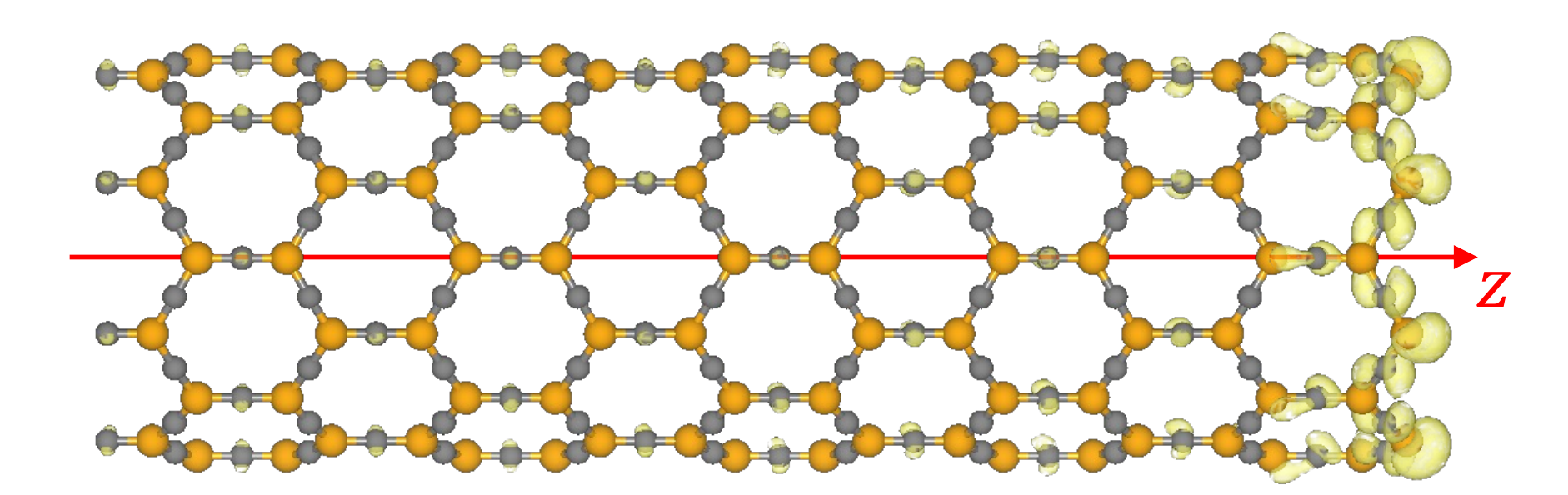}
}\label{fig:edge_state_homo}}\;
\subfloat[]{
{
\includegraphics[scale =0.4,trim={1cm 5cm 1cm 4cm},clip]{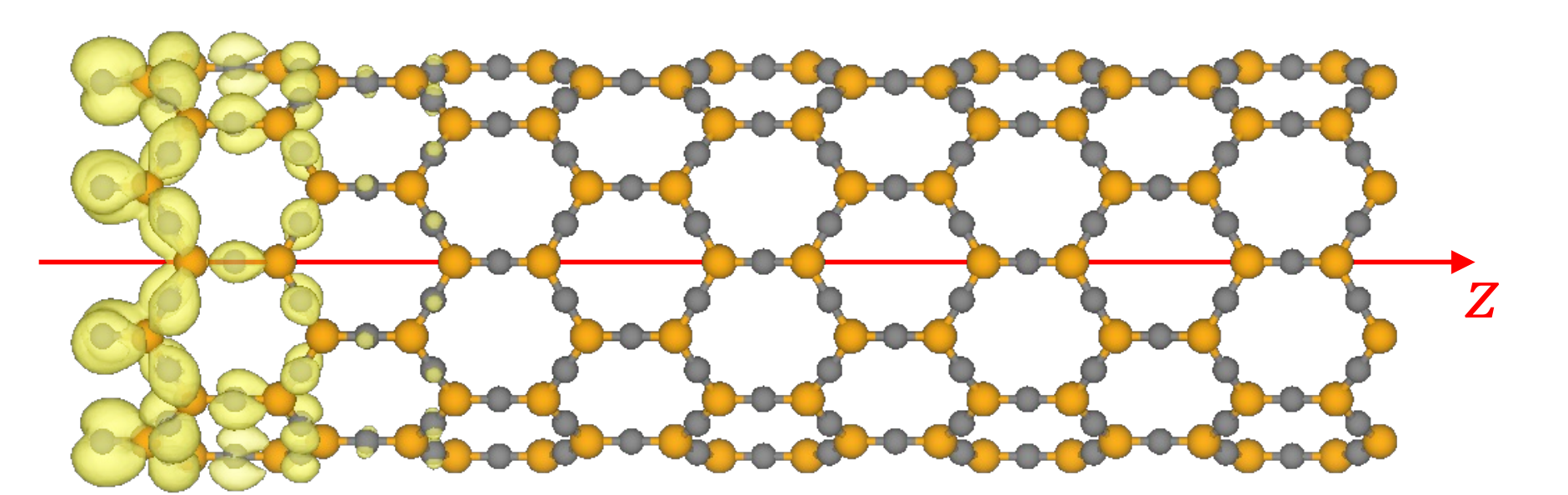}
}\label{fig:edge_state_lumo}}\;
\caption{Isosurfaces of wavefunction (absolute value squared) of finite $(8,0)$ zigzag \ce{P2C3}NT at the (a) valence band maximum and (b) conduction band minimum. The wavefunction has higher concentration at the end of the nanotube.}
\label{fig:edge_state}
\end{figure}

Recently, conventional CNTs have attracted attention from the perspective of 1D topological insulators, where the topology is characterized by the $\Z$ topological invariant (winding number) \citep{okuyama2019topological,izumida2016angular,okuyama2017topological,moca2020topologically,izumida2017topology}. The appearance of zero-energy edge states at the ends of finite-length CNTs of chirality $(n,m)$ depends on the integer $\mathfrak{N} = \gcd(n,m)$. Thus, only zigzag and chiral nanotubes with $\mathfrak{N}$ not divisible by 3 show edge states, and armchair CNTs remain topologically trivial. Motivated by these considerations and the fact that  \ce{P2C3}NTs have an underlying hexagonal lattice, we performed symmetry adapted first principles calculations and used $(8,0)$ zigzag \ce{P2C3}NTs as a prototypical example (both ends are zigzag-type). The edge states appearing at the valence band maximum (VBM) and the conduction band minimum (CBM) are shown in Fig. \ref{fig:edge_state_homo} \& \ref{fig:edge_state_lumo}. The end states are toward the right-hand side for the VBM and the left-hand side for the CBM. Notably in the studies described above, we do not explicitly compute a winding number for  \ce{P2C3}NTs; our interpretation of the edge states is based on the CNT analogy, Berry-phase arguments, and the finite-length wavefunction localization. We expect that these edge states will appear in other chiral \ce{P2C3}NTs, which invites further study. While it is well known that Dirac points induce edge states due to non-trivial Berry phases \citep{wakabayashi2009electronic,ando2005theory,ryu2002topological}, in \ce{P2C3}NTs the Dirac points intersect with the flat bands, thus suggesting both these electronic features are responsible for the edge states in this material. Our results show that \ce{P2C3}NTs are an exciting example of a quasi-1D nanostructure that supports topological behavior and electron transport at the edges. 

In summary, we have introduced \ce{P2C3}NTs formed through a roll-up construction of monolayer phosphorus carbide (\ce{P2C3}). Using symmetry adapted first principles calculations and other theoretical tools, we extensively characterized two types of nanotubes, armchair and zigzag. These nanotubes provide an unique platform where both Dirac fermions and strongly correlated states co-exist in a realistic 1D nanostructure. The orbital analysis shows that the electronic bands near the Fermi level are the direct sum of the band diagrams of  honeycomb splitgraph and Kagome lattices. \ce{P2C3}NTs develop  magnetic order on creating a carbon vacancy or by doping with hydrogen, and the magnetic behavior is highly controllable via strain in the latter case. The flat band states in both types of nanotubes are robust to small deformation. Under large tensile strains, the nanotubes undergo a structural transition process to a ``brick-wall'' phase, and we observed various fascinating electronic phenomena including Dirac cone annihilation and multiple metal-insulator transitions in the transition pathway. The finite nanotube simulations show topological features in the form of the localized edge states induced by Dirac points and flat bands. Calculations of structural properties suggest that \ce{P2C3}NTs are stable structures at the room temperature, and due to the relatively low bending energy of  \ce{P2C3} sheets, may be easily fabricated. Many novel types of carbon and phosphorus allotrope have been successfully synthesized in recent years, and it seems likely that \ce{P2C3} nanotubes can be grown and investigated experimentally in the near future. {Finally, further computational characterization, including explicit transport, many-body effects, finite temperature correlation calculations, as well as more reliable simulations of magnetic effects, form the scope of future work.}



\begin{suppinfo}
Discretization parameters for Helical DFT, determination of deformation energies and torsional/extensional stiffness values, tight-binding model for \ce{P2C3}NTs bands, band structures for some 2D \ce{P2C3} and some \ce{P2C3}NTs; projected density of states, magnetism studies on carbon vacancy,  magnified band diagrams of some structures shown in main text (PDF)
\end{suppinfo}

\begin{acknowledgement}
This work was primarily supported by grant DE-SC0023432 funded by the U.S. Department of Energy, Office of Science. This research used resources of the National Energy Research Scientific Computing Center, a DOE Office of Science User Facility supported by the Office of Science of the U.S. Department of Energy under Contract No.~DE-AC02-05CH11231, using NERSC awards BES-ERCAP0033206, BES-ERCAP0025205, BES-ERCAP0025168, and BES-ERCAP0028072. ASB acknowledges funding from UCLA’s Council on Research (COR) Faculty Research Grant and the Faculty Career Development Award from UCLA. Contributions by SS to this manuscript were made mainly during his time at the University of Minnesota. SS would like to thank Richard D. James (University of Minnesota) for helpful discussions and for providing support through the Vannevar Bush Faculty Fellowship (Grant No. N00014-19-1-2623) and the Air Force Defense Research Sciences Program Grant No. FA9550-24-1-0344. The authors acknowledge the use of the GPT-5 (OpenAI) model to polish the language and edit grammatical errors in some sections of this manuscript. The authors subsequently inspected, validated and edited the text generated by the AI model, before incorporation. The authors also acknowledge help from Kartikey Srivastava (UCLA) for help with a few simulations during the preparation of a revised version of the manuscript.
\end{acknowledgement}

\begin{center}
---
\end{center}

\clearpage
\renewcommand{\thefigure}{S\arabic{figure}}
\setcounter{figure}{0}
\section{Supporting Information (SI)}

\noindent \textbf{\underline{Discretization parameters for Helical DFT:}} To reduce computational burden, Helical DFT simulations were conducted in three successive phases, with increasing levels of discretization fineness \citep{yu2024carbon}. Initially, given a nanotube and applied strain parameters, structural relaxation was carried out using limited memory Broyden–Fletcher–Goldfarb–Shanno (LBFGS) algorithm \citep{liu1989limited}, and the force convergence criterion was set to $1$ mHa/bohr. For these simulations, the real space mesh spacing was set to $h=0.3$ bohr and $15\;\eta$ points were used to sample the helical reciprocal space; these parameters having been previously ascertained to result in chemically accurate energies and forces \citep{banerjee2021ab, yu2022density}. Relaxed structures were subsequently recomputed with more stringent discretization parameters ($h=0.25$ bohr and $21 \; \eta$-points), to evaluate the self-consistent fields and ground state energies. Finally, these self-consistent fields were used to set up the Kohn-Sham Hamiltonian and to then carry out a single (non-self-consistent) diagonalization step with a large number of reciprocal space points ($45$ $\eta$-points).  This last step was used to determine the band-diagrams and other related electronic properties.

\noindent \textbf{\underline{Determination of deformation energies and torsional/extensional stiffness values:}} Symmetry adapted simulations of nanotubes and their deformations have been described in detail in our previous contributions \citep{banerjee2021ab, yu2022density, yu2024carbon}. We summarize the key ideas here for the sake of completeness. For a nanotube with axis $\textbf{e\textsubscript{Z}}$, the symmetry group of the nanotube consists of the following collection of isometries (i.e., rotations and translations):
\beqs
\mathcal{G} = \Big\{ \Upsilon_{\zeta,\mu}=\big(\bfR_{(2\pi \zeta\alpha+\mu\Theta)}|
\,\zeta\tau \textbf{e\textsubscript{Z}}):
 \zeta \in \mathbb{Z},\mu=0,1,\dots,\mathfrak{N} - 1\Big\}\,.
\eeqs
Each symmetry operation $\Upsilon_{\zeta,\mu}$ is a screw transformation that consists of a rotation  about $\textbf{e\textsubscript{Z}}$ by the angle $2\pi \zeta\alpha+\mu\Theta$ (denoted via the action of the rotation matrix ${\textbf{R}}_{(2\pi \zeta\alpha+\mu\Theta)}$), along with simultaneous translation by $\zeta\tau$ about the same axis. The quantity $\mathfrak{N}$ is a natural number that captures cyclic symmetries in the nanotube, with the angle $\Theta=2\pi/\mathfrak{N}$ (i.e., $\mathfrak{N} = n$ for armchair $(n,n)$ and zigzag $(n,0)$ nanotubes). The scalar $\alpha$ is related to the applied or intrinsic twist in the nanotube, and the parameter $\tau$ is the pitch of the screw transformation symmetries of the nanotube. The amount of twist per unit length is $\beta = 2\pi\alpha/\tau$. To describe the complete nanotube, let $\calP = \{\bfr_1, \bfr_2,\ldots, \bfr_M: \bfr_i \in \rz^3\}$ denote the coordinates of the representative atoms in the symmetry adapted simulation cell. Then, the collection of coordinates of the entire structure can be expressed as:
\begin{align}
\mathcal{S} = \underset{\mu=0,1,\cdots,\mathfrak{N}-1}{\bigcup_{\zeta \in \mathbb{Z}}}\!\!\bigcup^{M}_{i=1} {\textbf{R}}_{(2\pi \zeta\alpha+\mu\Theta)}\mathbf{r}_{i}+\zeta\tau\,\textbf{e\textsubscript{Z}}\,.
\label{Eqn:sim_atoms}
\end{align}

For torsion simulations, we vary (in uniform steps) the parameter $\alpha$  described above. We use the limit of linear response for conventional CNTs \citep{Dumitrica_James_OMD} as the upper limit of imposed twist, going up to $\beta = 4.5^{\circ}$ of twist per nanometer. For each deformed configuration, the atomic forces are relaxed and the twisting energy per unit length of the nanotubes is computed in terms of the difference in the ground state free energy (per simulation cell) of the twisted and untwisted structures \citep{yu2024carbon}, i.e.:
\begin{align}
U_{\text{twist}}(\beta) = \frac{\mathfrak{N}}{\tau} \bigg({\calF}_{\substack{\text{Ground}\\ \text{State}}}(\calP^{**} , \calD, \calG|_{\beta}) - {\calF}_{\substack{\text{Ground}\\ \text{State}}}(\calP^{*}  , \calD, \calG|_{\beta=0})\bigg)\,.
\end{align}
In the equation above, $\calG|_{\beta}$ and $\calG|_{\beta = 0}$ denote the symmetry groups associated with the twisted and untwisted structures respectively,  and $\mathfrak{N}$ denotes the nanotube cyclic group order. Furthermore, $\calP^{**}$ and $\calP^{*}$ denote the collections of relaxed positions of the atoms in the symmetry adapted unit cell in each case. From these values of the deformation energy, the torsional stiffness is computed as:
\begin{align}
k_{\text{twist}} = \hpd{U_{\text{twist}}(\beta)}{\beta}{2}\bigg\rvert_{\beta = 0}\,.
\end{align}

\begin{figure}
    \centering
    \subfloat[]{\scalebox{0.35}
{
\begin{tikzpicture}
    \begin{axis}[
    width=\textwidth,
    xlabel={E-$\lambda_{F}$ (Ha)},
    ylabel={Projected density of states $\aleph_{T_{\text{e}}}(\cdot)$ ($\text{Ha}^{-1}$)},
    legend columns=2, 
    legend style={at={(0.7,0.83)},anchor=west,font=\sffamily\Large, fill=none,cells={anchor=west},row sep=1pt},
    y label style={xshift=-10pt},/tikz/column 2/.style={
                column sep=5pt},
    label style={font=\sffamily\Large},
    tick label style={font=\sffamily\Large},
    xmin=-0.17, xmax=0.18,
    ymin = -0.05, ymax=400,
    ytick={0,100,200,300,400,500,600},
    yticklabels={0,100,200,300,400,500,600},
    xtick={-0.2,-0.15,-0.1,-0.05,0,0.05,0.1,0.15,0.2},
    xticklabels={-0.2,-0.15,-0.1,-0.05,0,0.05,0.1,0.15,0.2},
    x tick label style={yshift=-5pt},
    y tick label style={xshift=-5pt},
    ]
    
    \addplot [line width=0.50mm,smooth,black] table[x index = 0, y index = 1]{./figures/Cpdos_p2c3.txt};
    \addlegendentry{\Large{$C_{s}$}}

     \addplot [line width=0.50mm,smooth,orange] table[x index = 0, y index = 1]{./figures/Ppdos_p2c3.txt};
    \addlegendentry{\Large{$P_{s}$}}
    
    \addplot [line width=0.50mm,smooth,red] table[x index = 0, y index = 3]{./figures/Cpdos_p2c3.txt};
    \addlegendentry{\Large{$C_{p_{xy}}$}}

    \addplot [line width=0.50mm,smooth,green] table[x index = 0, y index = 3]{./figures/Ppdos_p2c3.txt};
    \addlegendentry{\Large{$P_{p_{xy}}$}}
    
    \addplot [line width=0.80mm,smooth,blue] table[x index = 0, y index = 2]{./figures/Cpdos_p2c3.txt};
    \addlegendentry{\Large{$C_{p_z}$}}
    
        \addplot [line width=0.50mm,smooth,cyan] table[x index = 0, y index = 2]{./figures/Ppdos_p2c3.txt};
    \addlegendentry{\Large{$P_{p_z}$}}

\addplot[mark=none, black, dashed, ultra thick] coordinates{(0, -0.05) (0, 900)}; 
    \node[] at (axis cs: 0.02,250) {\Large{$\lambda_{\text{F}}$}};
    
    \end{axis}
\end{tikzpicture}
}\label{supp:fig:pdos_armchair}}\hspace{1cm}
\subfloat[]{\scalebox{0.35}
{
\begin{tikzpicture}
    \begin{axis}[
 width=\textwidth,
    xlabel={E-$\lambda_{F}$ (Ha)},
    ylabel={Projected density of states $\aleph_{T_{\text{e}}}(\cdot)$ ($\text{Ha}^{-1}$)},
    legend columns=2, 
    legend style={at={(0.7,0.83)},anchor=west,font=\sffamily\Large, fill=none,cells={anchor=west},row sep=1pt},
    y label style={xshift=-10pt},/tikz/column 2/.style={
                column sep=5pt},
    label style={font=\sffamily\Large},
    tick label style={font=\sffamily\Large},
    xmin=-0.17, xmax=0.2,
    ymin = -0.05, ymax=400,
    ytick={0,100,200,300,400,500,600,700},
    yticklabels={0,100,200,300,400,500,600,700},
    xtick={-0.2,-0.15,-0.1,-0.05,0,0.05,0.1,0.15},
    xticklabels={-0.2,-0.15,-0.1,-0.05,0,0.05,0.1,0.15},
    x tick label style={yshift=-5pt},
    y tick label style={xshift=-5pt},
    ]
    
 \addplot [line width=0.50mm,smooth,black] table[x index = 0, y index = 1]{./figures/Cpdos_p2c3_zz.txt};
    \addlegendentry{\Large{$C_{s}$}}

     \addplot [line width=0.50mm,smooth,orange] table[x index = 0, y index = 1]{./figures/Ppdos_p2c3_zz.txt};
    \addlegendentry{\Large{$P_{s}$}}
    
    \addplot [line width=0.50mm,smooth,red] table[x index = 0, y index = 3]{./figures/Cpdos_p2c3_zz.txt};
    \addlegendentry{\Large{$C_{p_{xy}}$}}

    \addplot [line width=0.50mm,smooth,green] table[x index = 0, y index = 3]{./figures/Ppdos_p2c3_zz.txt};
    \addlegendentry{\Large{$P_{p_{xy}}$}}
    
    \addplot [line width=0.80mm,smooth,blue] table[x index = 0, y index = 2]{./figures/Cpdos_p2c3_zz.txt};
    \addlegendentry{\Large{$C_{p_z}$}}
    
    \addplot [line width=0.50mm,smooth,cyan] table[x index = 0, y index = 2]{./figures/Ppdos_p2c3_zz.txt};
    \addlegendentry{\Large{$P_{p_z}$}}

\addplot[mark=none, black, dashed, ultra thick] coordinates{(0, -0.05) (0, 900)}; 
    \node[] at (axis cs: 0.02,250) {\Large{$\lambda_{\text{F}}$}};
    
    \end{axis}
\end{tikzpicture}
} \label{supp:fig:pdos_zigzag}
}\\
\subfloat[]{{
\includegraphics[scale =0.5,trim={0cm 4cm 0cm 5cm},clip]{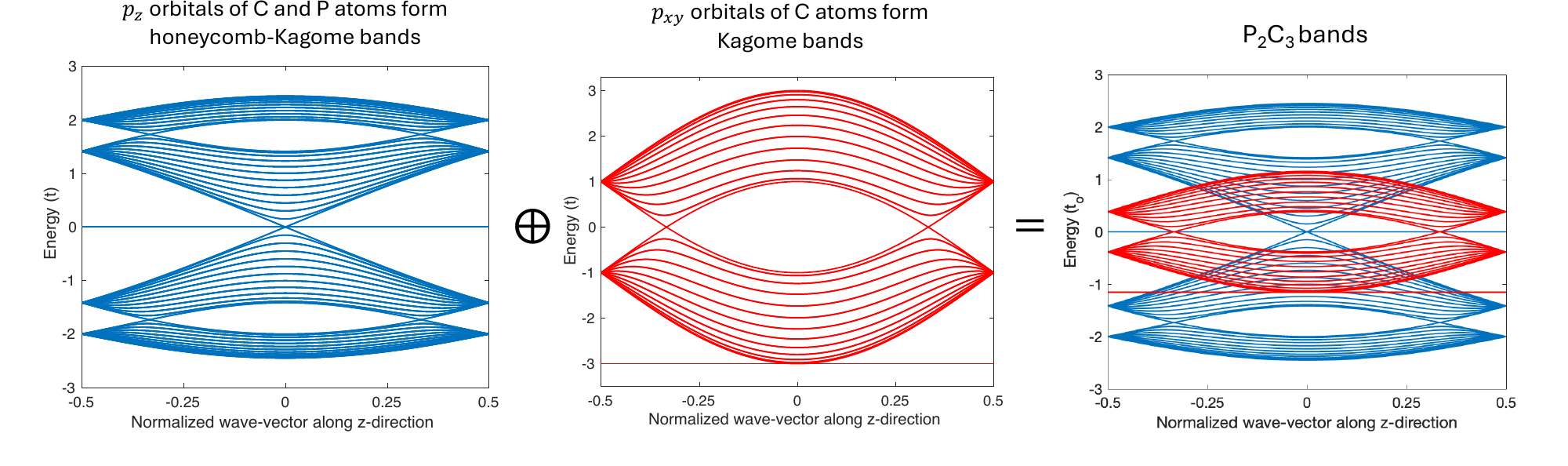}
}\label{supp:fig:mixing_bands}}
    \caption{ Projected density of states (pDOS) for (a) (9,9) armchair \ce{P2C3}NT and (b) (9,0) zigzag \ce{P2C3}NTs. (c) The representation of origin of bands in \ce{P2C3}NTs from a simple TB model. The honeycomb-Kagome type bands are due to the radially oriented $p_z$ orbitals of phosphorus and carbon atoms, whereas, in-plane $p_{xy}$ orbitals of carbon atoms form Kagome bands. The direct sum of these bands gives the bands of the \ce{P2C3} nanotube. The band diagram of armchair nanotube is shown as an example. Zigzag nanotube band structure can be constructed in a similar away.}
\end{figure}

For simulations involving axial stretch and compression, we proceed analogously. Given a value of the axial strain $\epsilon$, we modify the pitch of the screw transformation used to describe the nanotube, as $\tau = \tau_0(1+\epsilon)$. Here  $\tau_0$ denotes the equilibrium, undistorted value and in our simulations, we restricted $\epsilon$ to be between $\pm 3.3 \%$. We relax the atomic forces subsequently, and then compute the extensional energy per unit length of the nanotubes as the difference in the ground state free energy (per simulation cell), between stretched and unstretched configurations \citep{yu2024carbon}, i.e.:
\begin{align}
U_{\text{stretch}}(\epsilon) = \frac{\mathfrak{N}}{\tau_0} \bigg({\calF}_{\substack{\text{Ground}\\ \text{State}}}(\calP^{**} , \calD,  \calG|_{\tau=\tau_0(1+\epsilon)}) - 
{\calF}_{\substack{\text{Ground}\\ \text{State}}}(\calP^{*}  , \calD, \calG|_{\tau=\tau_0})\bigg)\,.
\end{align}
Here, $\calG|_{\tau=\tau_0(1+\epsilon)}$ and $\calG|_{\tau=\tau_0}$ denote the symmetry groups associated with the stretched and unstretched structures, respectively. Additionally, $\calP^{**}$ and $\calP^{*}$ denote the collections of relaxed positions of the atoms in the fundamental domain in each case. From this, we may calculate the stretching stiffness of the nanotubes as:
\begin{align}
k_{\text{stretch}} = \hpd{U_{\text{stretch}}(\epsilon)}{\epsilon}{2}\bigg\rvert_{\epsilon = 0}\,.
\end{align}
\noindent \textbf{\underline{Projected Density of States and Tight Binding (TB) Model:}} Fig.~\ref{supp:fig:pdos_armchair} \& \ref{supp:fig:pdos_zigzag} show the projected density of states (pDOS) plots for armchair and zigzag  \ce{P2C3}NTs. In both cases, the strong peak near the Fermi level is due to $p_z$ orbitals of carbon atoms, shown in dark blue. The red color peak in pDOS comes from $p_{xy}$ atomic orbitals of  carbon atoms which form the Kagome-type bands shown in the middle panel of Fig.~\ref{supp:fig:mixing_bands}. These pDOS plots are strongly suggestive that to a good approximation, the overall electronic structure of \ce{P2C3}NTs (see, e.g. Fig. $2$ of the main text)  is well described in terms of $p_z$ orbitals (arising from carbon and phosphorus atoms, and their hybridization) --- which result in honeycomb-kagome like bands, and $p_{xy}$ orbitals of the carbon atoms --- which result in pure kagome bands (Fig.~\ref{supp:fig:mixing_bands}). Notably, the honeycomb-kagome bands themselves can be understood in terms of pure honeycomb and kagome lattice bands in the sense of square-root topology formalism \citep{mizoguchi2020square,mizoguchi2021square,ma2020spin}. Interestingly, our interpretation of the the electronic structure of  \ce{P2C3}NTs (Fig.~\ref{supp:fig:mixing_bands}) and the subsequent construction of our TB model (described below) marks a departure from previous work \citep{huang2018double}, where planar \ce{P2C3} was interpreted to have ``double-Kagome'' bands. We believe our TB model results (see below) are closer to the first-principles electronic structure data, and replicates the observed effects of strain well.

We constructed a next-nearest neighbor (NNN) symmetry-adapted tight binding (TB) model to capture the salient features of the electronic structure of \ce{P2C3}NTs. We utilized the Dresselhaus approach \citep{dresselhaus2000carbon}, which entails developing a TB formulation for the flat sheet of \ce{P2C3}, followed by mapping the atoms of the two-dimensional lattice onto a cylinder, to apply boundary conditions suitable for the nanotube (see reference \citep{yu2024carbon} for further details of this approach). As mentioned above, the TB model considered here has contributions from two sets of orthogonal orbitals, i.e., three in-plane $p_{xy}$ orbitals of \ce{C} atoms, and three radially oriented $p_z$ orbitals of \ce{C} atoms, along with two more from \ce{P} atoms. To incorporate the influence of deformation on the nanotubes, we considered the NNN hopping for the $p_z$ honeycomb split graph bands. To explicitly write down the TB model, we note that since the interactions between the $p_{xy}$ and $p_z$ orbitals is negligible, the $8$-band TB Hamiltonian is written as direct sum of $p_{xy}$ Kagome bands and $p_z$ Honeycomb Kagome (HK) bands, i.e.:
\beq
\bfH = \sum_{i,\gamma}\varepsilon_{i\gamma}\mathsf{a}^\dagger_{i\gamma}\mathsf{a}_{i\gamma} + \sum_{\gamma}\sum_{\langle i,j\rangle}t_{(i\gamma,j\gamma)}\mathsf{a}^\dagger_{i\gamma}\mathsf{a}_{j\gamma} +\sum_{\gamma}\sum_{\langle\langle i,j \rangle\rangle}\tilde{t}_{(i\gamma,j\gamma)}\mathsf{a}^\dagger_{i\gamma}\mathsf{a}_{j\gamma} + \text{h.c.}\,.
\eeq
Here, the annhilation and creation operators are denoted by $\mathsf{a}_{i\gamma}$, $\mathsf{a}^\dagger_{i\gamma}$, respectively. The onsite energy of site $i$ and orbital $\gamma$ is $\varepsilon_{i\gamma}$, $t_{(i\gamma,j\gamma)}$ and $\tilde{t}_{(i\gamma,j\gamma)}$ are the hopping amplitudes between orbitals $\gamma$ of the nearest-neighbors (NNs) $\langle i,j\rangle$ and the next-nearest-neighbors (NNNs) $\langle\langle i,j \rangle\rangle$, respectively, and h.c. is the hermitian conjugate. The on-site energy of $p_z$ orbitals of carbon and phosphorous atoms is $\varepsilon_{\ce{C}_{p_z}} = 0$ and  $\varepsilon_{\ce{P}_{p_z}} = 0.05$eV, respectively. While, the NN hopping amplitude between $p_z$ orbitals both the types of atoms is $t_{\ce{CP}_{p_z}} = -2.6$eV. The interactions between three $p_{xy}$ orbitals for carbon is given by $t_{\ce{C}_{p_{xy}}} = 0.6$eV with the on-site energy $\varepsilon_{\ce{C}_{p_{xy}}} = -0.6$eV. The partially flat band of $p_{xy}$ character near $-0.08$ Ha is due to the consideration of NNN interaction of magnitude $\tilde{t}_{C_{p_{xy}}}= 0.1$eV between the $p_{xy}$ electrons. To incorporate the influence of deformation on the nanotubes, we also considered the NNN hopping amplitudes for $p_z$ HK, denoted as $\tilde{t}_{C_{p_{z}}} = 0.01$eV and  $\tilde{t}_{P_{p_{z}}} = 0.001$eV, respectively. The effect of deformation on the hopping parameter is given by:
\beq
t^\prime = t\exp\left[-\beta\left(\frac{|\delta_{i,j}|}{a_{i,j}}-1\right)\right].
\eeq
Here, $t$ is undeformed hopping parameter, $a_{i,j}$ is the distance between atom $i$ and $j$ and $\beta$ is Gr\"{u}neisen parameter \cite{mohiuddin2009uniaxial,mojarro2023strain} which is considered equal to 2 to match DFT results. The effect of strain on the atomic distance is given as $\delta_{i,j} = (\bfI + \epsilon)\bfa_{i,j}$, where $\epsilon$ is the strain matrix. The Poisson's ratio $\nu$ is  set to $0.165$. 

\begin{figure}[ht!]
\centering
\subfloat[]{
{
\includegraphics[scale =0.3,trim={1.cm 6.5cm 1.5cm 6cm},clip]{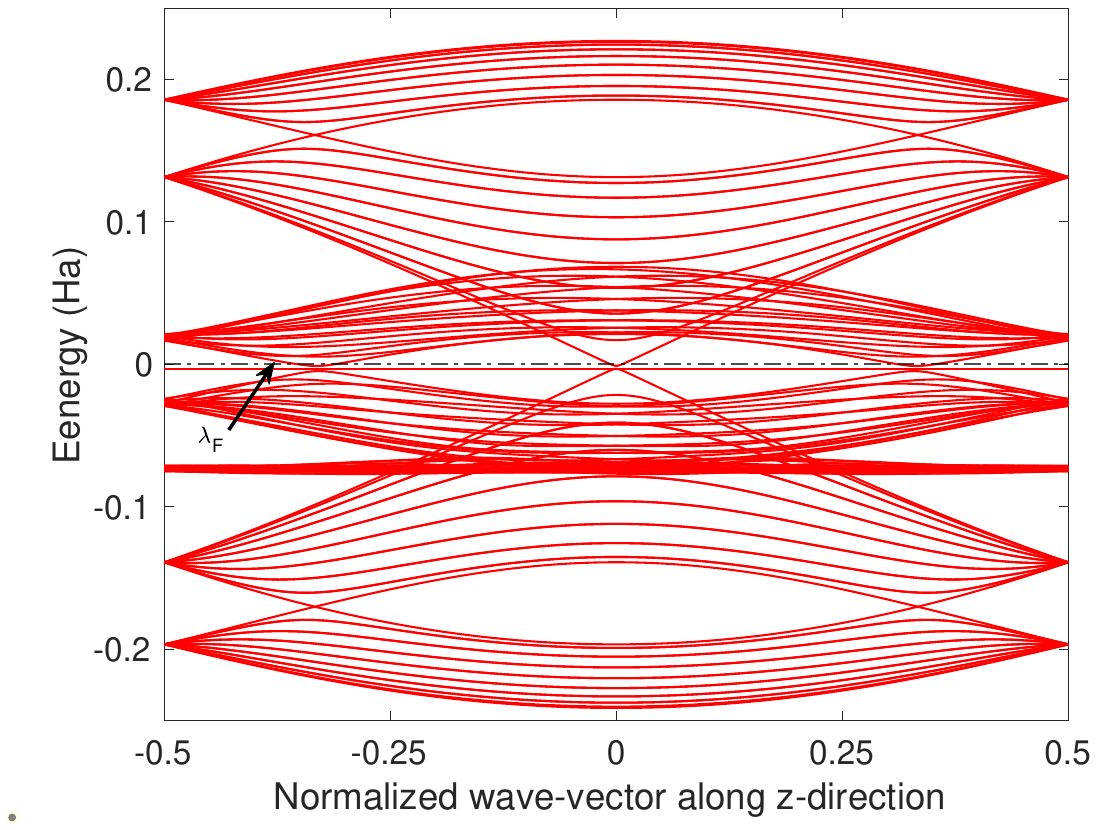}
}\label{supp:fig:TB_armchairfull}} \hspace{0.5cm}
\subfloat[]{
{
\includegraphics[scale =0.3,trim={1.cm 6.5cm 1.5cm 6cm},clip]{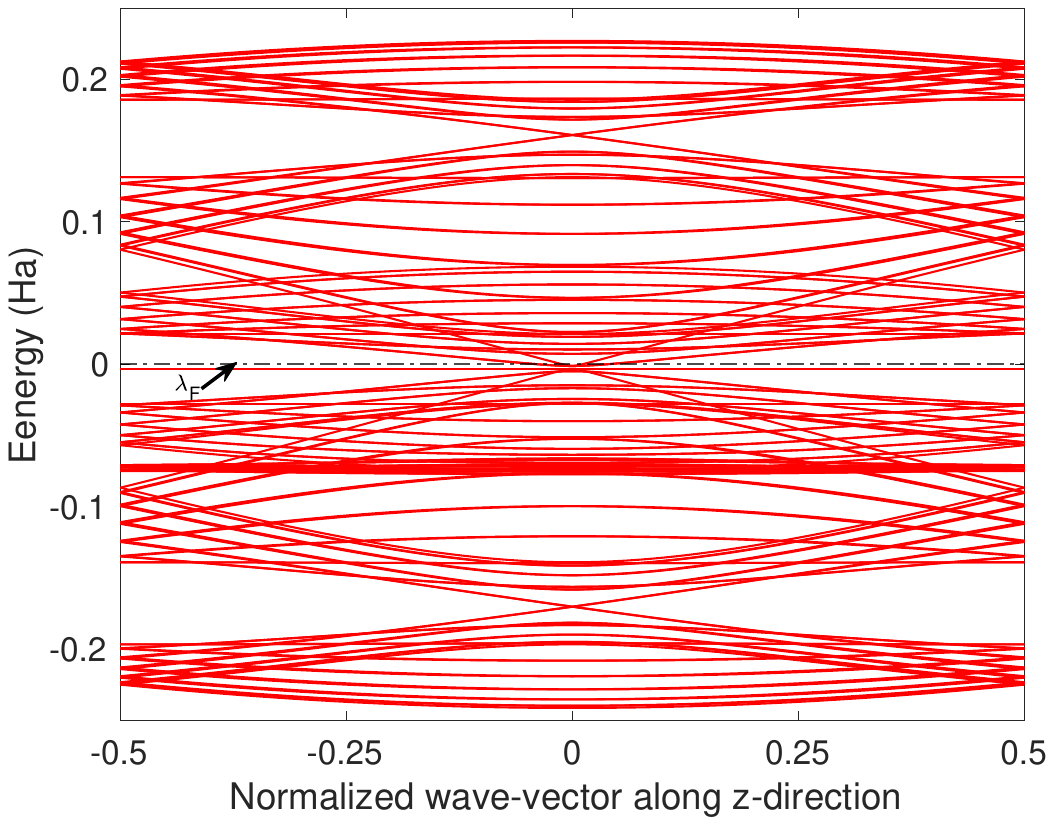}
}\label{supp:fig:TB_zigzagfull}}\\
\subfloat[]{{
\includegraphics[scale =0.3,trim={1.cm 6.5cm 1.5cm 6cm},clip]{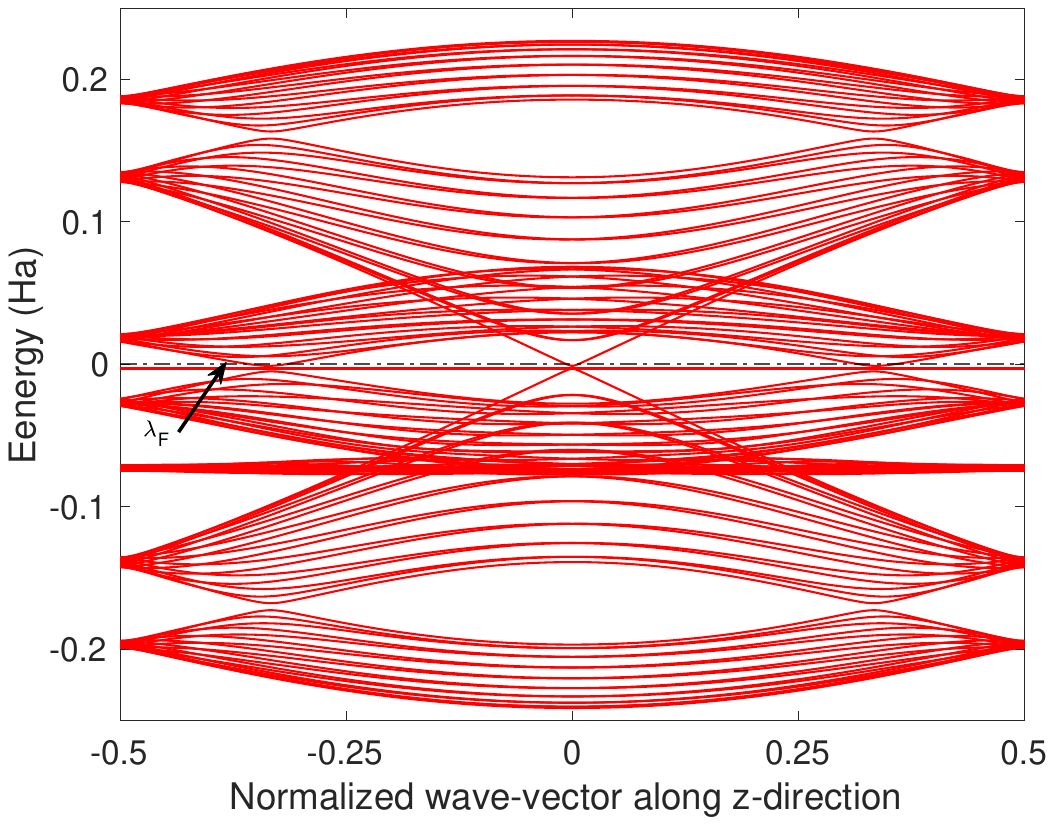}
}\label{supp:fig:TB_armchairfull_twist}} \hspace{0.5cm}
\subfloat[]{{
\includegraphics[scale =0.3,trim={1.cm 6.5cm 1.5cm 6cm},clip]{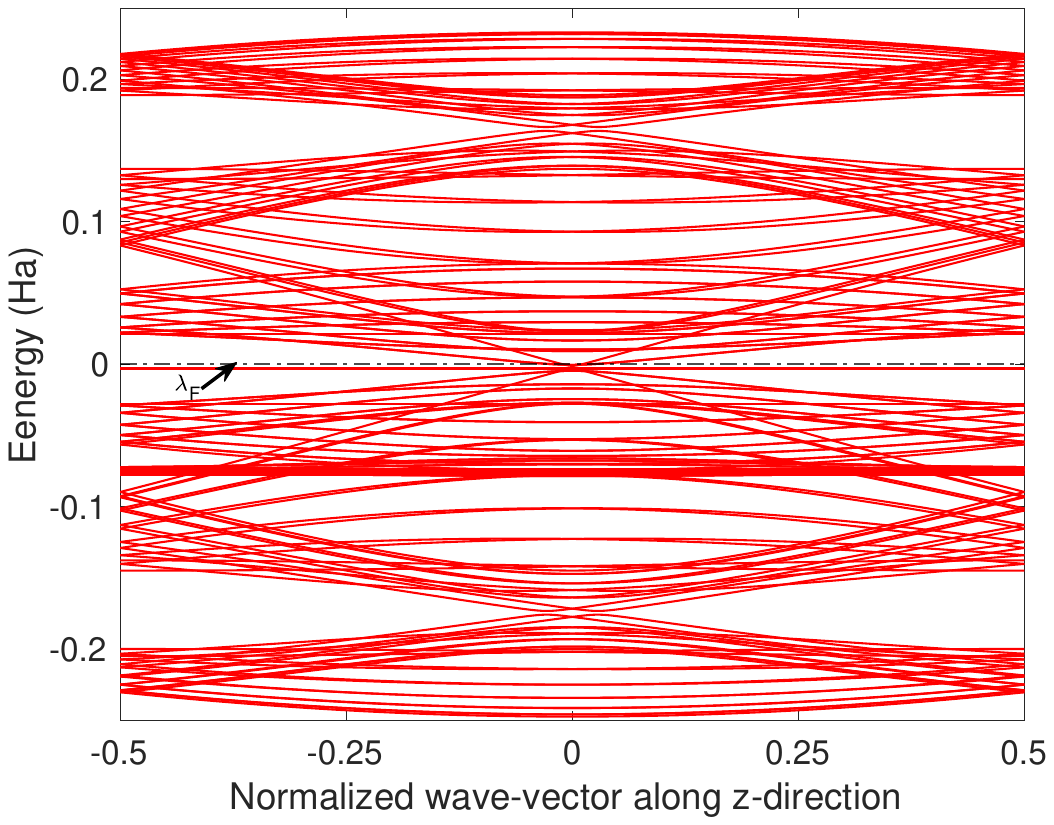}
}\label{supp:fig:TB_zigzag_twist}}
\caption{Tight binding band diagram of pristine and twisted $(9,9)$ armchair (a) and (c), and pristine and compressed $(12,0)$ zigzag (b) and (d) \ce{P2C3}NTs, respectively. The Fermi level $\lambda_F$ corresponds to the x-axis.}
\label{supp:fig:tight_binding_plots}
\end{figure}

The outcomes of our TB calculations for pristine \ce{P2C3}NTs are illustrated in Fig.~\ref{supp:fig:TB_armchairfull} \& \ref{supp:fig:TB_zigzagfull}, and for twisted armchair nanotube and uniaxially compressed zigzag nanotube are showcased in Fig.~\ref{supp:fig:TB_armchairfull_twist} \& \ref{supp:fig:TB_zigzag_twist}. It is evident from these figures that there is a remarkable qualitative agreement between these results and the first principles data presented elsewhere in the letter.
 

\begin{figure}[ht!]
    \centering
   \subfloat[]{
{
\includegraphics[scale =0.32,trim={7cm 0cm 7cm 0cm},clip]{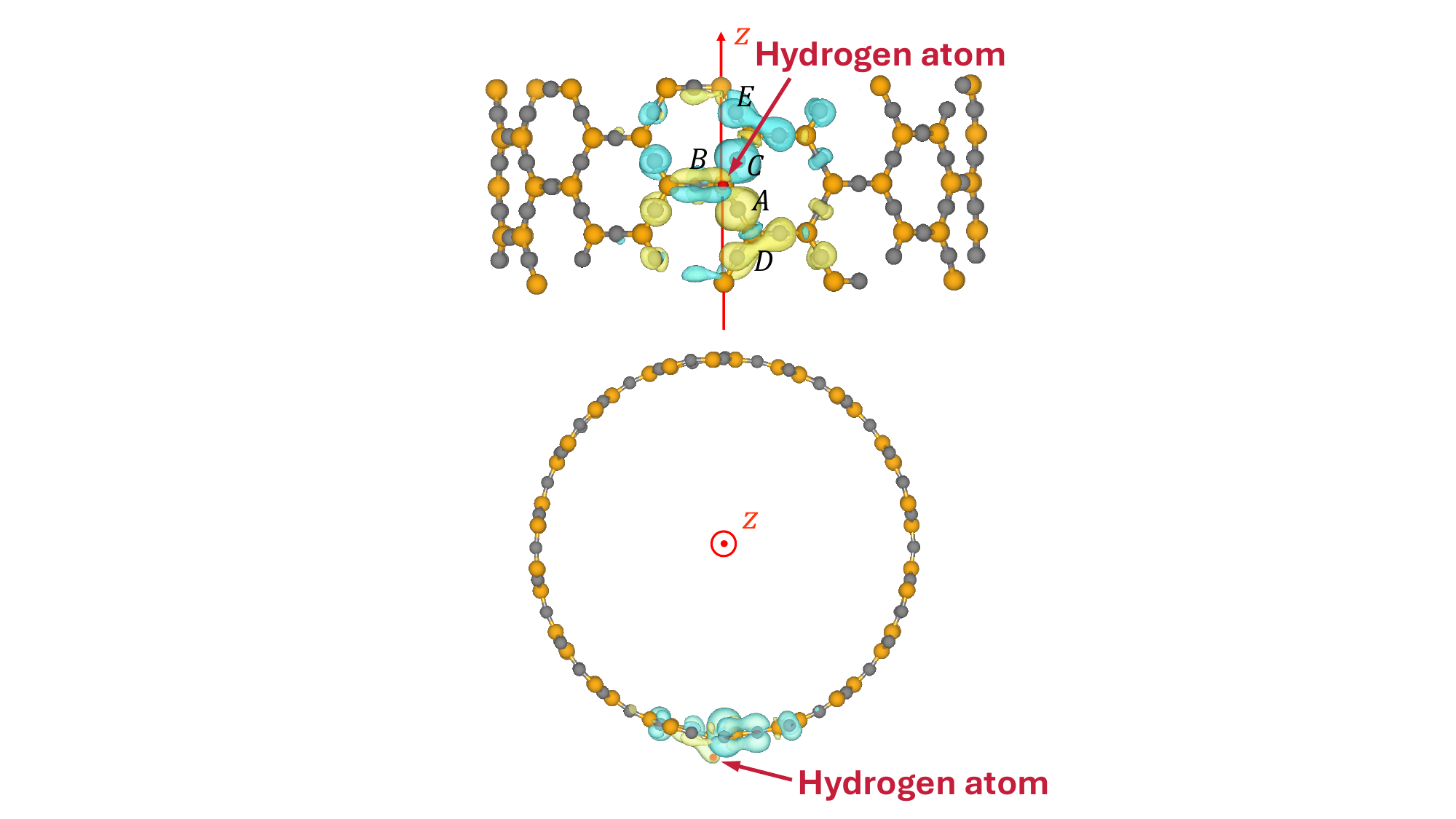}
}\label{supp:fig:p-H_density}}\;
 \subfloat[]{
{
\includegraphics[scale =0.3,trim={1cm 5.5cm 1.5cm 5.5cm},clip]{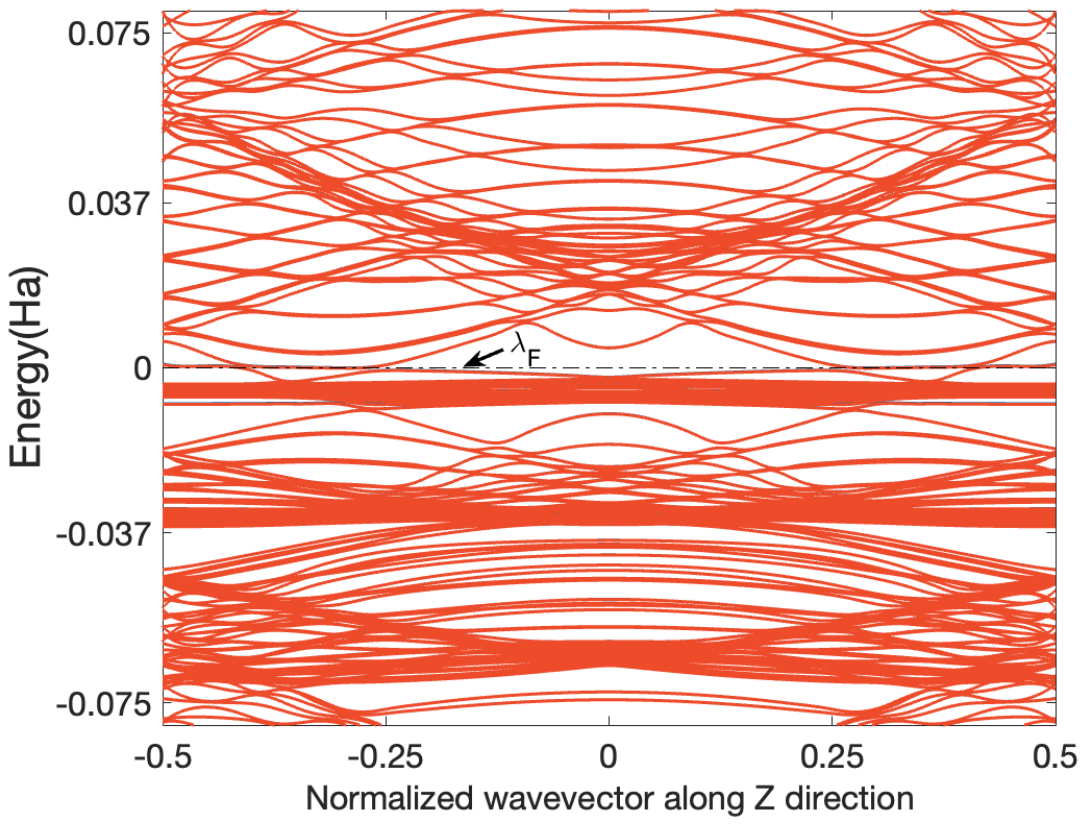}
}\label{supp:fig:P-H_band_diagram}}\\
   \subfloat[]{
{
\includegraphics[scale =0.3,trim={7cm 0cm 5.5cm 0cm},clip]{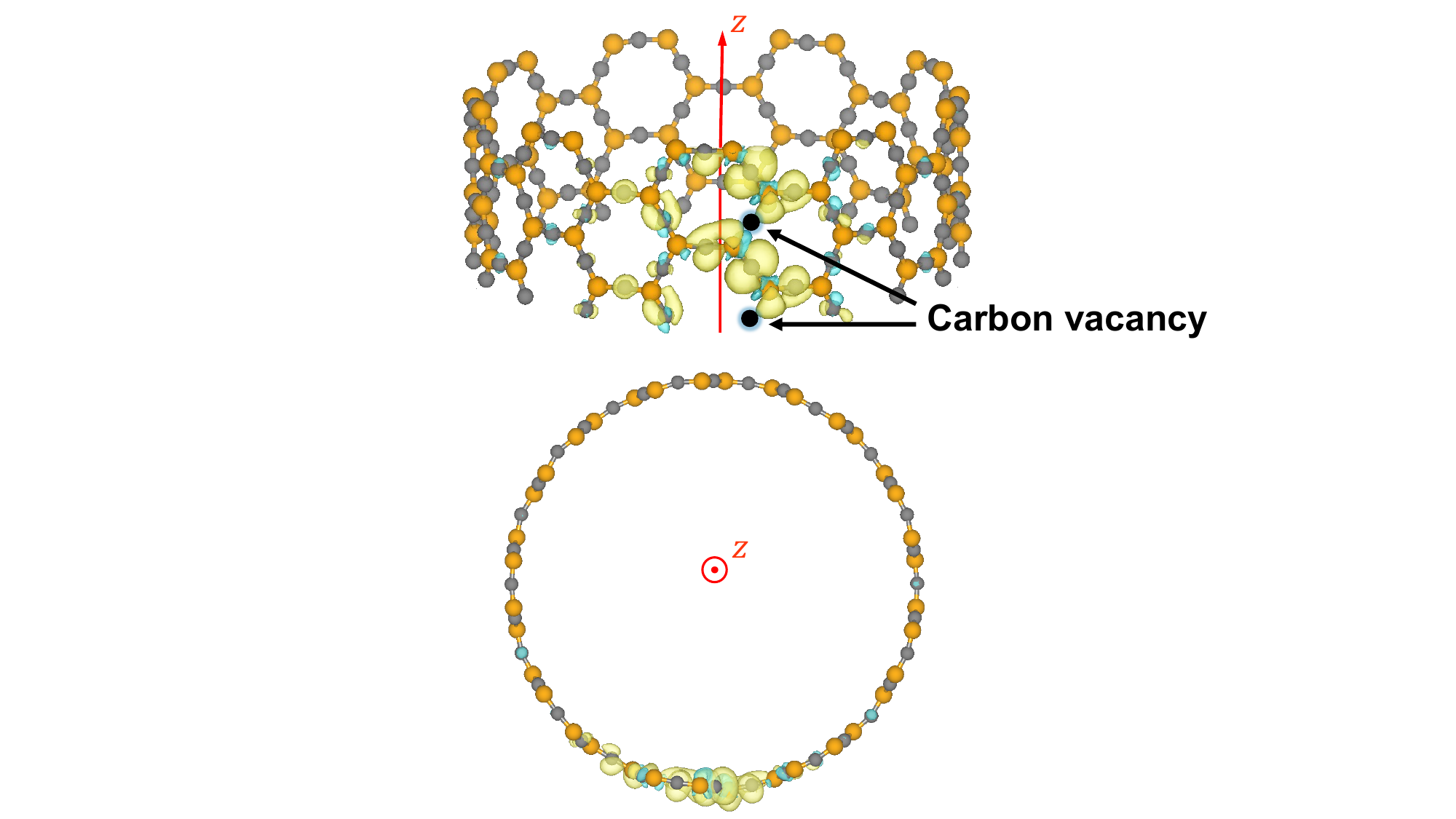}
}\label{supp:fig:c_vacancy_density}}\;
   \subfloat[]{
{
\includegraphics[scale =0.3,trim={1cm 5.5cm 1.5cm 5.5cm},clip]{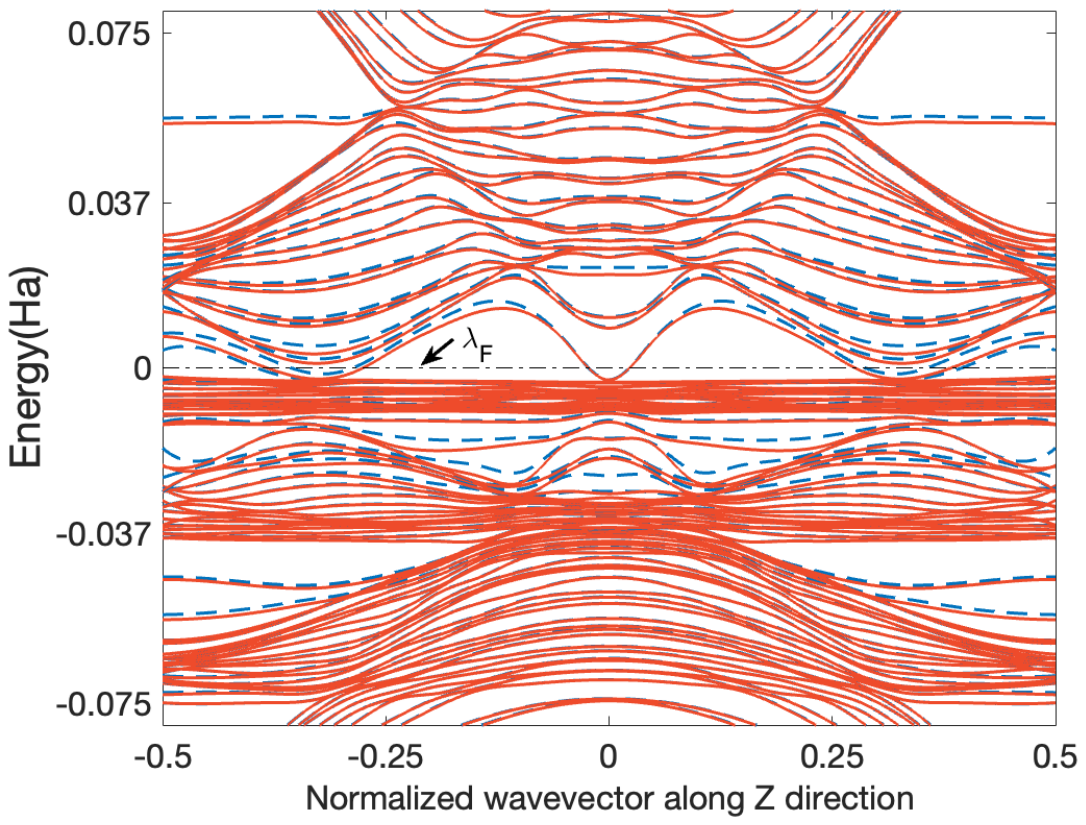}
}\label{supp:fig:c_vacancy_band_diagram}}
    \caption{Magnetization density isosurfaces for (a) hydrogenated $(9,9)$ armchair \ce{P2C3}NT, where the hydrogen atom (red color) is attached to the phosporous atom, and (c) carbon vacancy (shiny black) in $(9,9)$ armchair \ce{P2C3}NT (Two periodic unit cells are shown in the z-direction for clarity). The blue and yellow color clouds denotes the spin-down and spin-up electrons, respectively. (b) and (d) show the band diagram for the respective cases. Spin-up and spin-down channels are represented by solid red and dashed blue lines, respectively. The Fermi level $\lambda_F$ corresponds to the x-axis.}
    \label{supp:fig:magnetism}
\end{figure}

\noindent \textbf{\underline{Magnetism studies:}} Flat bands with the Coulomb interactions are often associated with magnetism. However, in many flat band materials the electrons remain spin-unpolarized. In the past, vacancy defect \cite{bhatt2022various,banhart2011structural} and hydrogenated graphene and CNTs have shown importance in inducing the magnetic order \cite{ma2004magnetic,yazyev2008magnetism,park2003magnetism,yang2005ferromagnetism}. In Fig.~\ref{supp:fig:magnetism}, we show two cases in a $(9,9)$ armchair \ce{P2C3}NT which exhibit magnetism: $(1)$ hydrogenated nanotube (Fig.~\ref{supp:fig:p-H_density}) where the hydrogen atom is adsorbed by the phosphorous atom (one hydrogen atom per two periodic layers in axial direction;  and $(2)$ one carbon vacancy per layer along the tube axis (Fig.~\ref{supp:fig:c_vacancy_density}). In both of cases, the nanotube  distorts in the radial direction and induces an anisotropy in the bond lengths and angles in the hexagonal plaquette which uplifts the degeneracy of the flat bands near the Fermi level (Fig.~\ref{supp:fig:P-H_band_diagram} \& \ref{supp:fig:c_vacancy_band_diagram}). In the first case, the nanotube has a total magnetic moment of $-0.0133 \; \mu_B$ where different spins distributions on carbon atoms make the tube ferrimagnetic-like. In particular, the spin-up (yellow) and spin-down (blue) clouds are mostly localized on carbon atoms A and C, respectively, with local magnetic moment on atom A being $0.109 \; \mu_B$  and on atom C is $-0.116 \; \mu_B$. Whereas, at atom B the local magnetic moment is low which is due to cancellation from both spins distributions around it. The rest of the contribution comes from the delocalized electrons at position D and E which resembles $p_{xy}$ orbitals of carbon atoms. The lower effective magnetization of the hydrogenated tube does not cause significant spin splitting in the bands (Fig.~\ref{supp:fig:P-H_band_diagram}). In the second case, the ferromagnetic-like character with total magnetic moment of $0.196 \; \mu_B$ separats spin-up and spin-down channels shown in Fig.~\ref{supp:fig:c_vacancy_band_diagram} as solid red and dashed blue lines, respectively. The dangling $\sigma$ and $\pi$ bonds near the vacancy polarizes the electrons causing spin-up clouds distributed largely on carbon atoms (Fig.~\ref{supp:fig:c_vacancy_density}). Due to the higher electronegativity the spins are mostly localized mostly on carbon atoms in both cases. \\

\begin{figure}[ht!]
\centering
\subfloat[]{{
\includegraphics[scale =0.25,trim={3cm 4cm 3.5cm 2cm},clip]{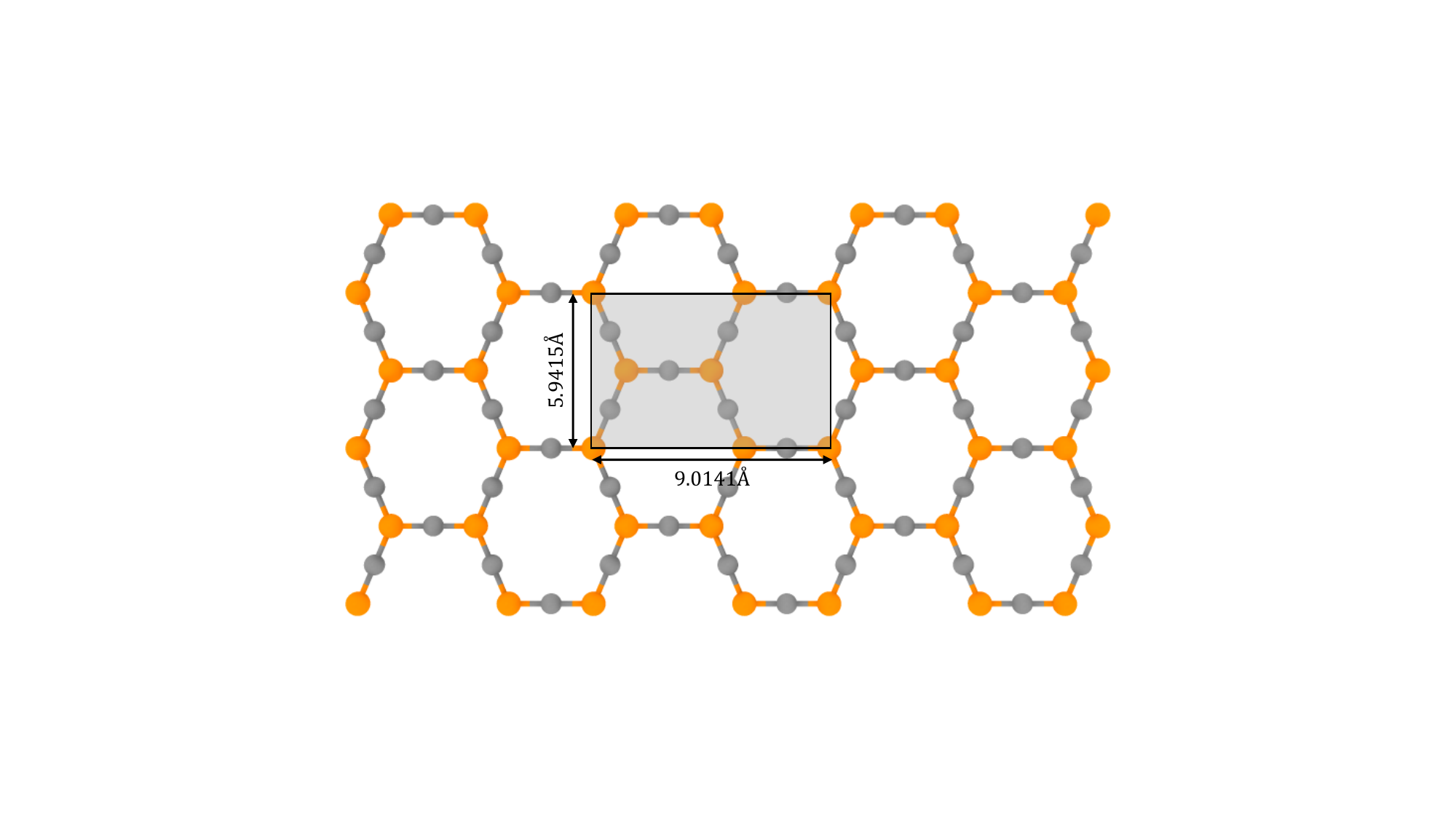}
}\label{supp:fig:112def_sheet}}\;
\subfloat[]{{
\includegraphics[scale =0.35,trim={3cm 8.5cm 3.5cm 7cm},clip]{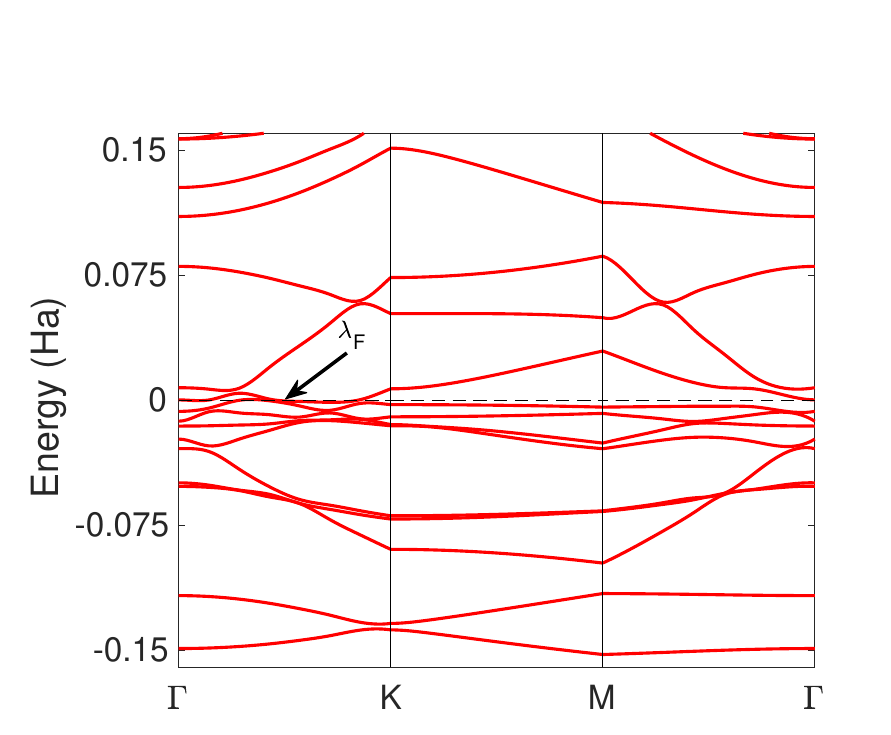}
}\label{supp:fig:112def_sheet_banddiag}}\\
\subfloat[]{{
\includegraphics[scale =0.25,trim={3cm 3.5cm 3.5cm 2cm},clip]{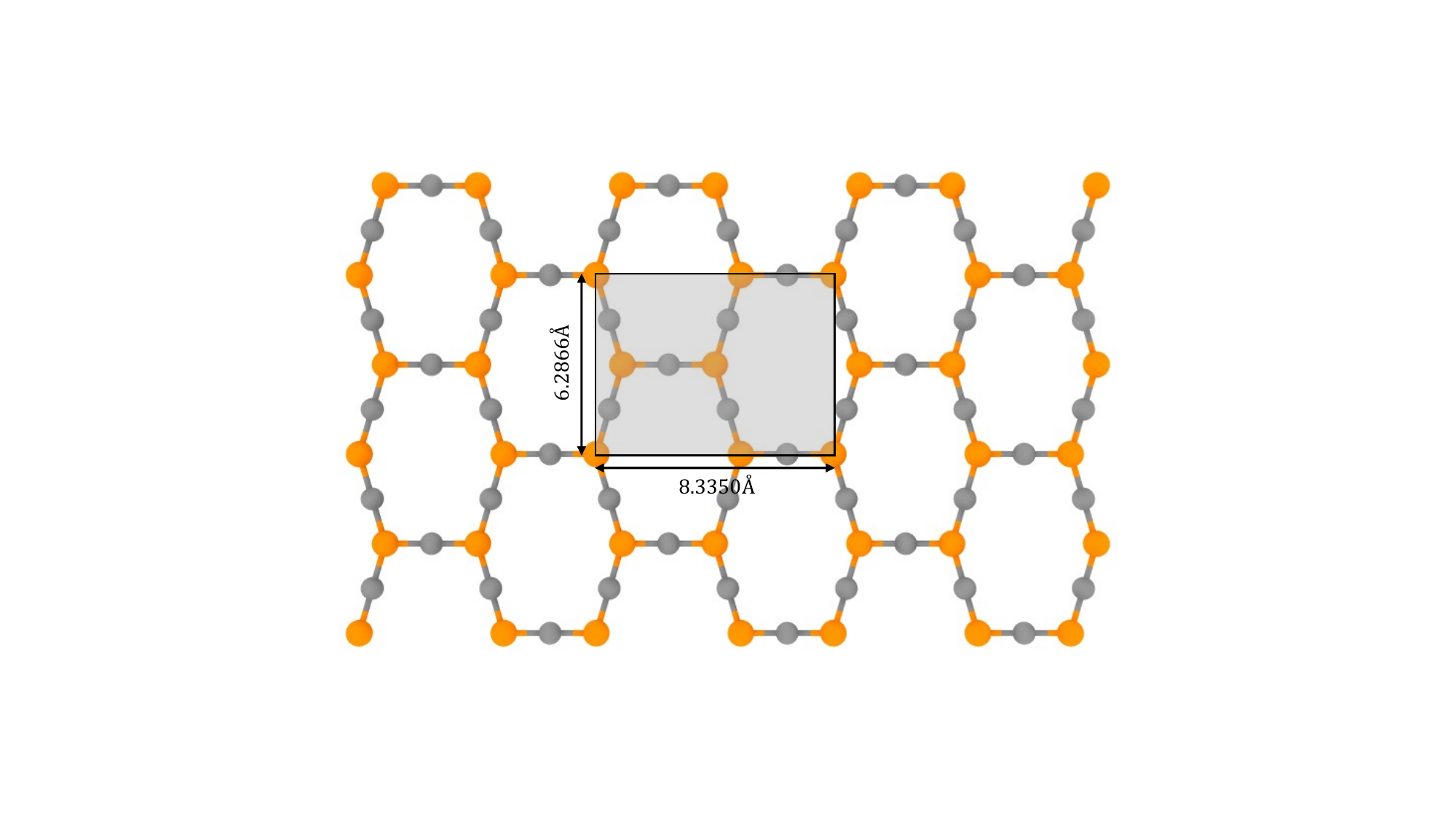}
}\label{supp:fig:106def_sheet}}\;
\subfloat[]{{
\includegraphics[scale =0.35,trim={3cm 9cm 3.5cm 7cm},clip]{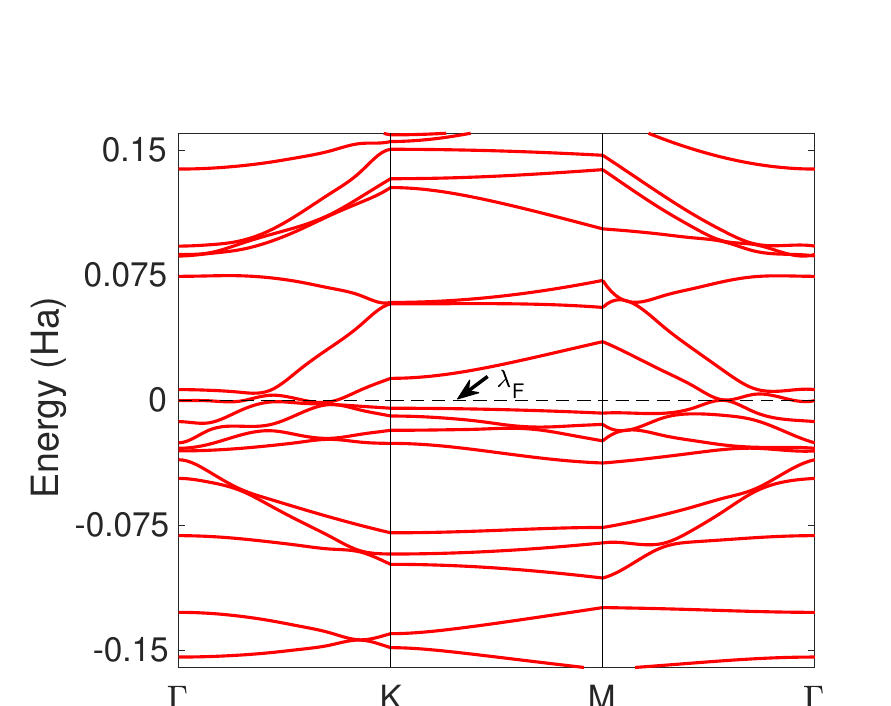}
}\label{supp:fig:106def_sheet_banddiag}} \\
\subfloat[]{{
\includegraphics[scale =0.25,trim={3cm 2.5cm 3.5cm 2cm},clip]{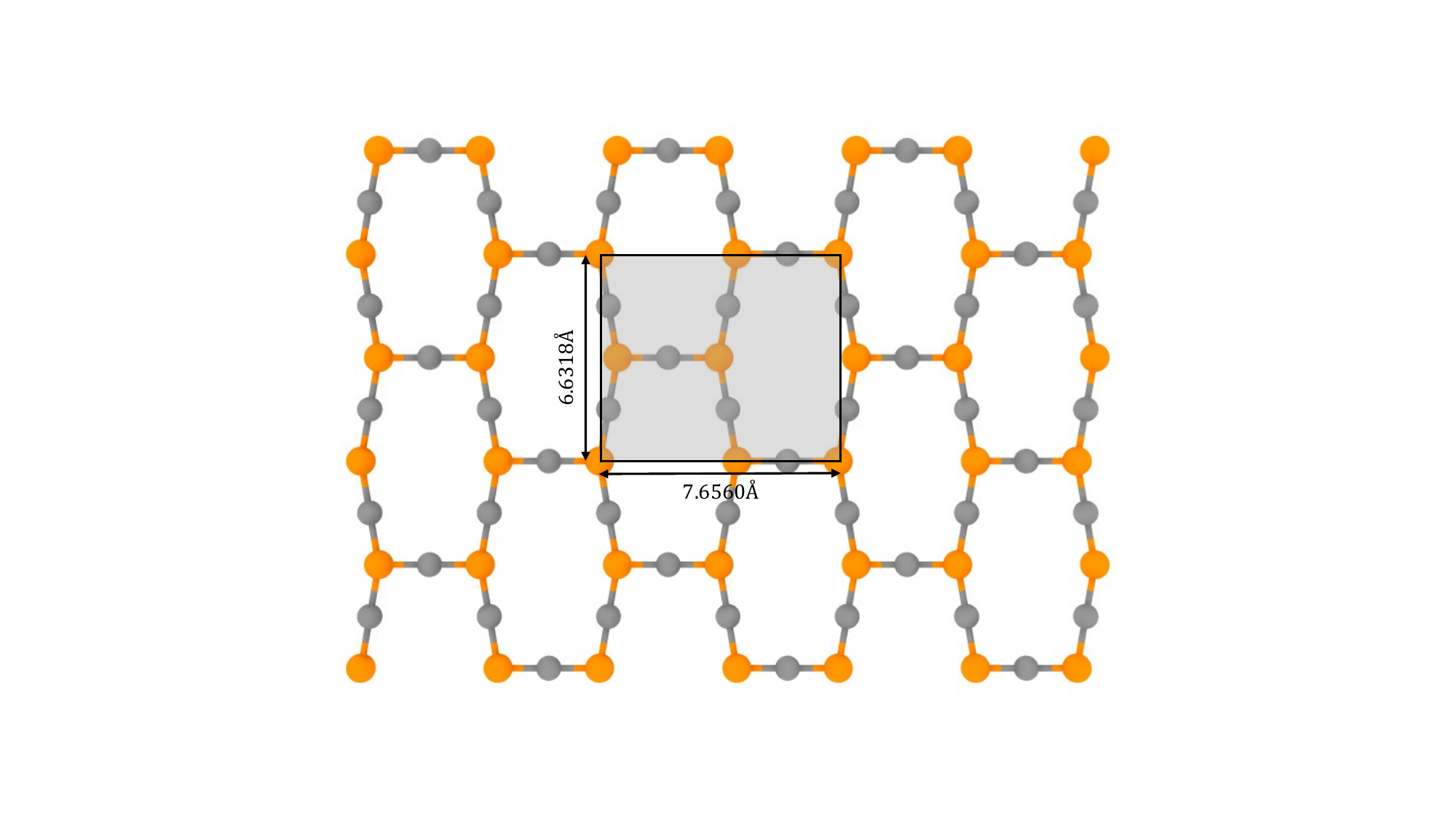}
}\label{supp:fig:100def_sheet}}\;
\subfloat[]{{
\includegraphics[scale =0.35,trim={3cm 9cm 3.5cm 7cm},clip]{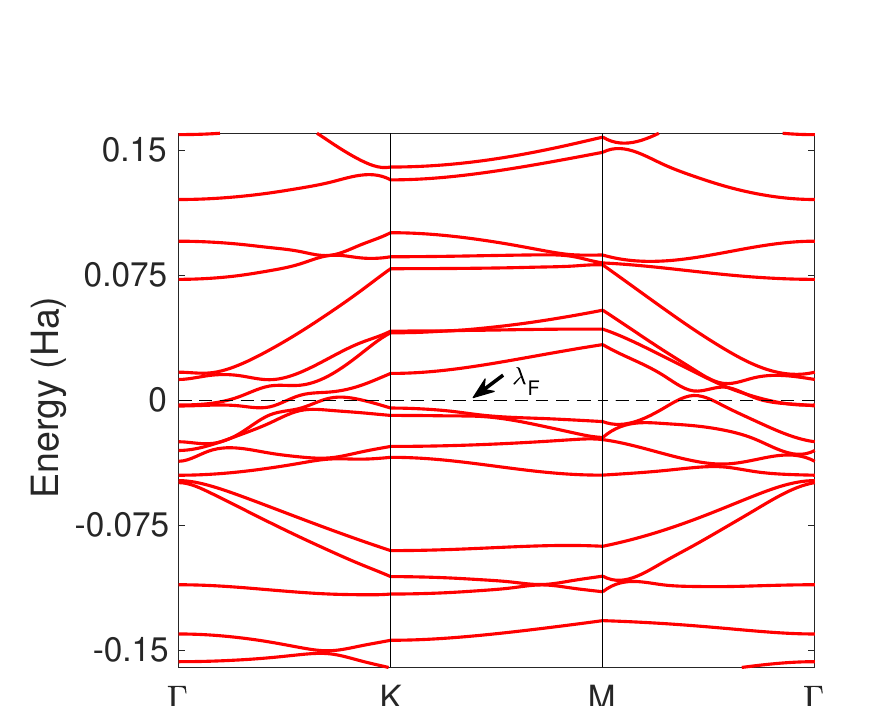}
}\label{supp:fig:100def_sheet_banddiag}} \\
\subfloat[]{{
\includegraphics[scale =0.25,trim={3cm 2cm 3.5cm 2cm},clip]{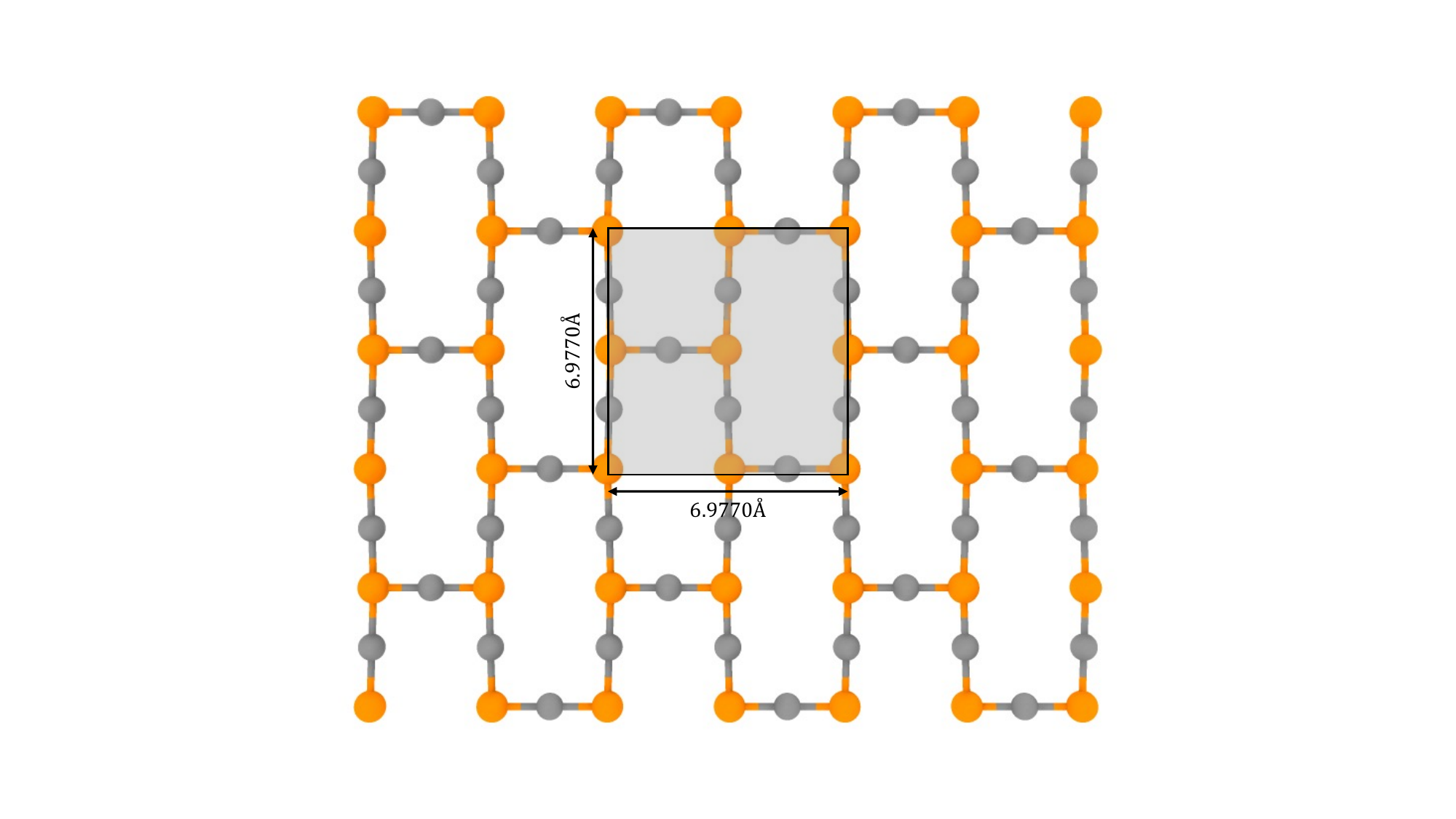}
}\label{supp:fig:90def_sheet}}\;
\subfloat[]{{
\includegraphics[scale =0.35,trim={3cm 9cm 3.5cm 7cm},clip]{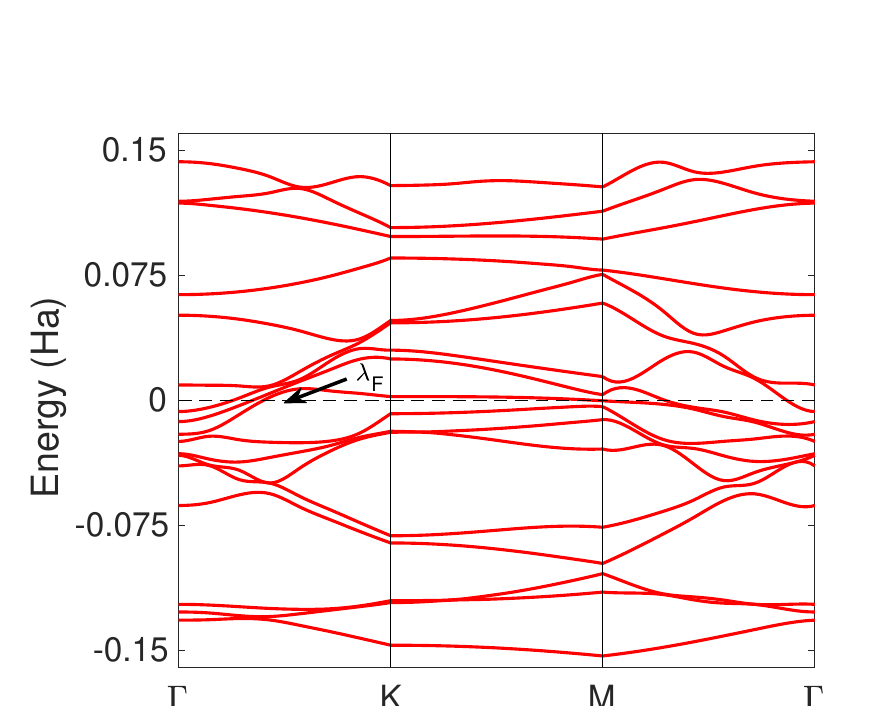}
}\label{supp:fig:90def_sheet_banddiag}}
\caption{Strain induced transition of 2D \ce{P2C3} lattice from pristine hexagonal shaped unit cell  to the square shaped unit cell (``brick-wall'' structure), along with electronic band diagrams along the transition pathway. (a) \& (b) $90^\circ$, (c) \& (d) $100^\circ $, (e) \& (f) $106^\circ$ and (g) \& (h) $112^\circ $. The Fermi level $\lambda_F$ corresponds to the x-axis.}
\label{supp:fig:indermediate_sheet}
\end{figure}

\noindent \textbf{\underline{Structural phase transition:}} To interpolate the phase transition path way under large distortion, we employ a ``freeze and relax'' strategy \citep{suwannakham2017dynamics}. Two prominent atomic structures (honeycomb and ``brick-wall'') are first relaxed through cell relaxation followed by atomic relaxation to ensure equilibrium structures as endpoints. Subsequently, we selectively freeze regions of the lattice anticipated to undergo minimal structural perturbation, while linearly interpolating the atomic positions of the remaining atoms to generate initial guesses for intermediate states. Subsequent to this, relaxation calculations are performed iteratively on the unfrozen degrees of freedom within each intermediate state, allowing for partial relaxation of the structure and then followed by full relaxation of all atoms. This approach ensures that the interpolation process focuses computational resources on regions of the lattice undergoing significant structural modifications, thereby facilitating the determination of an accurate pathway between the two endpoint structures. The transition pathway for 2D \ce{P2C3} sheets is highlighted in Fig.~\ref{supp:fig:indermediate_sheet}.
\\
\clearpage
\noindent \textbf{\underline{Strain engineering of 2D \ce{P2C3}:}} 
\begin{figure}[ht!]
\centering
\subfloat{
{
\includegraphics[scale =0.45,trim={0.5cm 0cm 0cm 0cm},clip
]{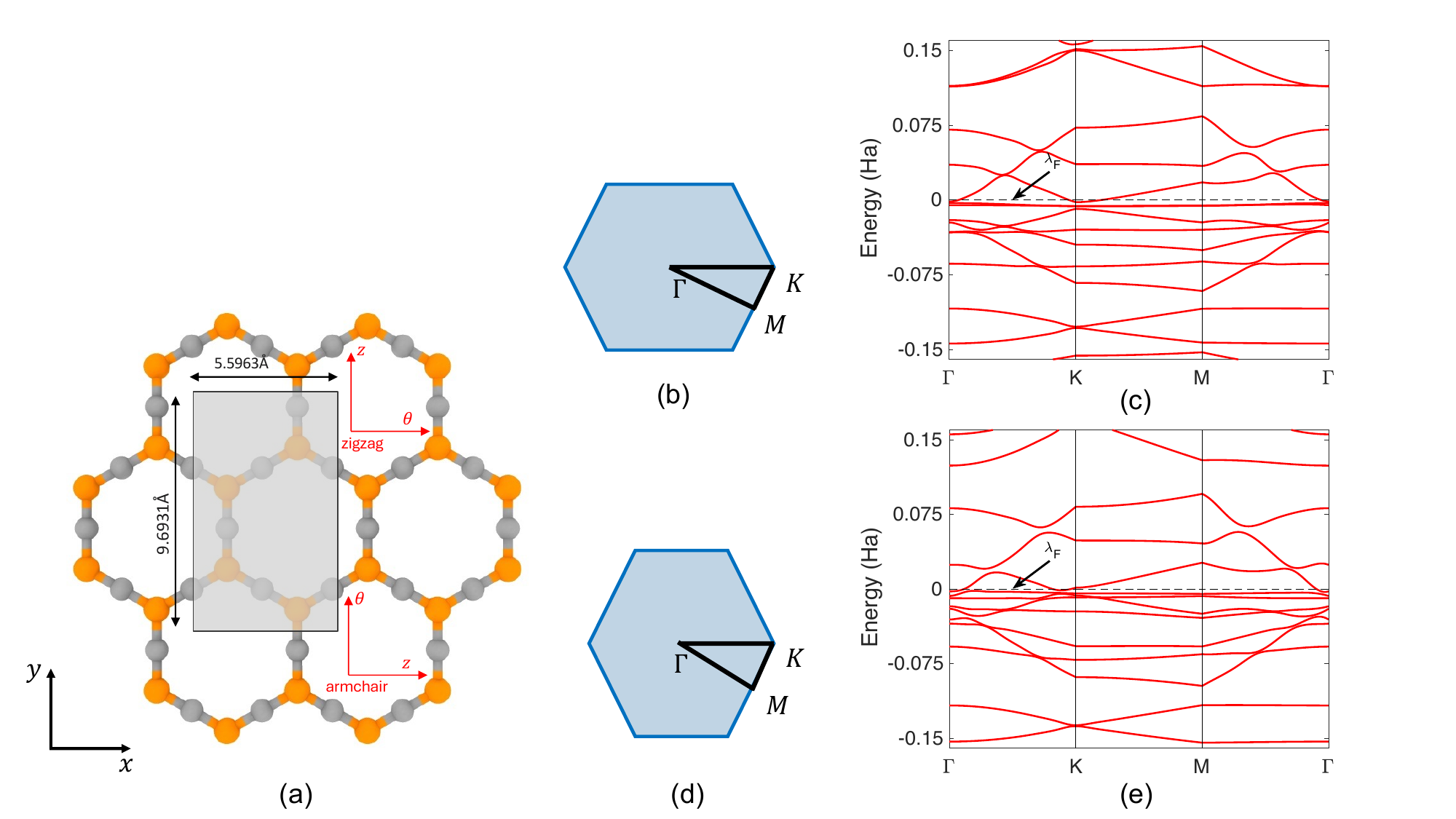}
}}
\caption{(a) Pristine 2D \ce{P2C3} lattice  $z$ is in the direction of nanotube's axis. The orange-color atoms are phosphorous, and the gray-color atoms are carbon. (b) and (c) show the band diagrams of 2D \ce{P2C3} lattice under tensile strain of $4 \%$ along the x-direction and the corresponding Brillouin zone path. (d) and (e) show the spectrum under compressive strain of $4 \%$ and the corresponding Brillouin zone path. The Fermi level $\lambda_F$ corresponds to the x-axis. }
\label{supp:fig:2d_band}
\end{figure}
\begin{figure}[ht!]
    \centering
\subfloat[]{
{
\includegraphics[scale =0.35,trim={3cm 8.5cm 3.5cm 7cm},clip]{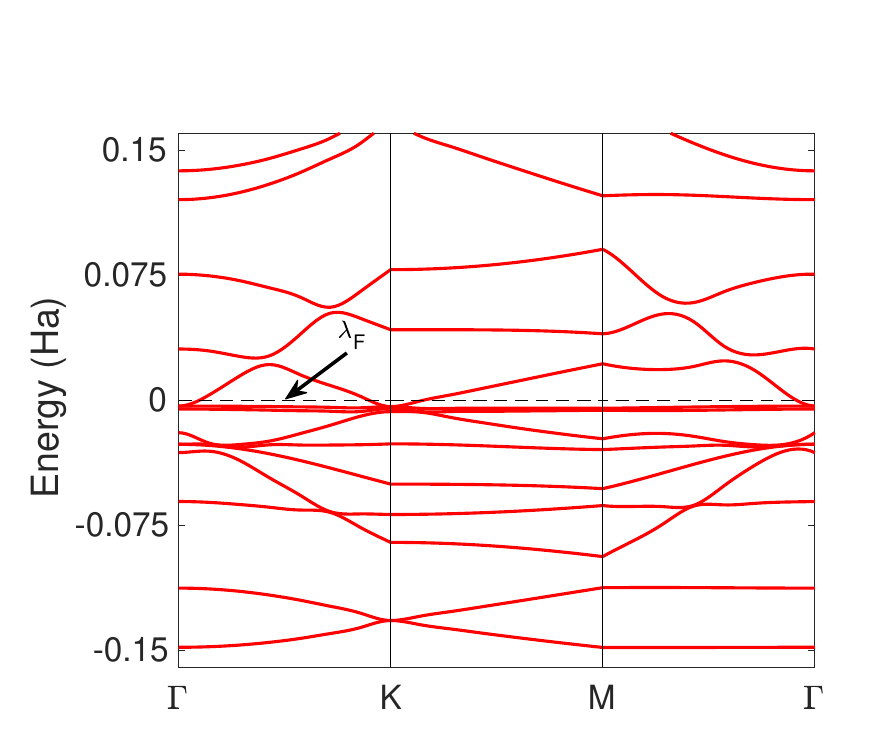}
}\label{supp:fig:2D_p2c3_pristine}}
\subfloat[]{
{
\includegraphics[scale =0.35,trim={3cm 8.5cm 3.5cm 7cm},clip]{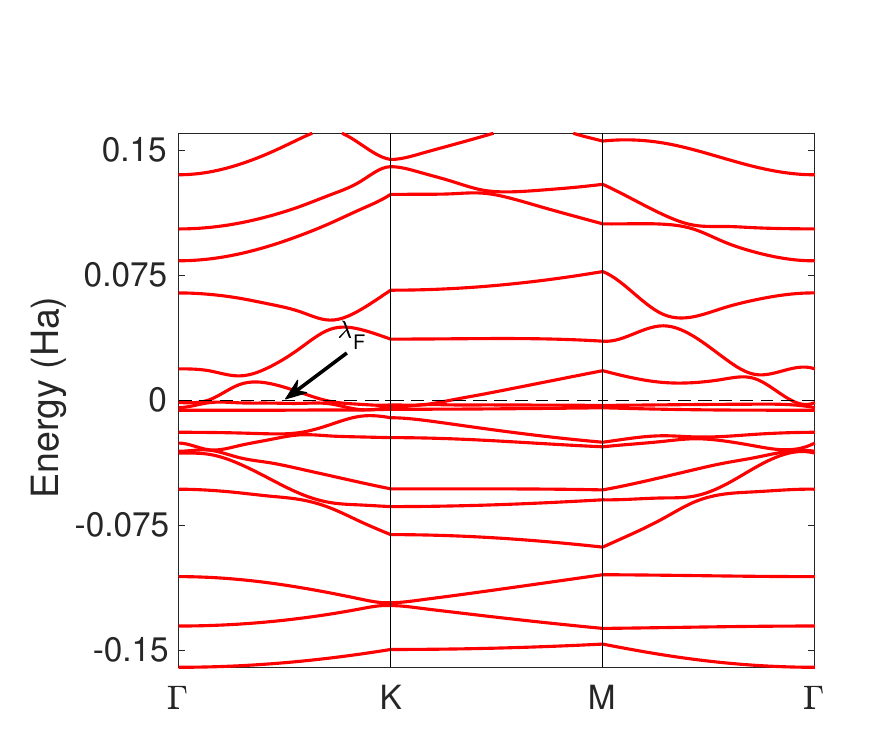}
}\label{supp:fig:2D_p2c3_pristine_shear}}
    \caption{Band diagram of pristine \ce{P2C3} 2D lattice (a) and under $4\%$ shear (b). The Fermi level $\lambda_F$ corresponds to the x-axis.}
    \label{supp:fig:2D_sheet_band}
\end{figure}

\clearpage
\noindent \textbf{\underline{Schematic of nanotube synthesis process}} 
\begin{figure}[ht!]
    \centering
    \subfloat[]{\includegraphics[scale=0.4]{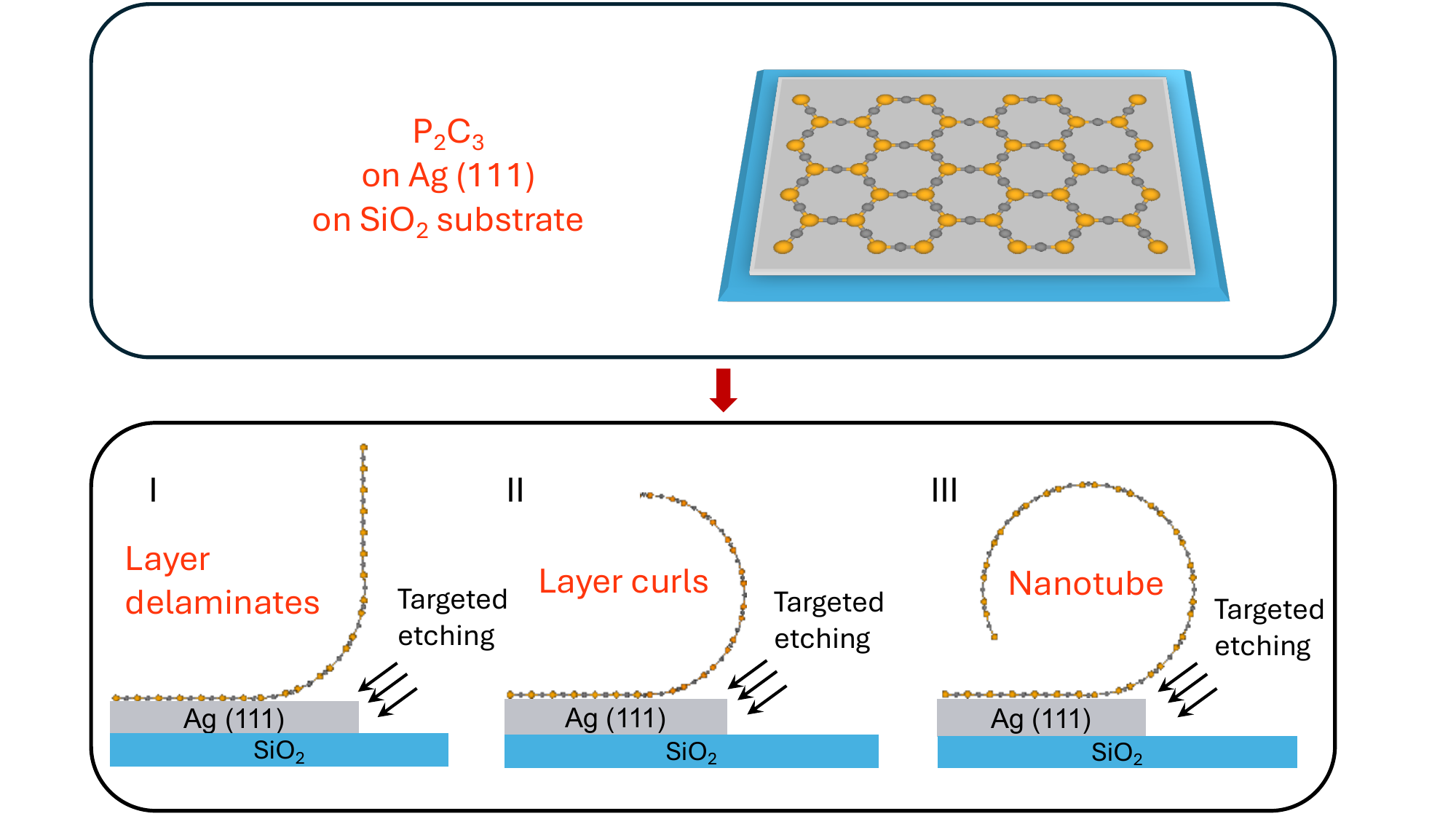}\label{supp:fig:synthes}} \;
    \caption{Possible route to synthesis of \ce{P2C3}NTs from 2D \ce{P2C3} sheets.}
    \label{supp:fig:synthesis_process}
\end{figure}

\noindent \textbf{\underline{Additional figures referenced in the main text:}} 
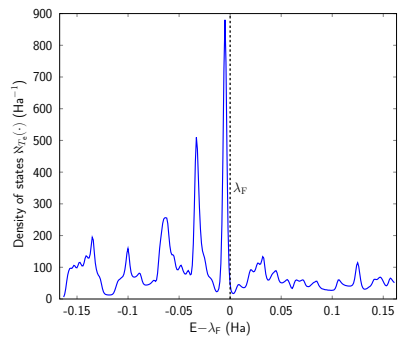
\begin{figure}

\subfloat[Electronic density of states for untwisted zigzag \ce{P2C3}NT.]{\scalebox{0.3}
{\begin{tikzpicture} 
\begin{axis}[
width=\textwidth,
xlabel={E$-\lambda_\text{F}$ (Ha)},
ylabel={Density of states $\aleph_{T_{\text{e}}}(\cdot)$ ($\text{Ha}^{-1}$)},
legend pos = {north west}, 
legend style={at={(0.125,0.90)},anchor=west,font=\sffamily, fill=none,cells={anchor=west},row sep=1pt},
y label style={xshift=-10pt},
x label style={xshift=-10pt},
label style={font=\sffamily\Large},
tick label style={font=\sffamily\Large},
xmin=-0.167, xmax=0.163,
ymin = -0.05, ymax=900,
ytick={0,100,200,300,400,500,600,700,800,900},
yticklabels={0,100,200,300,400,500,600,700,800,900},
xtick={-0.15,-0.1,-0.05,0,0.05,0.1,0.15},
xticklabels={-0.15,-0.1,-0.05,0,0.05,0.1,0.15},
x tick label style={yshift=-5pt},
y tick label style={xshift=-5pt},
]
\addplot[line width=0.50mm, blue,] table[x index=0,y index=1]{./figures/zigzag_go12_21kpts_DOS.txt};
\addplot[mark=none, black, dashed, ultra thick] coordinates{(0, -0.05) (0, 900)}; 
\node[] at (axis cs: 0.010,350) {\Large{$\lambda_{\text{F}}$}};
\end{axis}
\end{tikzpicture}}
\label{supp:fig:DOS_untwist_zigzag}}\\
\caption{ Electronic density of states (eDOS) of \ce{P2C3}NT showing sharp peak near $\lambda_F$. }
\label{supp:fig:full_band_zigzag_untwisted}
\end{figure}

\begin{figure}[ht!]
\subfloat[]{
{
\includegraphics[scale =0.3,trim={1.8cm 6.5cm 1.5cm 6cm},clip]{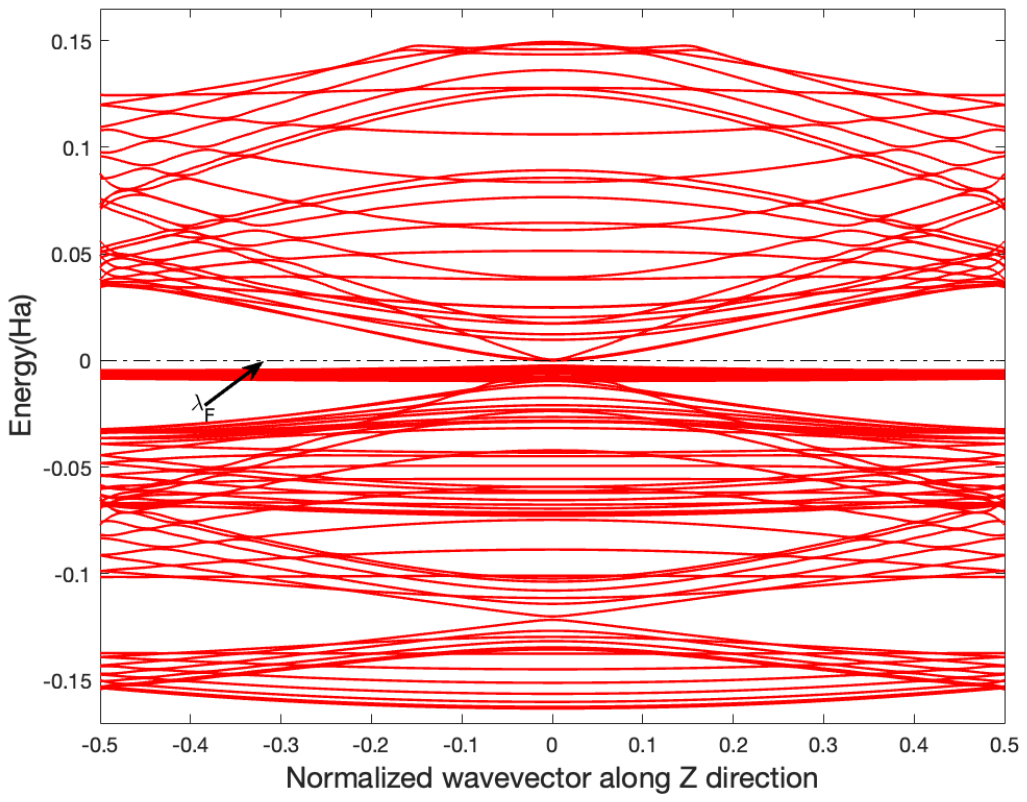}
}\label{supp:fig:banddiag_zigzagfull_strained}} \hspace{0.5cm}
\subfloat[]{\scalebox{0.29}
{\begin{tikzpicture} 
\begin{axis}[
width=\textwidth,
xlabel={E$-\lambda_\text{F}$ (Ha)},
ylabel={Density of states $\aleph_{T_{\text{e}}}(\cdot)$ ($\text{Ha}^{-1}$)},
legend style={at={(0.060,0.82)},anchor=west,font=\sffamily, fill=none,cells={anchor=west},row sep=1pt},
y label style={xshift=-10pt},
label style={font=\sffamily\Large},
tick label style={font=\sffamily\Large},
xmin=-0.167, xmax=0.163,
ymin = -0.05, ymax=900,
ytick={0,100,200,300,400,500,600,700,800,900},
yticklabels={0,100,200,300,400,500,600,700,800,900},
xtick={-0.15,-0.1,-0.05,0,0.05,0.1,0.15},
xticklabels={-0.15,-0.1,-0.05,0,0.05,0.1,0.15},
x tick label style={yshift=-5pt},
y tick label style={xshift=-5pt},
]
\addplot[line width=0.90mm, blue,] table[x index=0,y index=1]{./figures/zigzag_go12_21kpts_DOS.txt};
\addplot[line width=0.70mm, red,] table[x index=0,y index=1]{./figures/zigzag_strain_neg_0p6_go12_dos.txt};
\addplot[line width=0.50mm, cyan,] table[x index=0,y index=1]{./figures/zigzag_strain_neg_0p2_go12_dos.txt};
\addplot[line width=0.30mm, black,] table[x index=0,y index=1]{./figures/zigzag_strain_pos_0p2_go12_dos.txt};
\addplot[line width=0.20mm, green,] table[x index=0,y index=1]{./figures/zigzag_strain_pos_0p6_go12_dos.txt};
\addplot[mark=none, black, dashed, ultra thick] coordinates{(0, -0.05) (0, 900)}; 
\node[] at (axis cs: 0.010,350) {\Large{$\lambda_{\text{F}}$}};
\legend{\Large{$\;$No axial strain}, \Large{$\;$ $-3.28\%$ axial strain},\Large{$\;$ $-1.09\%$ axial strain},\Large{$\;$ $+1.09\%$ axial strain},\Large{$\;$ $+3.28\%$ axial strain}};
\end{axis}
\end{tikzpicture}}
\label{supp:fig:DOS_strained_zigzag}}\\
\subfloat[]{\includegraphics[scale=0.2,trim={1.8cm 6.5cm 1.5cm 6cm},clip]{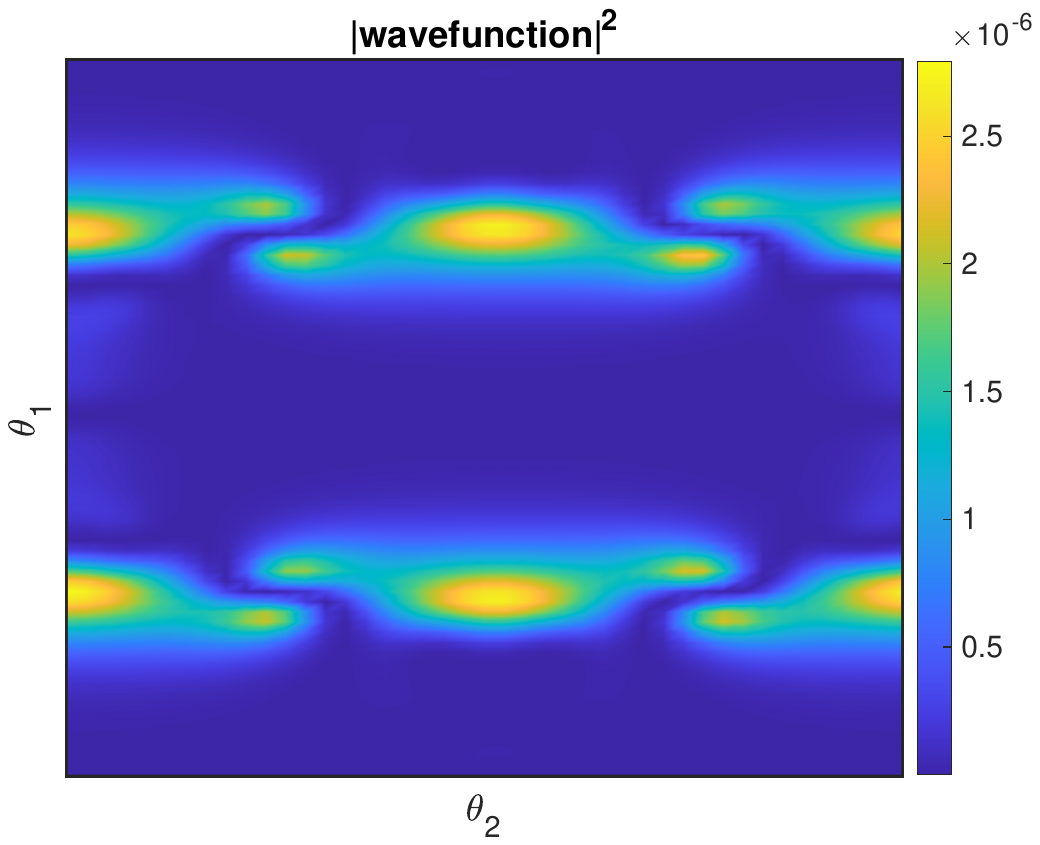}\label{supp:fig:zigzag_comp_wve2_dicpt}} \;
\subfloat[]{\includegraphics[scale=0.2,trim={1.8cm 6.5cm 1.5cm 6cm},clip]{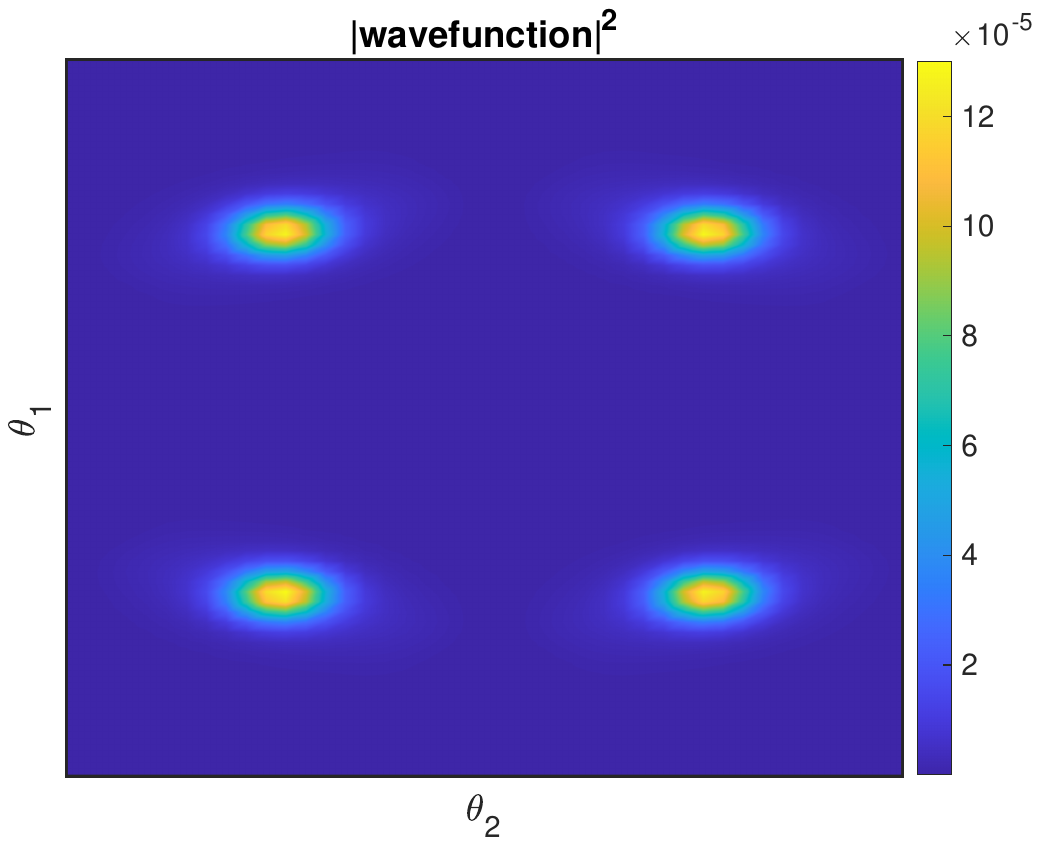}\label{supp:fig:zigzag_comp_wve2_ftbnd}} \;
\subfloat[]{\includegraphics[scale=0.2,trim={1.8cm 6.5cm 1.5cm 6cm},clip]{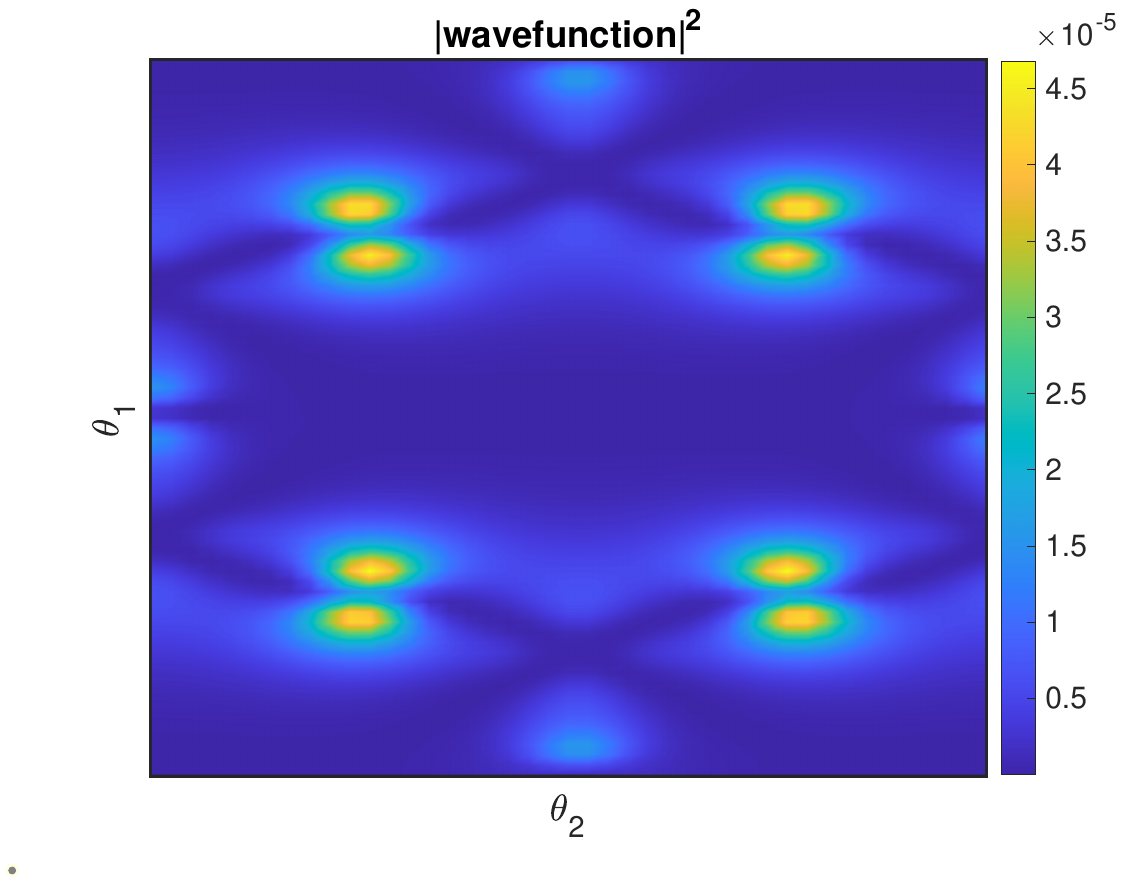}\label{supp:fig:zigzag_comp_wve2_pxy}} \;
\subfloat[]{\includegraphics[scale=0.2,trim={1.8cm 6.5cm 1.5cm 6cm},clip]{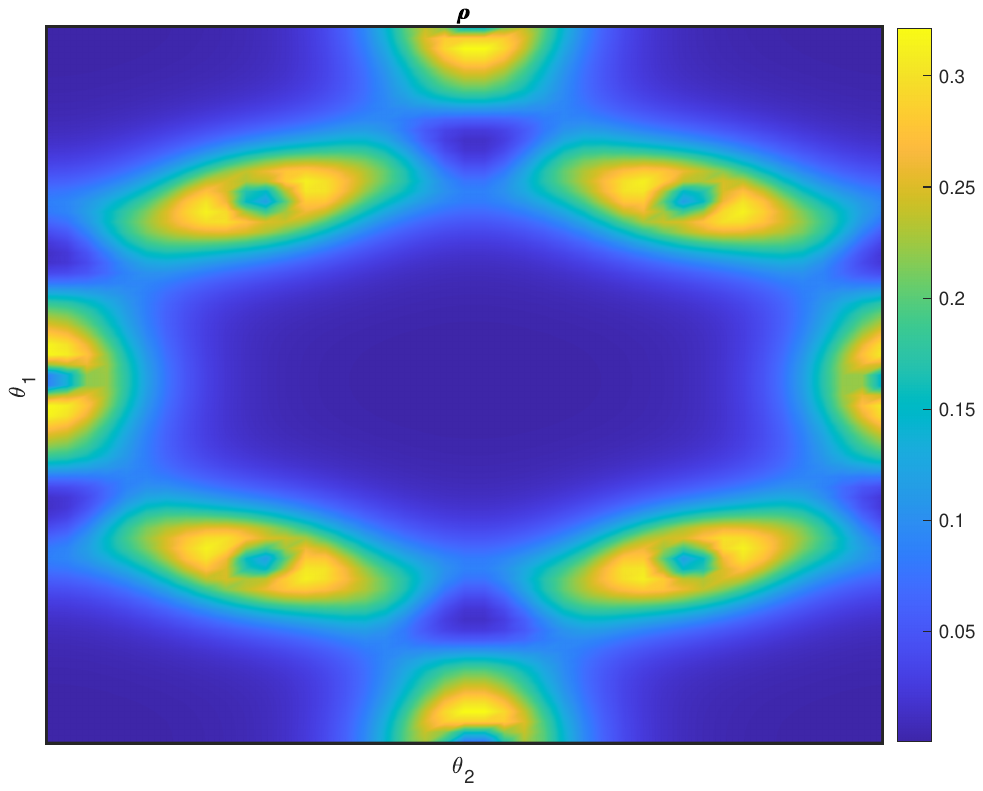}\label{supp:fig:edensity_strained}}
\caption{ (a) Band diagram of $(12,0)$ compressed zigzag \ce{P2C3}NT by $3.28\%$. (c) Electronic density of states (eDOS) plot for different strains. From (d) to (f) shows the electronic states (square of absolute value of wavefunction) associated with $D_1$, $D_2$ and $D_3$ points shown in (b). (g) Electronic density $\rho$. A slice of electronic fields at an average radial distance of atoms in computational domain is shown in terms of helical coordinates.  A slice of electronic fields at an average radial distance of the atoms in the computational domain is shown in each case. $\theta_1,\theta_2$ denote helical coordinates that parametrize the tube surface at a fixed radial distance.}
\label{supp:fig:full_band_zigzag_strained}
\end{figure}
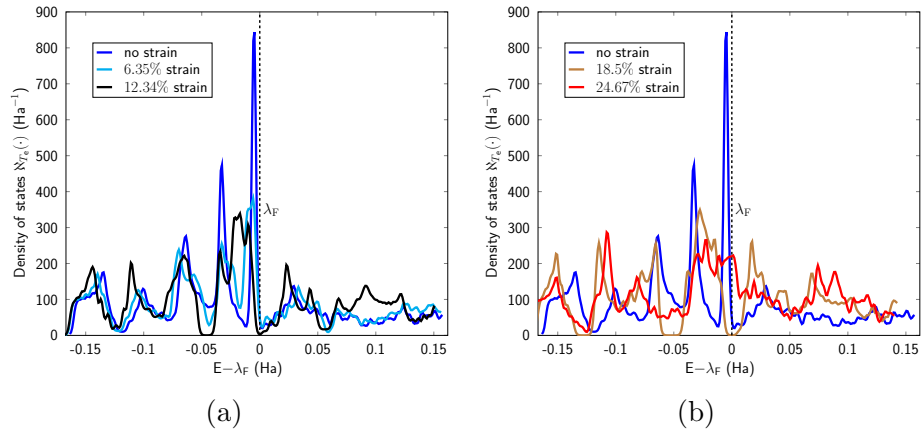
\begin{figure}[ht!]
\centering
\subfloat[]{\scalebox{0.34}
{\begin{tikzpicture} 
\begin{axis}[
width=\textwidth,
xlabel={E$-\lambda_\text{F}$ (Ha)},
ylabel={Density of states $\aleph_{T_{\text{e}}}(\cdot)$ ($\text{Ha}^{-1}$)},
legend pos = {north west}, 
legend style={at={(0.070,0.82)},anchor=west,font=\sffamily, fill=none,cells={anchor=west},row sep=1pt},
y label style={xshift=-10pt},
x label style={xshift=-10pt},
label style={font=\sffamily\Large},
tick label style={font=\sffamily\Large},
xmin=-0.167, xmax=0.162,
ymin = -0.05, ymax=900,
ytick={0,100,200,300,400,500,600,700,800,900},
yticklabels={0,100,200,300,400,500,600,700,800,900},
xtick={-0.15,-0.1,-0.05,0,0.05,0.1,0.15},
xticklabels={-0.15,-0.1,-0.05,0,0.05,0.1,0.15},
x tick label style={yshift=-5pt},
y tick label style={xshift=-5pt},
]
\addplot[line width=0.90mm, blue,] table[x index=0,y index=1]{./figures/armchair_go9_21kpts_DOS.txt};
\addplot[line width=0.90mm, cyan,] table[x index=0,y index=1]{./figures/DOS_113.txt};
\addplot[line width=0.90mm, black,] table[x index=0,y index=1]{./figures/DOS_106.txt};
\addplot[mark=none, black, dashed, ultra thick] coordinates{(0, -0.05) (0, 900)}; 
\node[] at (axis cs: 0.010,350) {\Large{$\lambda_{\text{F}}$}};
\legend{\Large{$\;$ no strain},\Large{$\;$ $6.35\%$ strain}, \Large{$\;$ $12.34\%$ strain}};
\end{axis}
\end{tikzpicture}}
\label{supp:fig:DOS_twist_armchir}}\;
\subfloat[]{\scalebox{0.34}
{\begin{tikzpicture} 
\begin{axis}[
width=\textwidth,
xlabel={E$-\lambda_\text{F}$ (Ha)},
ylabel={Density of states $\aleph_{T_{\text{e}}}(\cdot)$ ($\text{Ha}^{-1}$)},
legend pos = {north west}, 
legend style={at={(0.07,0.82)},anchor=west,font=\sffamily, fill=none,cells={anchor=west},row sep=1pt},
y label style={xshift=-10pt},
x label style={xshift=-10pt},
label style={font=\sffamily\Large},
tick label style={font=\sffamily\Large},
xmin=-0.167, xmax=0.162,
ymin = -0.05, ymax=900,
ytick={0,100,200,300,400,500,600,700,800,900},
yticklabels={0,100,200,300,400,500,600,700,800,900},
xtick={-0.15,-0.1,-0.05,0,0.05,0.1,0.15},
xticklabels={-0.15,-0.1,-0.05,0,0.05,0.1,0.15},
x tick label style={yshift=-5pt},
y tick label style={xshift=-5pt},
]
\addplot[line width=0.90mm, blue,] table[x index=0,y index=1]{./figures/armchair_go9_21kpts_DOS.txt};
\addplot[line width=0.90mm, brown,] table[x index=0,y index=1]{./figures/DOS_100.txt};
\addplot[line width=0.90mm, red,] table[x index=0,y index=1]{./figures/DOS_90.txt};
\addplot[mark=none, black, dashed, ultra thick] coordinates{(0, -0.05) (0, 900)}; 
\node[] at (axis cs: 0.010,350) {\Large{$\lambda_{\text{F}}$}};
\legend{\Large{$\;$ no strain},\Large{$\;$ $18.5\%$ strain },\Large{$\;$ $24.67\%$ strain}};
\end{axis}
\end{tikzpicture}}
\label{supp:fig:DOS_twist_armchir_second}}
\label{supp:fig:indermediate_tubes}
\caption{Electronic density of states (eDOS) for nanotubes under structural transition. (a) shows comparison between nanotubes with no strain, $6.35\%$ and $12.34\%$ strains. (b) shows comparison between nanotubes with no stain, $18.5\%$ and $24.67\%$ strains. Figures of transition nanotubes are in the main text.}
\end{figure}

\clearpage
\noindent \textbf{\underline{Annotated and magnified views of band diagrams from the main text:}} 

\begin{figure}[ht!]
    \centering
\subfloat[]{
{
\includegraphics[scale =0.35]{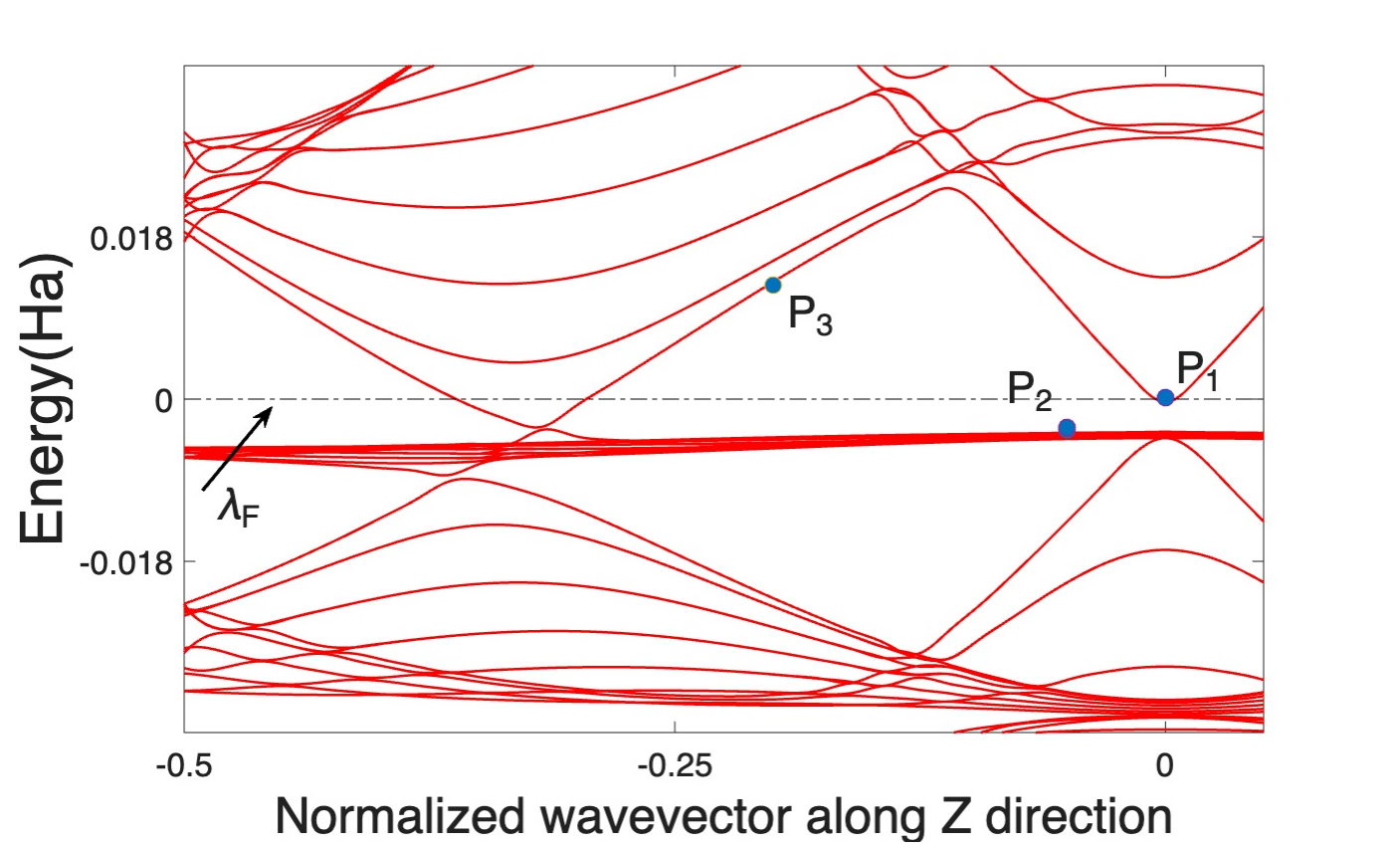}
}\label{supp:fig:azoom}}
\subfloat[]{
{
\includegraphics[scale =0.35]{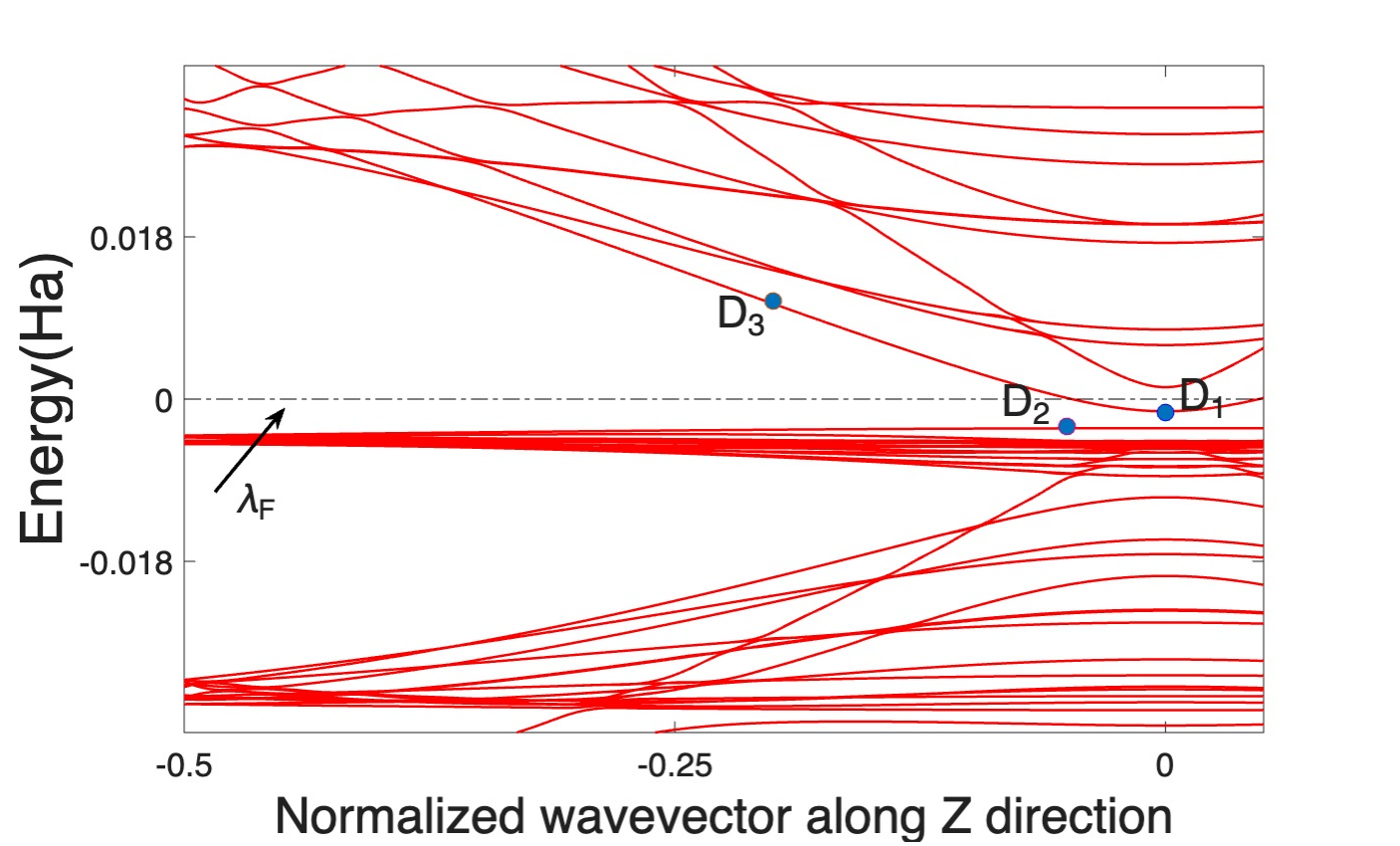}
}\label{supp:fig:zzoom}}
    \caption{Magnified Band diagram of pristine \ce{P2C3}NTs shown in Fig.~2 of main text (a) armchair $(9,9)$ nanotube (b) zigzag $(12,0)$ nanotube. }
    \label{supp:fig:ssss}
\end{figure}

\begin{figure}[ht!]
    \centering
\subfloat[]{
{
\includegraphics[scale =0.35]{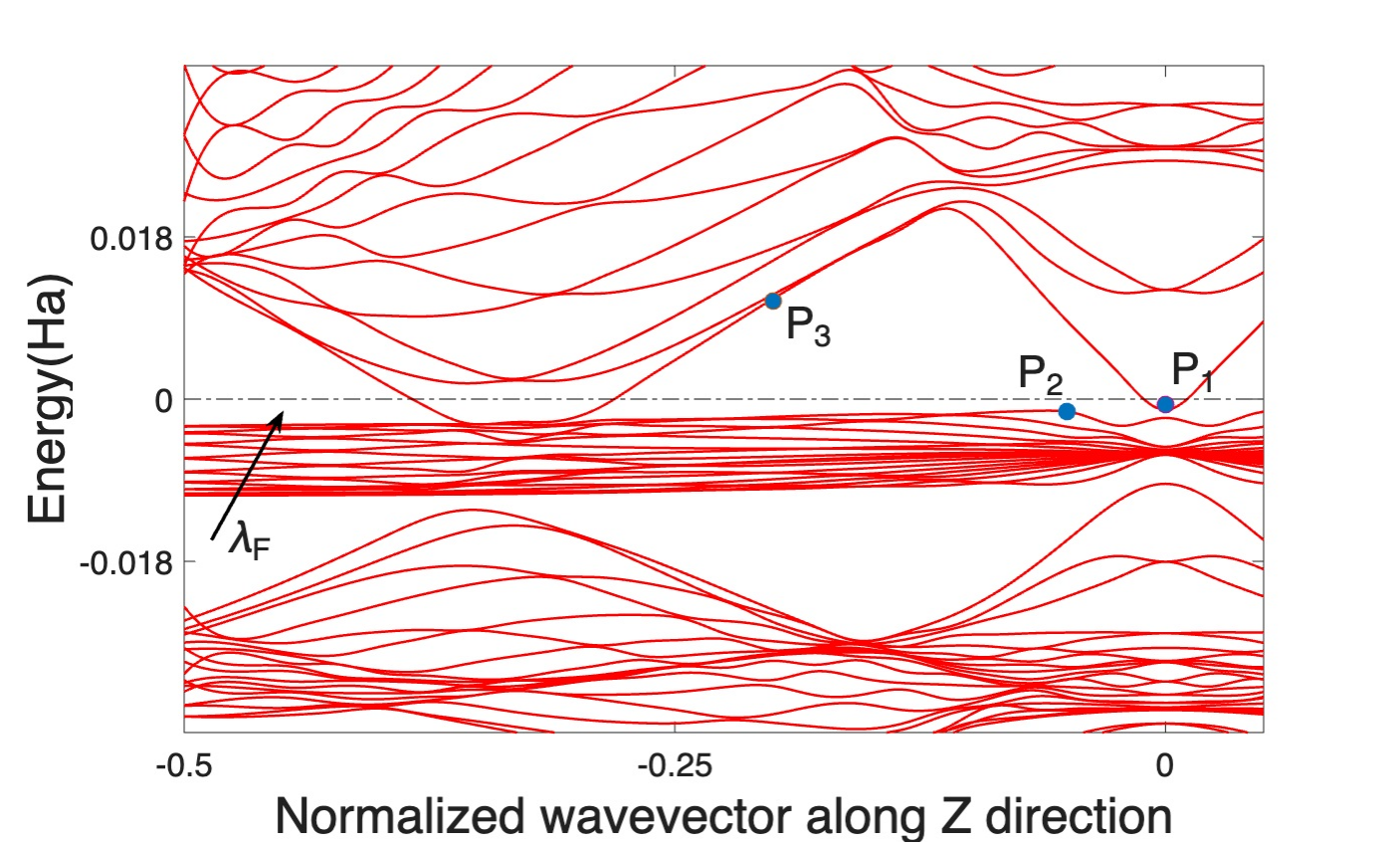}
}\label{supp:fig:azoomtw}}
\subfloat[]{
{
\includegraphics[scale =0.35]{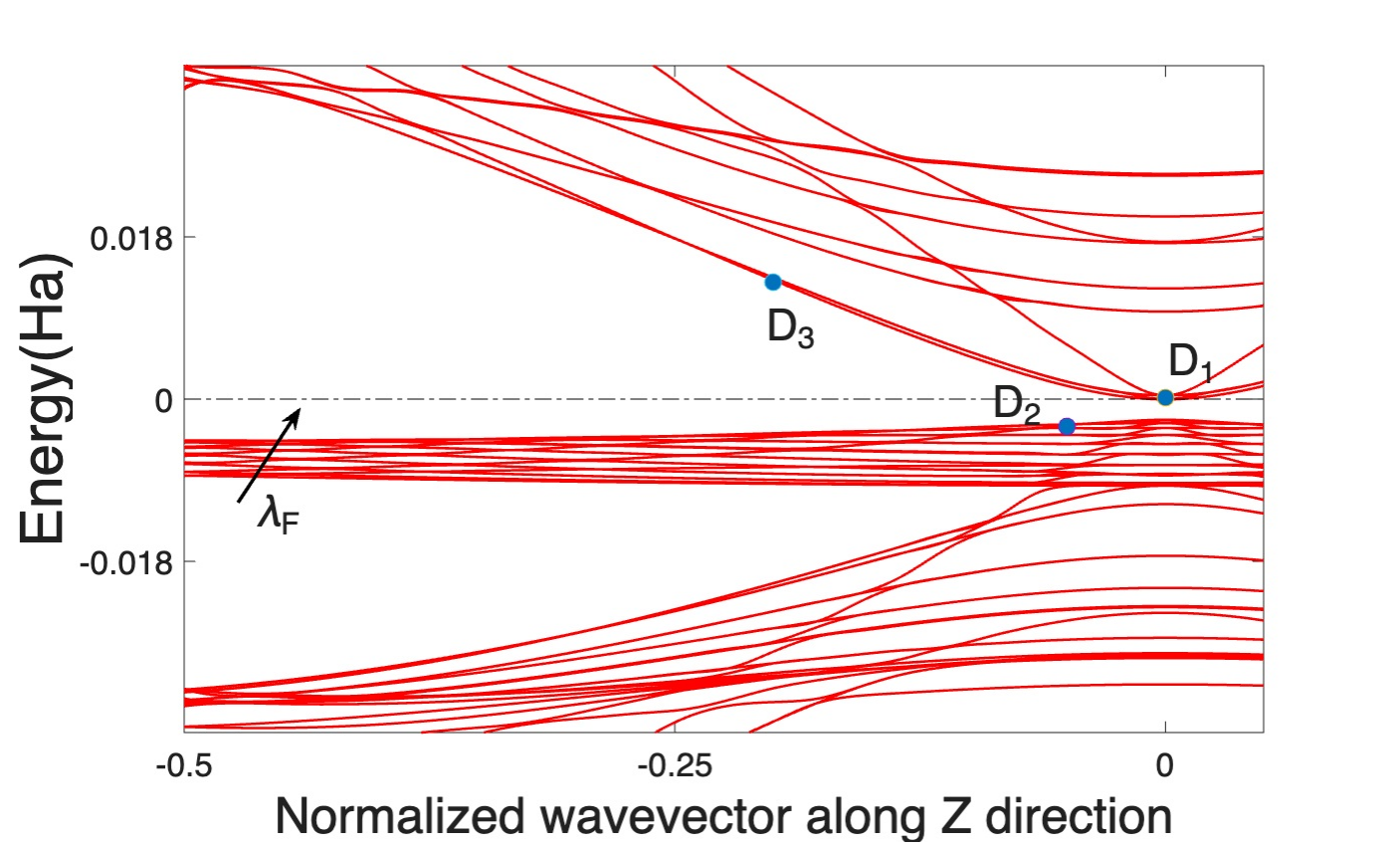}
}\label{supp:fig:azoomdtw}}
    \caption{Magnified Band diagram of (a) twisted armchair \ce{P2C3}NT (Fig.~4, main text)  (b) compressed  zigzag \ce{P2C3}NT (Fig.~S9, SI) .}
    \label{supp:fig:ssssass}
\end{figure}

\begin{figure}[ht!]
    \centering
\subfloat[]{
{
\includegraphics[scale =0.35]{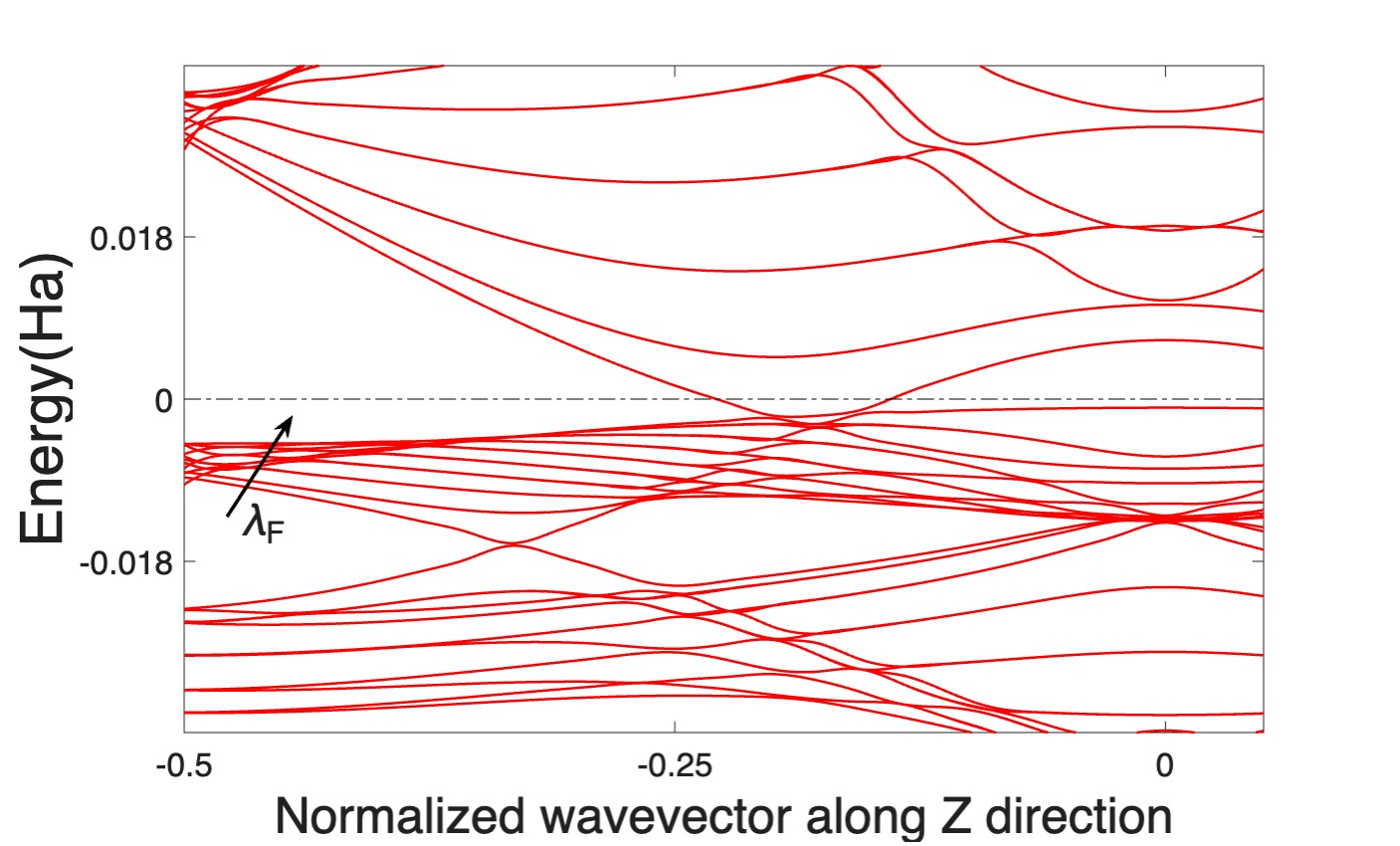}
}\label{supp:fig:azooma1}}
\subfloat[]{
{
\includegraphics[scale =0.35]{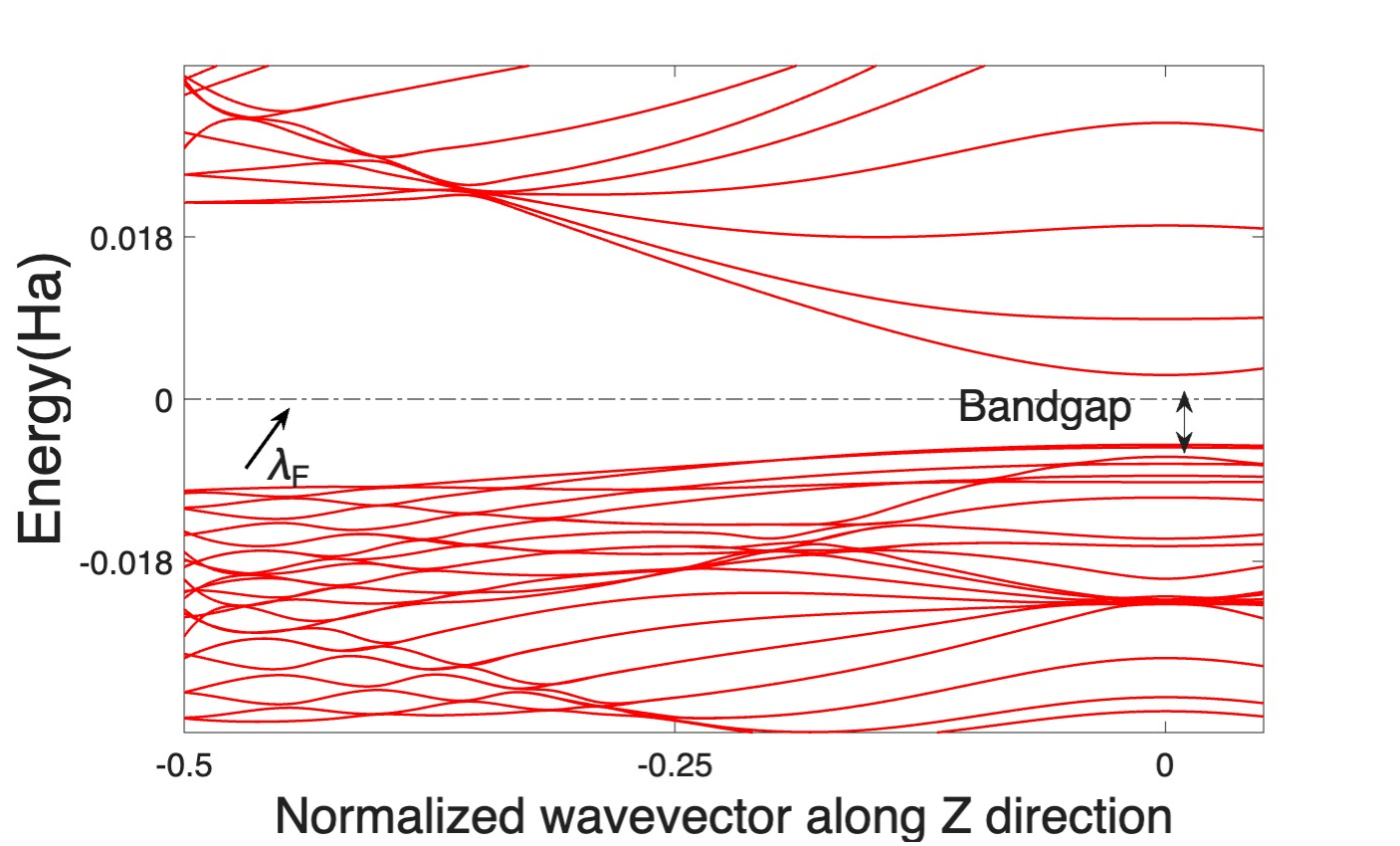}
}\label{supp:fig:zzooms12}} \\
\subfloat[]{
{
\includegraphics[scale =0.35]{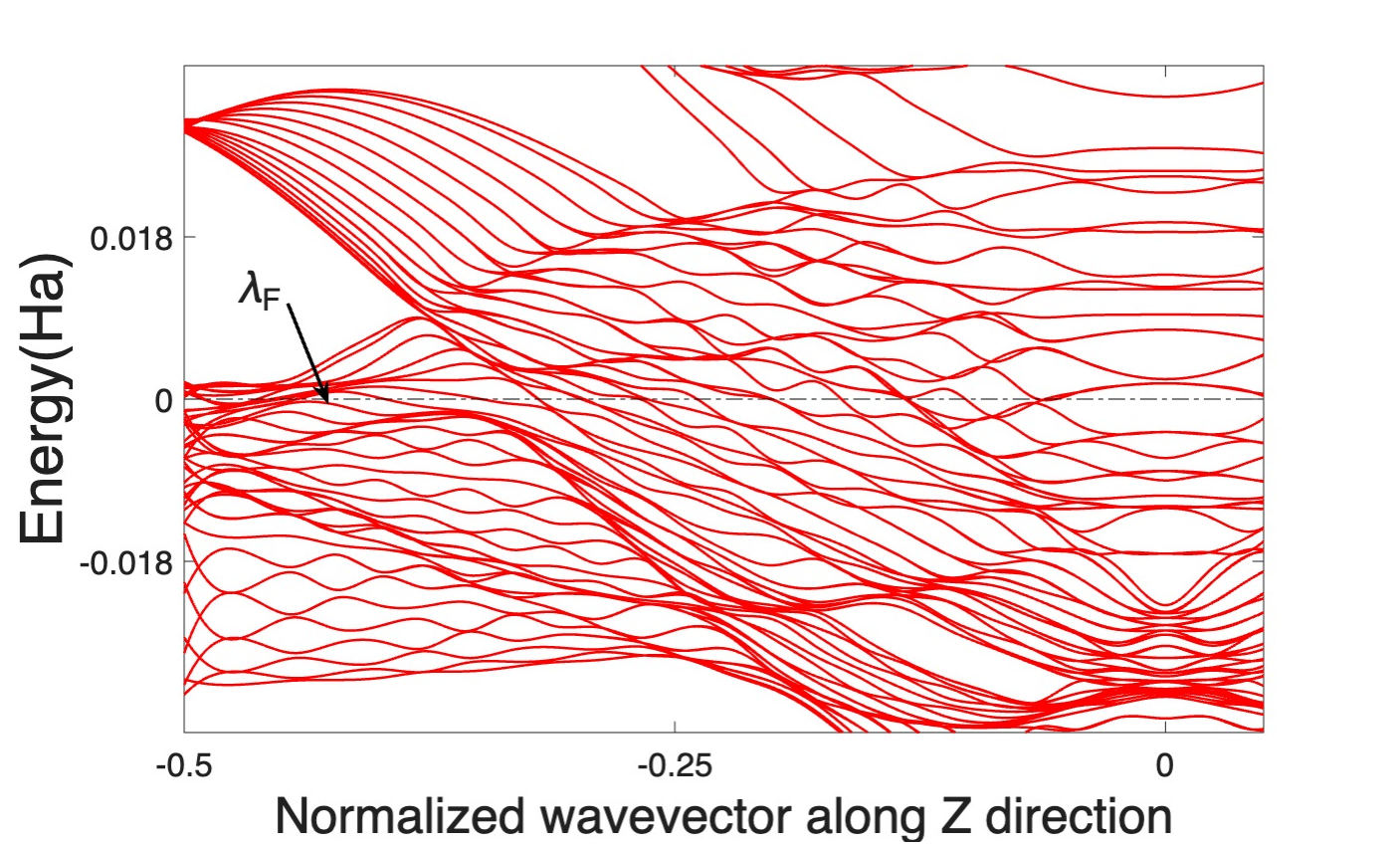}
}\label{supp:fig:zzooms12w}}
\subfloat[]{
{
\includegraphics[scale =0.35]{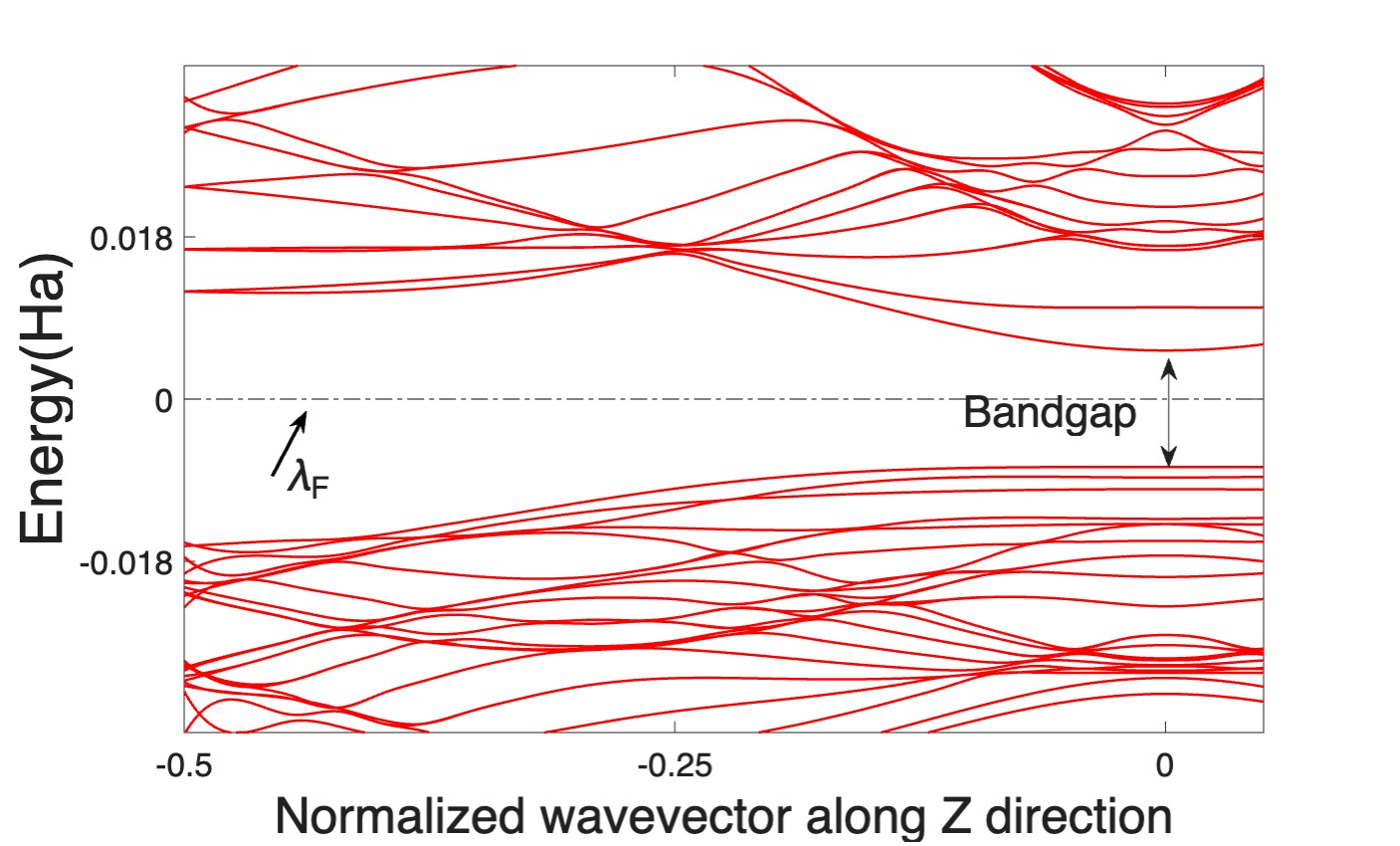}
}\label{supp:fig:zzooms12e}}
    \caption{Magnified band diagrams of Fig.~5 of main text (effect of applied strain).  (a) intermediate $6.35\%$,(b) intermediate $12.34\%$, (c) intermediate $18\%$ and (d) brick wall $24.67\%$. }
    \label{supp:fig:ssscds}
\end{figure}

\begin{figure}[ht!]
    \centering
    \includegraphics[scale=1.8]{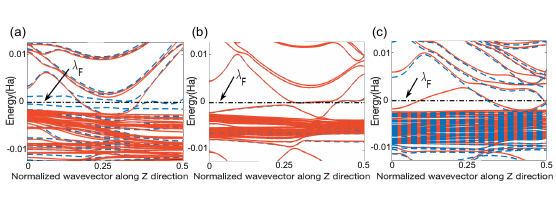}
    \caption{The zoomed-in band diagram around the Fermi level from Fig.~3 of the main text is shown for the (a) $-4\%$ strain, (b) unstrained, and (c) $+4\%$ strain cases, respectively. Spin-up and spin-down channels are represented by solid red and dashed blue lines, respectively.}
    \label{supp:fig:zoom_mag_band}
\end{figure}

\clearpage
\bibliography{Main}

\end{document}

%% file: preamble.tex
\newcommand{\rz}{\mathbb{R}}

\newcommand{\bfa}{{\bf a}}

\newcommand{\bfr}{{\bf r}}

\newcommand{\bfH}{{\bf H}}
\newcommand{\bfI}{{\bf I}}

\newcommand{\bfR}{{\bf R}}

\newcommand{\beq}{\begin{equation}}
\newcommand{\eeq}{\end{equation}}
\newcommand{\beqs}{\begin{eqnarray}}
\newcommand{\eeqs}{\end{eqnarray}}
\newcommand{\beql}{\begin{equation} \label}

\newcommand{\calD}{{\cal D}}

\newcommand{\calF}{{\cal F}}
\newcommand{\calG}{{\cal G}}

\newcommand{\calP}{{\cal P}}

\newcommand{\Z}{\mathbb{Z}}

\newcommand{\hpd}[3]{\frac{\partial^{#3}#1}{\partial#2^{#3}}}

\let\oldFootnote\footnote
\newcommand\nextToken\relax

\renewcommand\footnote[1]{%
    \oldFootnote{#1}\futurelet\nextToken\isFootnote}

\newcommand\isFootnote{%
    \ifx\footnote\nextToken\textsuperscript{,}\fi}